\documentclass[review, 12pt]{elsarticle}

\usepackage{amssymb}
\usepackage{bm}
\usepackage[colorlinks=true, linkcolor=blue, anchorcolor=blue,citecolor=blue]{hyperref}

\usepackage{amsmath}
\usepackage{caption}
\usepackage{subcaption}

\usepackage{cleveref}
\usepackage{multirow}
\usepackage{color,soul}
\usepackage[table]{xcolor}

\usepackage[switch]{}

\usepackage{lineno}

\newcommand\G{{\bf g}}

\newcommand\N{{\bf n}}

\newcommand\F{{\bf f}}

\newcommand\U{{\bf u}}

\newcommand\X{{\bf x}}

\newcommand\be{\begin{equation}}
\newcommand\nd{\end{equation}}
\newcommand\bed{\begin{displaymath}}
\newcommand\ndd{\end{displaymath}}

\newcommand\ba{\begin{array}}
\newcommand\ea{\end{array}}
\newcommand\bea{\begin{eqnarray}}
\newcommand\nda{\end{eqnarray}}

\begin{document}
\sethlcolor{yellow}
\setstcolor{red}
\soulregister\cite7 
\soulregister\ref7

\begin{frontmatter}

\title{A Three-dimensional Edge-Based Interface Tracking (EBIT) Method for Multiphase-flow Simulations}

\author[a]{Jieyun Pan\corref{cor1}}
\ead{yi.pan@sorbonne-universite.fr}

\author[a,b]{Tian Long}
\ead{tian.long@xjtu.edu.cn}

\author[c]{Ruben Scardovelli}
\ead{ruben.scardovelli@unibo.it}

\author[a]{St\'{e}phane Popinet\footnote{}}
\ead{popinet@basilisk.fr}

\author[a,d]{St\'{e}phane Zaleski}
\ead{stephane.zaleski@sorbonne-universite.fr}

\cortext[cor1]{Corresponding author}
\address[a]{Sorbonne Universit\'{e} and CNRS, Institut Jean Le Rond d'Alembert UMR 7190, F-75005 Paris, France}
\address[b]{State Key Laboratory for Strength and Vibration of Mechanical Structures, School of Aerospace, Xi'an Jiaotong University, Xi'an, 710049, PR China}
\address[c]{ DIN - Lab. di Montecuccolino, Universit\`a di Bologna, I-40136 Bologna, Italy}
\address[d]{Institut Universitaire de France, Paris, France}

\begin{abstract}

The Edge-Based Interface Tracking (EBIT) method is a novel Front-Tracking method in which the markers are located on grid edges, and their
connectivity is implicitly represented by a color vertex field. 
This localized representation simplifies topology changes and allows for automatic parallelization. 

In this work, we propose a simplified extension of EBIT to three
dimensions (3D). The directional split scheme used for interface advection in the 
two-dimensional (2D) version is generalized to 3D by decomposing each 3D directional advection step into 2D
advection problems on the two corresponding cube faces.
The same dimension-reduction strategy is applied to connectivity representation, allowing the main 2D algorithms and data structures to be reused and thereby reducing the algorithmic and implementation complexity.

For coupling with the Navier--Stokes equations, volume fractions are reconstructed from the EBIT markers and color vertex field using the Front2VOF geometric method. The resulting volume fractions are then used to compute the fluid properties and surface-tension forces with the Height-Function method.
The 3D EBIT method has been implemented in the free Basilisk framework, with the documented source code available in the online repository.
The method is verified against five benchmark cases: translation with uniform velocity, 3D deformation test, oscillating drop, rising bubble, and bubble merging. The results show good agreement with reference solutions and with the Volume-of-Fluid (VOF) method implemented in Basilisk,
demonstrating the effectiveness of the simplified extension. In addition, the 3D EBIT method exhibits excellent weak scalability, highlighting its potential advantage over traditional front-tracking methods for large-scale parallel computations.
\end{abstract}

\begin{keyword}
Two-phase flows \sep Front-Tracking \sep Volume-of-Fluid  
\end{keyword}

\end{frontmatter}


\section{Introduction}
Multiphase flows, which involve a wide range of scales, are ubiquitous in nature and engineering, appearing in phenomena such as breaking ocean waves,  industrial processes, heat exchange and transport in nuclear reactors, and atomizing liquid jets in combustion engines. 
Numerical simulations provide an essential tool for investigating these complex flows,
which are often difficult to reproduce in lab experiments.
Over the past decades, tremendous progress has been made in both numerical algorithms and computing hardware.
Notable advancements include adaptive mesh refinement (AMR) based on quadtree and octree meshes,
which allow for higher resolution only in regions containing small-scale 
flow structures or high gradients of physical variables,
as well as the advent of exascale supercomputers and graphics processing units (GPUs) for scientific computing.
These achievements make Direct Numerical Simulations (DNS) of certain multiscale multiphase flows increasingly affordable. Obtaining the statistical properties of small-scale dynamics through such simulations is of great importance, as these statistics provide the basis for developing subgrid-scale models for Euler--Euler simulations in realistic industrial scenarios.

Accurate prediction of the interface motion is one of the most crucial challenges
in the field of multiphase flow simulations. It can be broadly divided into two
problems: the kinematic problem, which determines the motion of the interface 
separating the fluid phases, given the velocity field and the rate of phase change,
and the dynamic problem, which solves the momentum and energy conservation equations based on the fluid properties.
For the kinematic problem, several methods have been developed that
are generally classified as Front-Capturing methods, including
Volume-of-Fluid (VOF)
\cite{Hirt_1981_39, Brackbill_1992_100, Lafaurie_1994_113, Scardovelli_1999_31},
Level-Set (LS) \cite{Osher_1988_79, Osher_2001_169},
and Front-Tracking methods \cite{Unverdi_1992_100, Tryggvason_2011_book}. 

In Front-Capturing methods, a tracer or marker function $f   (\X,t)$ 
is integrated in time with a prescribed velocity field $\U(\X,t)$.
The function $f$ may be a Heaviside function in VOF methods
or a smooth distance function in LS methods.
The localized feature of the tracer function facilitates the parallelization, making these methods generally computationally efficient.
However, a key limitation of these methods lies in their difficulty in resolving
subgrid-scale (SGS) structures, since the interface elements are not explicitly tracked
but rather reconstructed from the tracer function.

Various algorithms have been proposed to address this limitation,
primarily within the framework of VOF methods.
The R2P method \cite{Chiodi_2020_phd, Han_2024_519}
reconstructs the interfaces within a single cell with two planes of arbitrary
relative orientation to capture a variety of SGS structures, such as thin films
and the closure of sheet rims.
The Moment-of-Fluid (MOF) method \cite{Dyadechko_2008_227, Jemison_2015_285, Shashkov_2023_479, Shashkov_2023_494, Chiodi_2025_528, Hergibo_2025_530} computes and evolves in time
the zeroth, first, and second moments of the fragment of material within
a computational cell. This richer geometric representation provides
more accurate reconstructions of SGS.

In Front-Tracking methods, the interface or ``front'' is represented by a set of
Lagrangian markers and their connectivity. These markers may be connected with
straight line segments \cite{Unverdi_1992_100} or global splines \cite{Popinet_1999_30},
and are advected with a given velocity field. Their connectivity must be updated
when topology changes occur, such as coalescence or breakup.
An introduction to the most popular Front-Tracking methods can be found in \cite{Tryggvason_2011_book}.
These methods provide a direct and accurate approach for predicting SGS dynamics.
Furthermore, they enable a straightforward distinction between slender objects
which share a similar volume fraction distribution, such as unbroken thin ligaments,
strings of small particles, and broken ligaments. This distinction is of great
relevance when analyzing statistically highly fragmented flows \cite{Chirco_2022_467}. 
Furthermore, geometrical properties of the fluid structure, such as skeletons \cite{chen2022characterizing}, are crucial for developing reduced-order models,
and are most naturally represented by Front-Tracking methods.
However, the global connectivity information used to represent the interface 
makes parallel computing
much more challenging when compared to Front-Capturing methods, in particular
in the presence of topology changes.

In addition to the SGS interface structures, multiscale features also arise at another level.
In flows that involve strong heat and mass transfer, such as nucleate boiling at
high Jakob numbers \cite{Bures_2024_271, Torres_2024_497, Long_2025_1020} or the dissolution of CO$_2$ bubbles in water \cite{Aboulhasanzadeh_2012_75,  Aboulhasanzadeh_2013_101, Weiner_2017_347, Claassen_2019_66, Wang_2026_196}, an extremely thin thermal or concentration
boundary layer forms at the interface. Unlike SGS structures, which are generally
confined to limited regions, these thin boundary layers often cover extensive portions of the interfaces.
Moreover, their thickness is generally several orders of magnitude smaller
than that of the surrounding fluid structures.
These characteristics make DNS of such flows extremely challenging, even with AMR, and necessitate the integration of subgrid boundary layer models.

Front-Capturing and Front-Tracking methods present different levels of complexity
when coupled with surface evolution equations, such as boundary layer models for multiscale simulations \cite{Weiner_2017_347, Aboulhasanzadeh_2012_75,  Aboulhasanzadeh_2013_101, Claassen_2019_66} and surfactant transport equations \cite{James_2004_201, Renardy_2002_21, Jain_2024_515, Farsoiya_2024_9, Muradoglu_2008_227, Muradoglu_2014_274}.
In VOF methods, the heat or mass concentration profiles 
within boundary layers must be evolved consistently
with the volume-fraction advection scheme \cite{Weiner_2017_347} due to the implicit representation of interfaces.
In contrast, FT methods provide an explicit Lagrangian representation of the interface. Surface quantities can therefore be attached directly to interfacial markers or elements and evolved along with the moving interface.
This leads to a more natural formulation of governing equations for interfacial
concentration profiles and facilitates the discretization of surface differential operators \cite{Aboulhasanzadeh_2012_75,  Aboulhasanzadeh_2013_101, Claassen_2019_66}.
The advantage is particularly clear for surfactant transport problems, where
surface advection--diffusion terms and the surface gradients are naturally defined on the tracked interface \cite{Muradoglu_2008_227, Muradoglu_2014_274}.

Several efforts have been made to combine the strengths of global methods,
such as Front-Tracking, with those of local methods, such as VOF or LS.
The combination of the VOF method with marker points allows a smooth representation
of interfaces without discontinuities \cite{Aulisa_2003_188, Aulisa_2004_197}
at cell faces, or tracking of SGS structures \cite{Lopez_2005_208}.

Hybrid approaches that combine Front-Tracking and LS methods include
the Level Contour Reconstruction Method (LCRM) \cite{Shin_2002_180, Shin_2005_203, Shin_2007_21}, developed for structured meshes,
and the hybrid LEvel set/fronNT tracking method (LENT) \cite{Maric_2015_113},
designed for unstructured meshes. 
In these methods, the interface is tracked by a set of disconnected triangular
elements rather than a collection of marker points.
During the reconstruction phase, an LS function is computed from the geometric information of the elements and is subsequently used to generate a new set of triangles. 
In the reconstruction procedure, topology changes are handled automatically,
and the resulting triangular elements are naturally distributed across processors,
which facilitates parallelization.
In LCRM, no connectivity information is stored, whereas LENT retains
neighboring relationships between two elements, which are recovered during the
localized reconstruction step. As a result, these methods avoid the need to
communicate complex connectivity information, as required in 
traditional front-tracking methods, thereby further simplifying parallel implementation.
The reconstruction process can also be performed using a purely geometric
approach without resorting to a LS function, which leads to the Local Front Reconstruction Method (LFRM) \cite{Shin_2011_230}.

We have recently presented a similar method, the Edge-Based Interface-Tracking (EBIT) method \cite{Chirco_2022_95, Pan_2024_508, Pan_2026_557}. 
In EBIT, the position of the interface is tracked by marker points located on the
edges of an Eulerian grid, with connectivity information implicitly represented by a color vertex field. 
Moreover, markers in the EBIT method are bound to the Eulerian grid by a local reconstruction of the interface at every time step. Therefore, the Eulerian grid and Lagrangian markers can be distributed to different processors by the same routine, allowing for automatic parallelization.

The core algorithm for interface advection has been discussed in \cite{Chirco_2022_95, Pan_2024_508, Pan_2026_557}, 
while the coupling algorithm between the EBIT method and the Navier--Stokes solver was 
presented in \cite{Pan_2024_508} for the 2D case, along with techniques for
improving the accuracy of mass conservation and handling topology changes.
The 2D EBIT method has been implemented in the free Basilisk platform, 
demonstrating its automatic parallelization when coupled with quadtree meshes.

Note that a variant of the EBIT method on triangular meshes has been proposed by Wang et 
al.~\cite{Wang_2025_520} in 2D.
In their approach, up to two markers are allowed per edge of a triangular cell,
enabling the tracking of sub-grid ligament structures. An area correction algorithm
with an additional in-cell marker is presented to improve the accuracy of mass conservation.
However, this approach does not guarantee continuity of the interface across cell faces, limiting its compatibility with advanced surface tension models, such as the integral method \cite{Popinet_1999_30, Abu_Al_Saud_2018_371}, 
known to be both momentum-conserving and well-balanced.

For 2D EBIT methods, both split \cite{Chirco_2022_95,
Pan_2024_508}
and unsplit \cite{Wang_2025_520, Pan_2026_557} advection schemes have been developed.
The unsplit schemes resemble traditional FT methods, offering
easier coupling with high-order time-integration schemes for better accuracy of 
mass conservation.
However, we found that the split scheme, which allows full reuse of the existing 2D reconstruction and topology-change algorithms, facilitates the extension to 3D.

In this work, we present the detailed algorithms required to extend the EBIT method to 3D.
Using a split scheme, we decompose the 3D advection problem into a
sequence of 2D problems and leverage the existing 2D techniques for interface
reconstruction and topology identification. This simplified strategy avoids
the development of complex 3D reconstruction algorithms, but introduces
additional topology ambiguities, which are nevertheless shown to be consistent
with the current marker layout. 
We adopt the original coupling method
with the Navier--Stokes equations for multiphase flows, in which the
volume enclosed by the interface within each cubic cell is computed using a
geometric Front2VOF (F2V) method. 
Subsequently, the curvature and surface tension are calculated 
using the Height-Function method based on the resulting volume fraction field.

The new 3D EBIT method has been implemented in Basilisk and remains
fully compatible with octree-based AMR.
This work represents the first extension of the EBIT
method to 3D, and its objective is to propose a succinct and efficient strategy for extending
the EBIT method to 3D without incurring significant algorithmic complexity.

The paper is organized as follows: the kinematics of the EBIT method is described in Section \ref{Numerical method}.
This includes a brief review of the 2D advection algorithm and the color vertex 
field used to implicitly represent connectivity, together with their extension to 3D.
The topology ambiguities arising from this simplified extension strategy are
also discussed.
The coupling algorithm between the EBIT interface description and the multiphase fluid dynamics is then presented. 
In Section \ref{Numerical results and discussion}, the EBIT method is verified through a series of benchmark tests.
The results obtained with the combined EBIT method are presented and compared with those from the Piecewise Linear Interface Calculation Volume-of-Fluid (PLIC-VOF) method available in Basilisk \cite{Popinet_2009_228}.

\section{Numerical method} \label{Numerical method}
In this section, we first briefly revisit the essential concepts of the 2D EBIT
method and the interface advection scheme, as the extension to 3D builds 
directly on these algorithms.
We then present the 3D EBIT method in detail, followed by a discussion of
the topology ambiguities which result from this straightforward generalization.

\subsection{The 2D EBIT method}

\begin{figure}[!htb]
\begin{center}
\begin{tabular}{cc}
\includegraphics[width=0.45\textwidth]{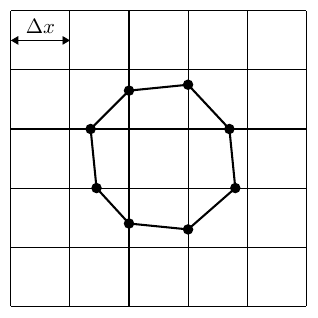} &
\includegraphics[width=0.45\textwidth]{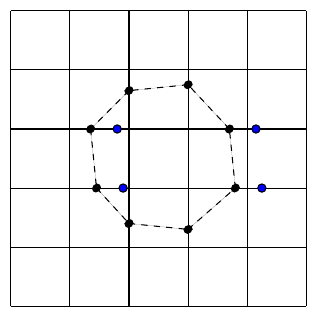}\\
(a) & (b) \\[6pt]
\includegraphics[width=0.45\textwidth]{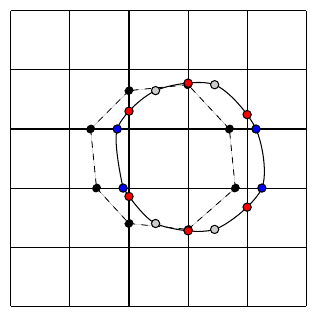} &
\includegraphics[width=0.45\textwidth]{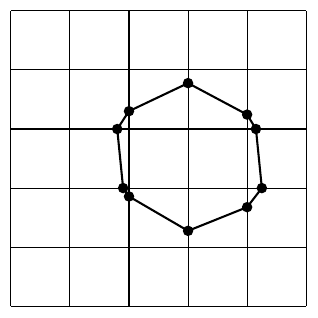}\\
(c) & (d) 
\end{tabular}
\end{center}
\caption{One-dimensional advection of the EBIT method along
the $x$-axis: (a) initial interface line; (b) advection of
the markers on the grid lines aligned with the velocity component
(blue points); (c) advection of the unaligned markers (gray points)
and computation of the intersections with the grid lines (red points);
(d) interface line after the 1D advection.}
\label{Fig_EBIT_advection}
\end{figure}

In the EBIT method, the interface is represented by a set of marker points placed on the grid lines. In order to keep the markers on the grid lines, it is necessary to compute the intersection points between the interface line and the grid lines at the end of each advection step. 
The equation of motion for a marker point at position $\X_i$ is
\begin{gather}
\frac{d \X_i}{dt} = \U_i,
\label{Eq_marker}
\end{gather}
which is discretized by a first-order explicit Euler method
\begin{gather}
\X^{n + 1}_i= \X^{n}_i + \U^{n}_i \Delta t,
\label{Eq_marker_dis}
\end{gather}
where the velocity $\U^{n}_i$ at the marker position $\X^{n}_i$ is calculated by a bilinear interpolation.

For a multi-dimensional problem, a split method is used to advect the interface (see Fig.~\ref{Fig_EBIT_advection}), which has been described in \cite{Chirco_2022_95, Pan_2024_508}. The marker points placed on the grid lines that are aligned with the velocity component of the 1D advection are called \textit{aligned markers}, while the remaining ones are called \textit{unaligned markers}. Starting from the initial position at time step $n$, the new position of the aligned markers is obtained by Eq.~\eqref{Eq_marker_dis} (blue points of Fig.~\ref{Fig_EBIT_advection}b). To compute the new unaligned markers, we first advect them using again Eq.~\eqref{Eq_marker_dis}, obtaining in this way the gray points of Fig.~\ref{Fig_EBIT_advection}c. Finally, the new position of the unaligned markers (red points of Fig.~\ref{Fig_EBIT_advection}d) is obtained by a circle fit method to improve the accuracy of the mass conservation.

\begin{figure}[!htb]
\begin{center}
\begin{tabular}{cc}
\includegraphics[width=0.47\textwidth]{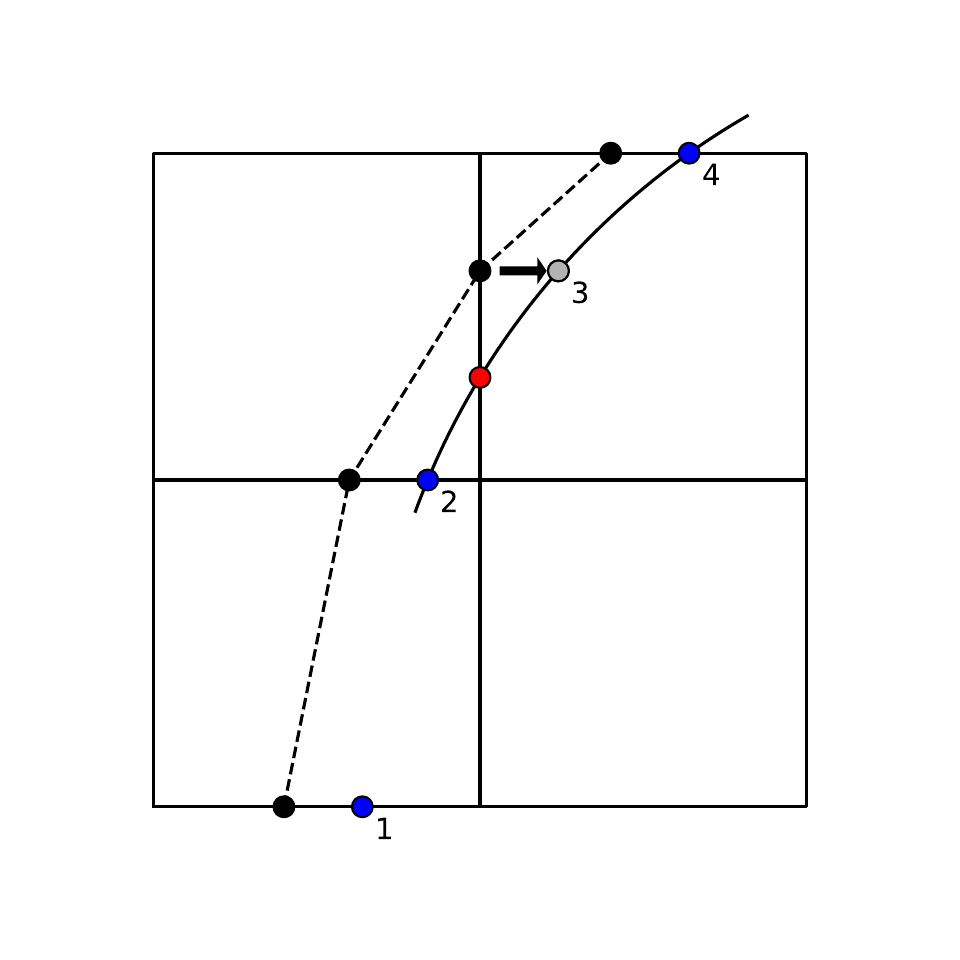} &
\includegraphics[width=0.47\textwidth]{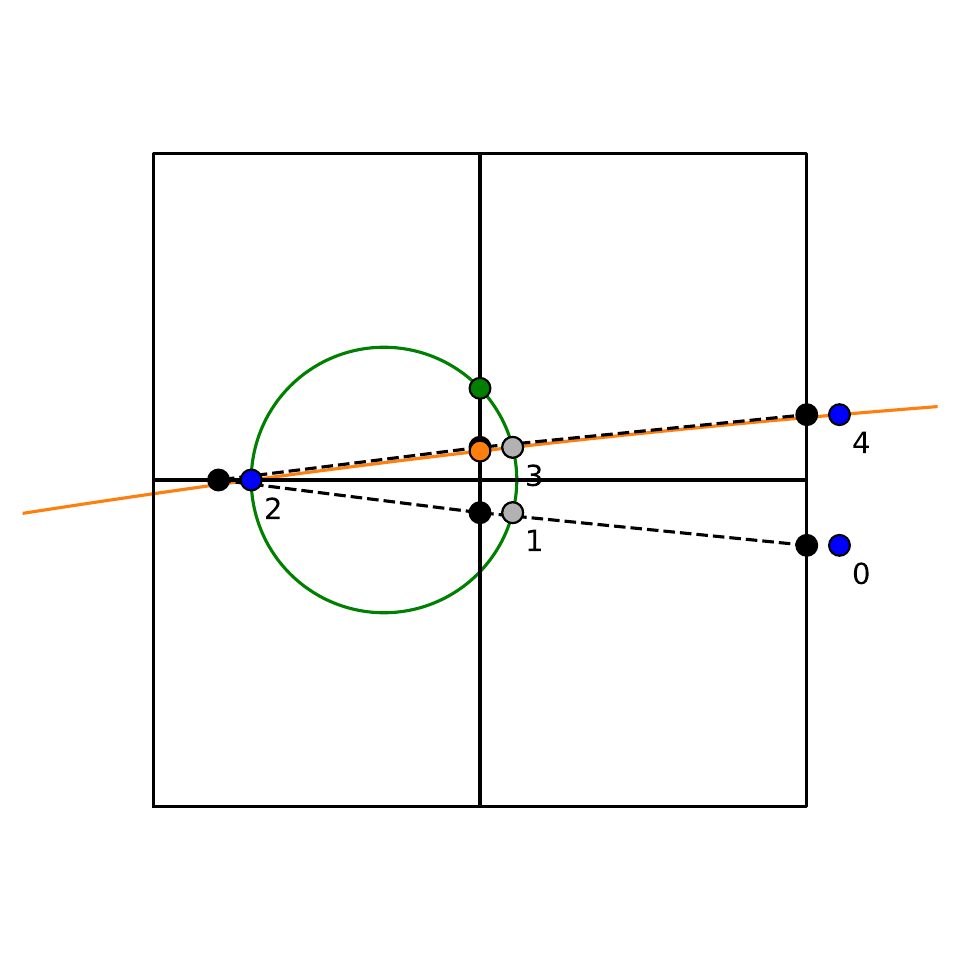}\\
(a) & (b) 
\end{tabular}
\end{center}
\caption{Circle fit for determining the position of the unaligned marker (red point): (a) general case; (b) special case at the tip of a ligament, where the circle with the larger radius (orange solid line) is used to compute the intersection point.}
\label{Fig_circle_fit}
\end{figure}

Generally, the position of an unaligned marker is determined using the average of the results from two circle fits (1-2-3 and 2-3-4 in Fig.~\ref{Fig_circle_fit}a).
However, this simple averaging may artificially produce a bulbous shape near
the tip of thin ligaments, as illustrated by the green circle in
Fig.~\ref{Fig_circle_fit}b, thereby significantly degrading the accuracy in such cases.
In our previous work on the unsplit advection scheme \cite{Pan_2026_557}, we proposed an ad hoc criterion to select the appropriate circle fit result. 
Specifically, when the ratio between the radii of the two fitted circles, $r_{\max} / r_{\min}$, exceeds a threshold value of $10$, the circle with the larger radius is used to compute the intersection point that determines the new position of the unaligned marker. 
The threshold value was determined based on numerical experiments with kinematic test cases. A lower threshold may result in less accurate reconstructions of a smooth interface, whereas a larger value may limit the ability of the algorithm to detect tip structures with strong curvature variations.
A detailed sensitivity analysis that supports this choice is provided in Ref.~\cite{Pan_2026_557}.

\subsection{Color Vertex}

\begin{figure}
\centering
\includegraphics[width=\textwidth]{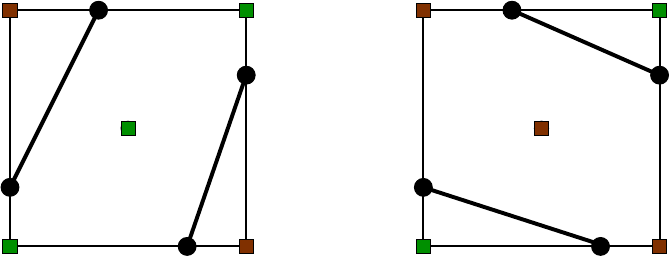}
\caption{Two color vertex configurations (brown and green squares) to select a different connectivity 
in the same set of markers.}
\label{Fig_color_vertex}
\end{figure}

In the EBIT method, marker connectivity is implicitly represented by
a color vertex field defined at both the cell corners and the cell center \cite{Singh_2007_224}.
In the 2D case shown in Fig.~\ref{Fig_color_vertex}, the value of the color vertex field indicates the fluid phase in the corresponding region of the cell. 
Furthermore, the distribution of color vertex values within a cell is used to resolve
ambiguous configurations in 2D problems, when there are four markers, one on
each cell edge. 
In other words, the distribution of color vertex values within each cell uniquely defines the corresponding topological configuration, thereby enabling unambiguous reconstruction of the interface segments.

\begin{figure}
\centering
\begin{subfigure}[b]{\textwidth}
\centering
\includegraphics[width=\textwidth]{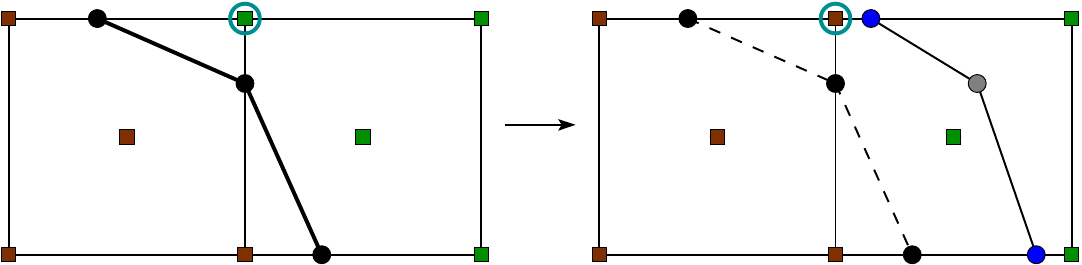}
\end{subfigure}
\caption{Update of the color vertex value on a cell corner: its value changes, 
from green to brown, as the marker is advected across the grid-line intersection.}
\label{Fig_color_vertex_update_corner}
\end{figure}

As the interface is advected, the color vertex field should also be updated 
accordingly to ensure that the implicit connectivity information is retained. 
The schematic, which shows the updating rule for a color vertex located on a cell corner, is presented in Fig.~\ref{Fig_color_vertex_update_corner}.
A more comprehensive algorithm for updating the color vertex field at the cell center is provided in \cite{Pan_2024_508}

\begin{figure}[!htb]
\centering
\begin{tabular}{cc}
\multicolumn{2}{c}{\includegraphics[width=0.9\textwidth]{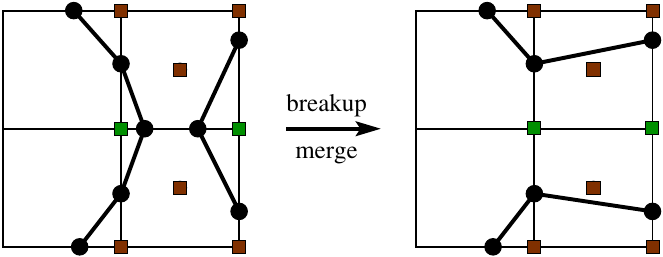}}\\
\multicolumn{2}{c}{(a)}\\
\includegraphics[width=0.45\textwidth]{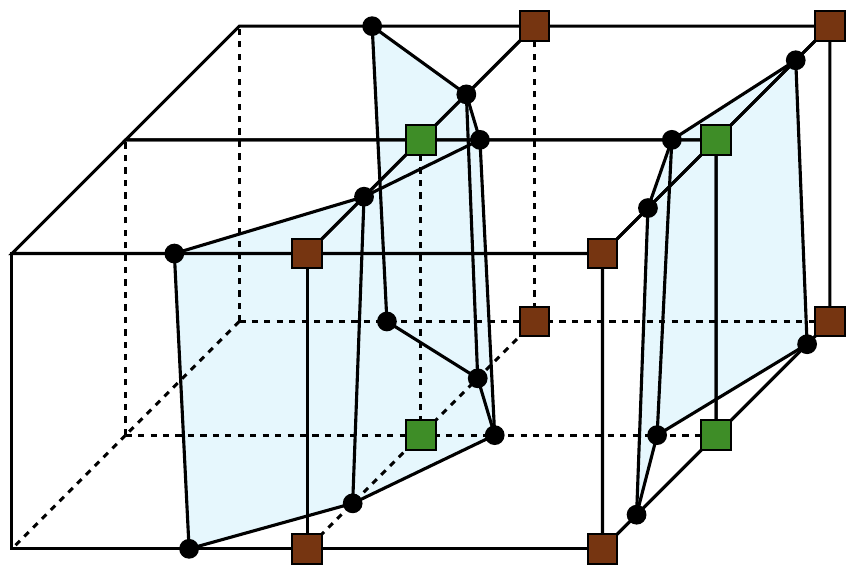}&
\includegraphics[width=0.45\textwidth]{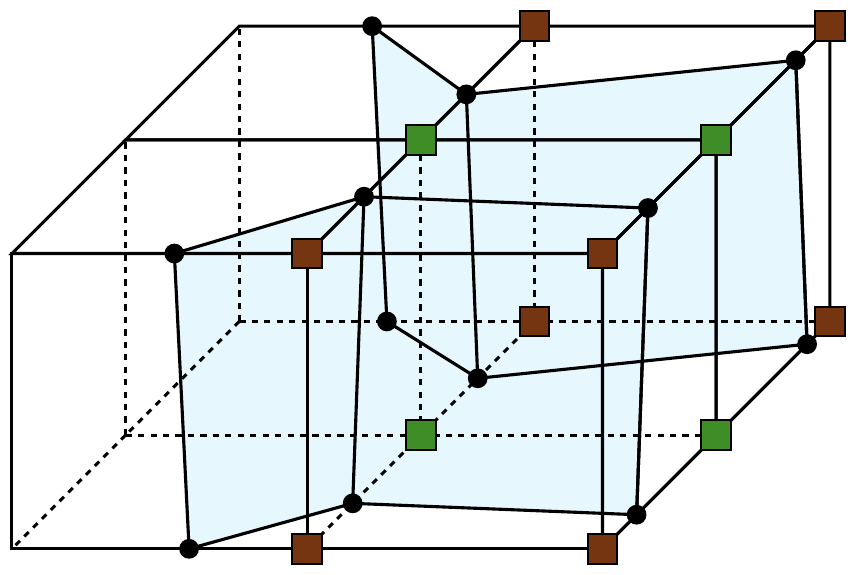}\\
\multicolumn{2}{c}{(b)}
\end{tabular}
\label{Fig_topo_change_2}
\caption{Topology change mechanism: (a) 2D case; (b) 3D topology change obtained by applying the 2D topology change on the cube faces (the colored central vertex is omitted for clarity).}
\label{Fig_topo_change}
\end{figure}

Additionally, the color vertex field provides an automatic mechanism 
for topology change in the EBIT method.
In the present EBIT method, only one marker is allowed on each cell edge,
which is indicated by different values of the color vertex field at the two edge endpoints. 
Consequently, when two markers move onto the same edge, as shown in
Fig.~\ref{Fig_topo_change}a, both markers are automatically removed,
because the color vertex values at the corresponding endpoints become
identical, indicating that no marker should exist on that edge. 
Moreover, the ``surviving'' markers within the cell will be reconnected automatically, see again Fig.~\ref{Fig_topo_change}a. This reconnection process naturally produces ligament breakup or droplet coalescence during
interface advection. As a direct consequence, the volume occupied by the reference phase either decreases or increases, and droplets or bubbles smaller than the grid size tend to be removed. 

It should be noted that this mechanism yields orientation-dependent topology changes, since only interfaces approaching each other along directions aligned
with the grid lines trigger a topology change. Interfaces approaching along diagonal directions, in contrast, do not induce such changes. This behavior results from the current restriction of allowing only one marker per cell edge, which is adopted to maintain algorithmic simplicity.
The symmetry of topology change in the EBIT method could be improved by relaxing the restriction on the number of markers per edge. This extension would
also enable the representation of SGS interfacial structures and allow 
automatic topology changes to be governed by a physically based criterion \cite{Chirco_2022_467}.

\subsection{Extension to 3D}

\begin{figure}[!hbt]
\begin{center}
\begin{tabular}{c}
\includegraphics[width=0.9\textwidth]{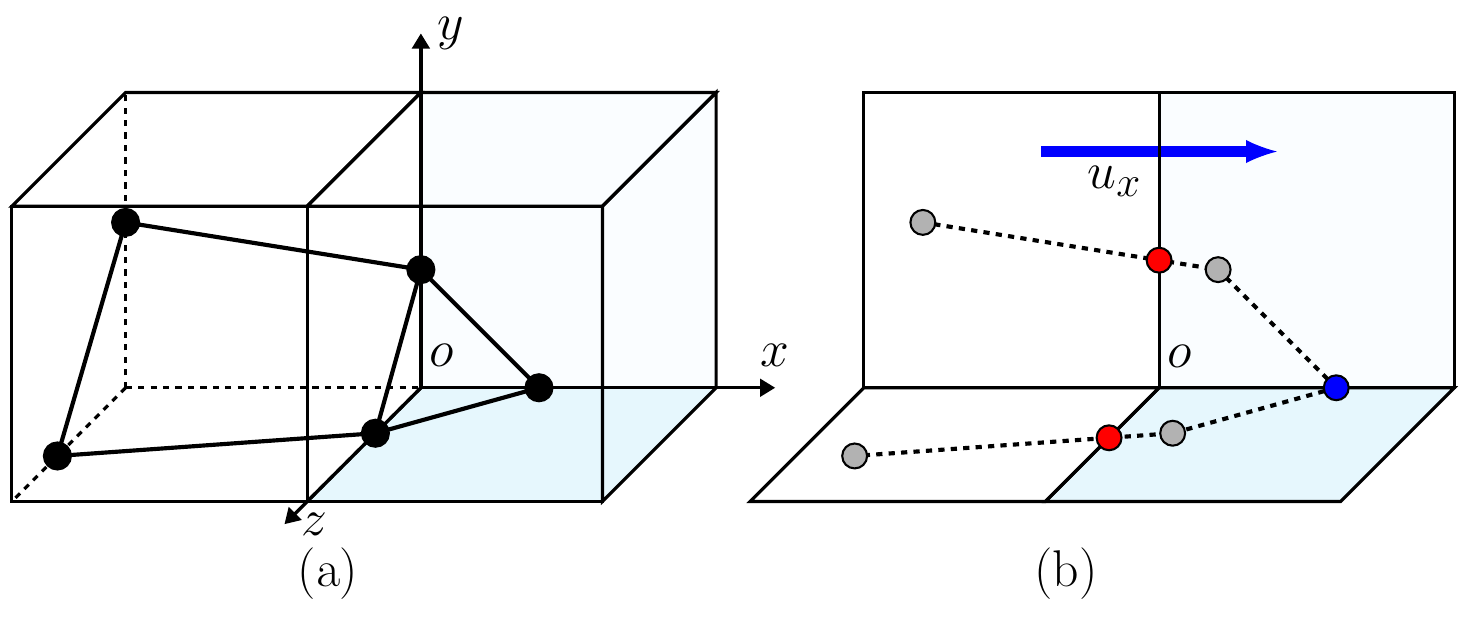}\\
\end{tabular}
\end{center}
\caption{One-dimensional advection of the 3D EBIT method:
 (a) initial interface; (b) decomposition of the 3D advection problem.}
\label{Fig_EBIT_3D}
\end{figure}

In the 3D EBIT method, the split scheme is retained to simplify the implementation and enable direct reuse of the subroutines originally developed for the 2D version.
For the 1D advection along the $x$-direction, as shown in Fig.~\ref{Fig_EBIT_3D}a, the 3D problem is decomposed into two simpler 2D problems on the cell faces highlighted in cyan.
In this case, the markers located on the edges aligned with
the $y$- and $z$-directions are classified as unaligned markers. 
Their reconstruction follows the same procedure illustrated in Fig.~\ref{Fig_EBIT_advection}, applied face-by-face.
This requires only the previously developed 2D circle fit method,
eliminating the need for a complex 3D fitting method.
The reconstruction process for $x$-direction advection is exhibited in Fig.~\ref{Fig_EBIT_3D}b,
including the advection direction and the corresponding marker points
(\url{https://basilisk.fr/sandbox/jieyun/src/ebit-3d.h\#one-dimensional-advection}).

\begin{figure}[!hbt]
\begin{center}
\begin{tabular}{cc}
\includegraphics[width=0.4\textwidth]{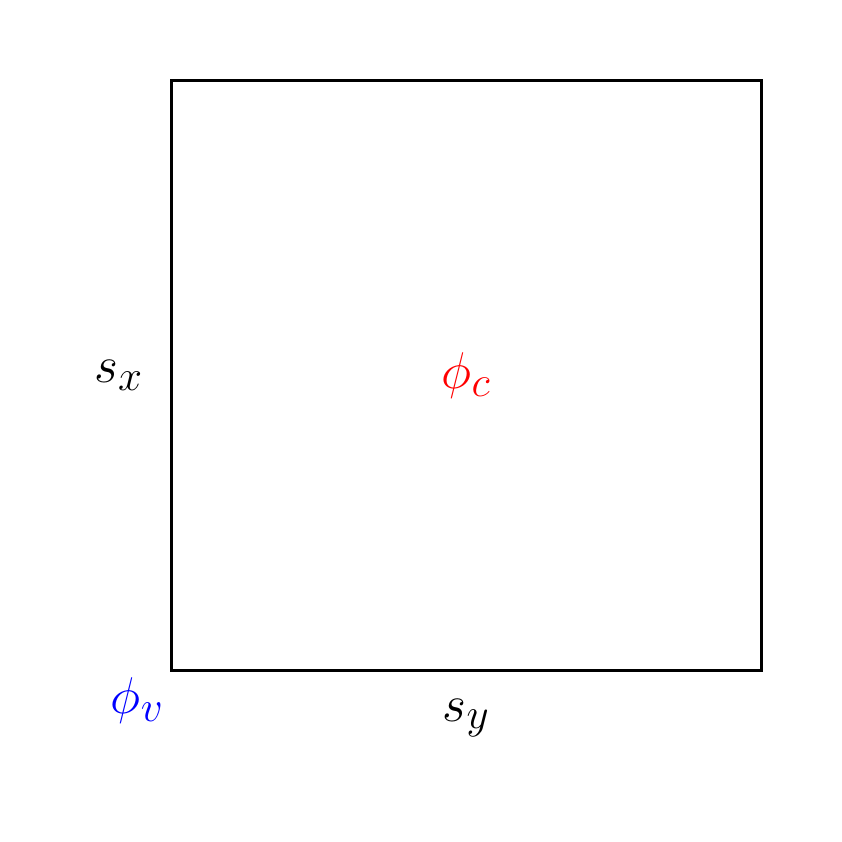}&
\includegraphics[width=0.4\textwidth]{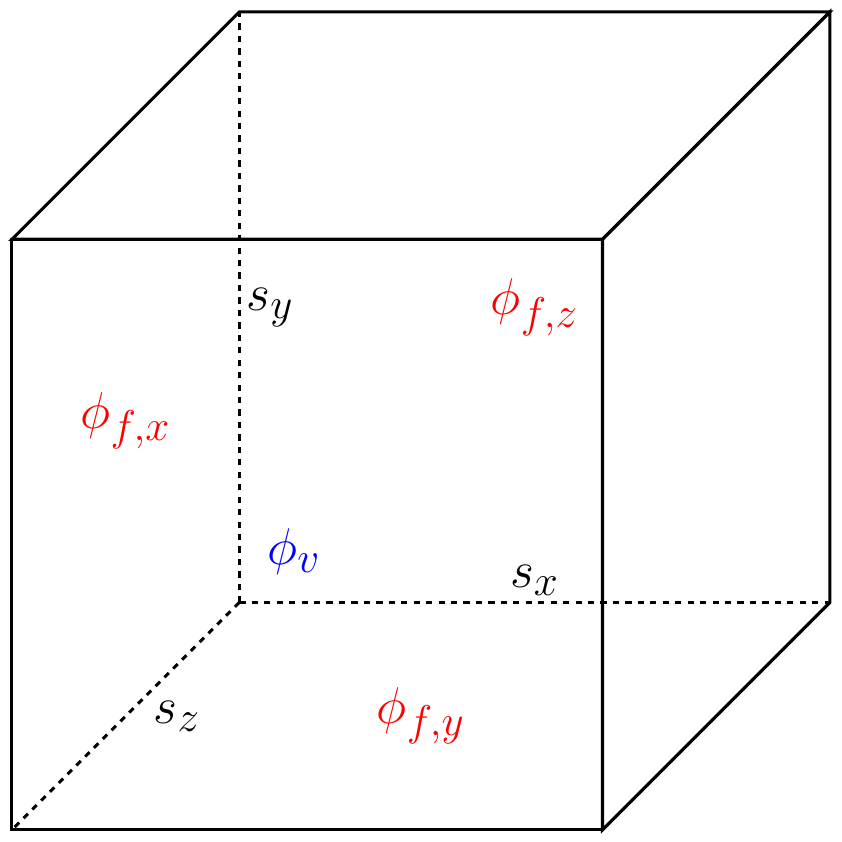}\\
(a) & (b)
\end{tabular}
\end{center}
\caption{Variables used to track the interface within a given cell $(i, j)$ in 2D and $(i, j, k)$ in 3D in the EBIT method. Cell indices are omitted in the figure for clarity. (a) In 2D, one cell-face field ($s_x, s_y$) stores the marker positions, while one cell-vertex field $\phi_v$ and one cell-centered field $\phi_c$ encode the interface connectivity; (b) In 3D, one cell-edge field ($s_x, s_y, s_z$) stores the marker positions, while one cell-vertex field $\phi_v$ and one cell-face field ($\phi_{f,x}, \phi_{f,y}, \phi_{f,z}$) encode the interface connectivity.}
\label{Fig_EBIT_data_structure}
\end{figure}

In 3D, the distribution of color vertex values on the six faces of a cubic cell defines the topology configuration within the cell.
Once the interface lines on each face are reconstructed, they are grouped to
identify closed loops, which represent interface segments within the cell
(\url{https://basilisk.fr/sandbox/jieyun/src/ebit-3d.h\#interface-segments-reconstruction}).
Since the color vertex field in 3D is defined only at the cell corners and
face centers, identical to the 2D arrangement, and a split scheme is employed for marker advection, the original algorithm for evolving the color vertex field can be applied directly on the cube faces.
This close correspondence between the 2D and 3D data structures is summarized in Fig.~\ref{Fig_EBIT_data_structure}, which illustrates the variables used to track the interface in the EBIT method.

It is worth noting that the topology of 3D interfacial structures within a cell
cannot, in general, be uniquely defined based solely on the facewise connectivity,
introducing additional topology ambiguities in 3D.
Nevertheless, the present EBIT method automatically selects one of the
admissible configurations, thereby enabling a deterministic interface
reconstruction. These topology ambiguities are discussed in more detail in the following section.

Based on this facewise representation of 3D interface segments, topology
changes in 3D are handled using a face-by-face procedure.
Specifically, topology changes are first applied independently on each cube face
according to the criterion introduced for the 2D case, for example, on the upper and 
lower cube faces shown in Fig.~\ref{Fig_topo_change}b. Once the updated interface
lines on all faces have been reconstructed, the corresponding 3D interface configuration within the
cell is then obtained (\url{https://basilisk.fr/sandbox/jieyun/src/ebit-3d.h\#topology-change}).

\begin{figure}[hbt!]
\begin{center}
\begin{tabular}{cc}
\includegraphics[width=0.45\textwidth]{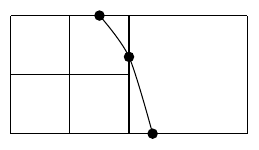}&
\includegraphics[width=0.45\textwidth]{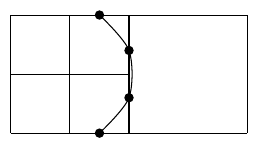}\\
(a) & (b)\\
\includegraphics[width=0.45\textwidth]{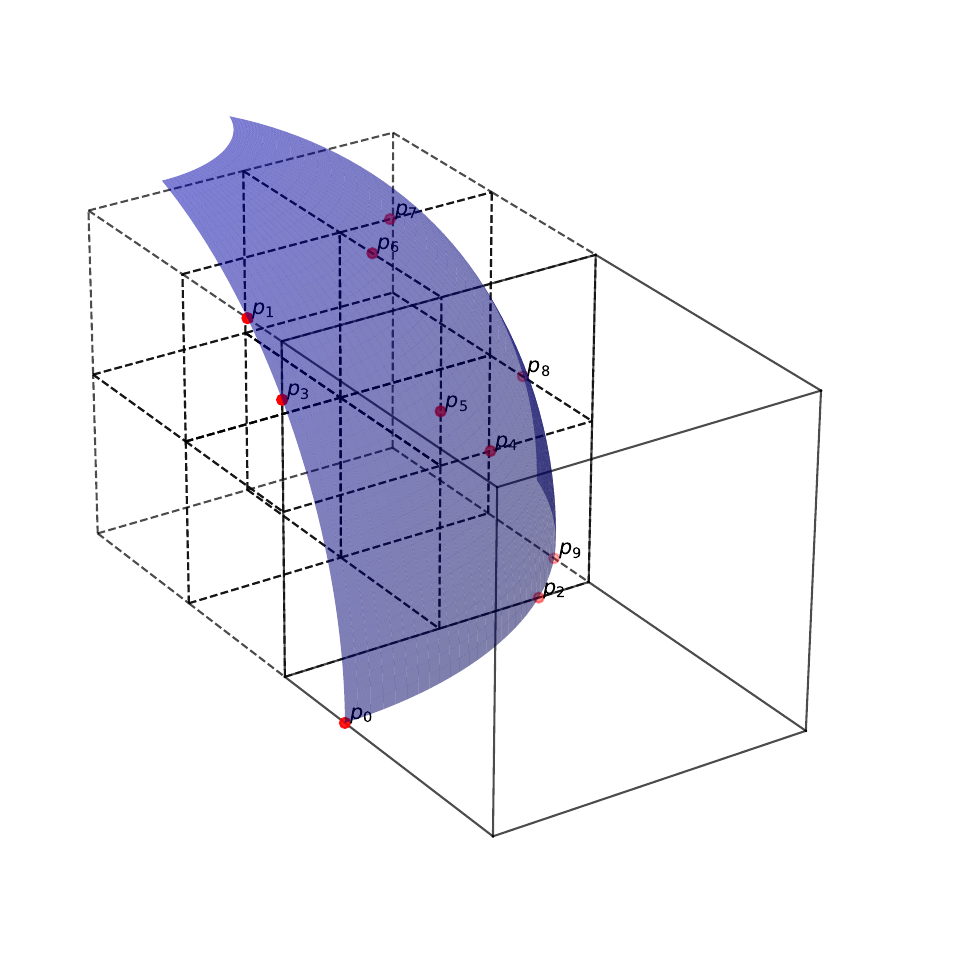}&
\includegraphics[width=0.45\textwidth]{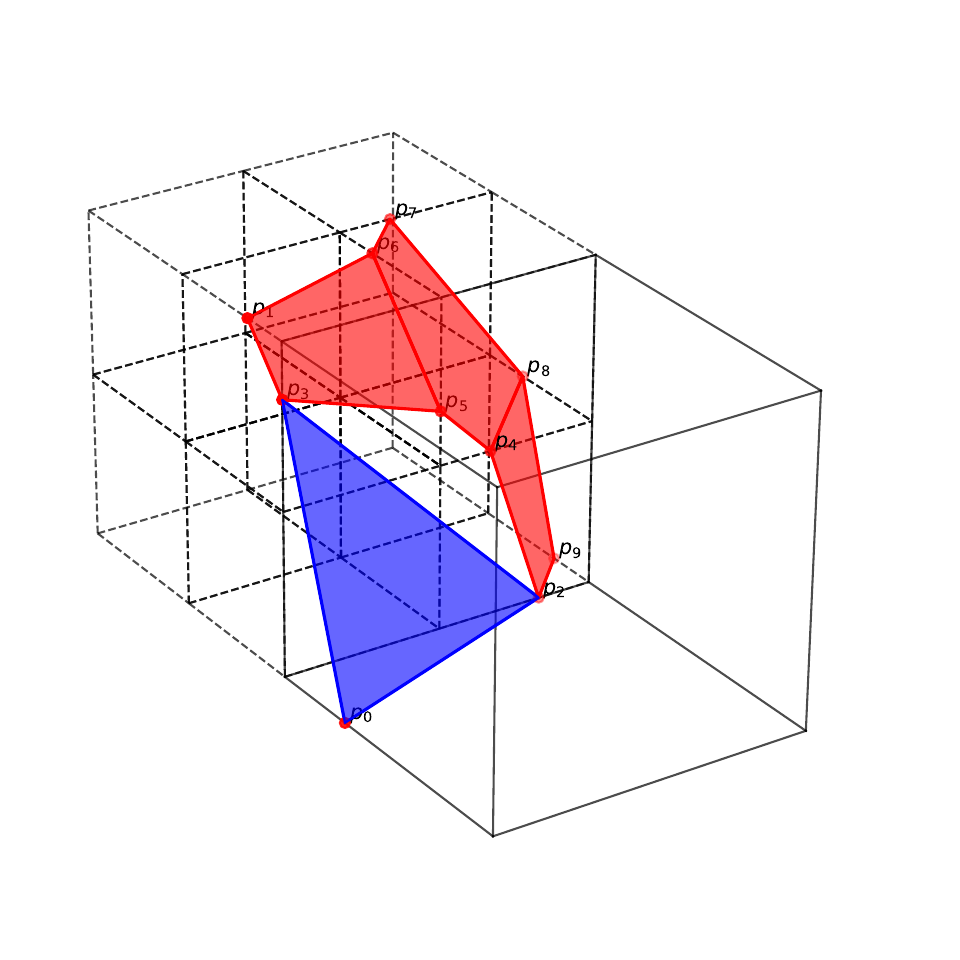}\\
(c) & (d)
\end{tabular}
\end{center}
\caption{Interface reconstruction at refinement boundaries on quadtree/octree meshes. (a) Consistent and (b) inconsistent 2D configurations. (c) A smooth 3D interface and corresponding EBIT markers; (d) reconstructed interface segments on the fine (red) and coarse (blue) sides, showing a gap across the refinement boundary.}
\label{Fig_EBIT_AMR}
\end{figure}
A key advantage of the EBIT method is its seamless coupling
with tree-based AMR \cite{Popinet_2015_302}, 
particularly in conjunction with MPI-based parallelization, which is crucial for efficient multiscale simulations.

Due to the restriction on the number of markers per edge, mesh refinement near interfacial cells must be handled carefully to avoid inconsistencies across refinement boundaries. In 2D, this issue can be identified directly from the marker configuration. For example, the coarse cell in Fig.~\ref{Fig_EBIT_AMR}b violates the one-marker-per-edge constraint, whereas the configuration in Fig.~\ref{Fig_EBIT_AMR}a yields a valid number of markers for cells on both sides of the refinement boundary.

In 3D, however, additional difficulties arise even when the marker-count constraint is satisfied. As illustrated in Fig.~\ref{Fig_EBIT_AMR}d, independently reconstructing the interface segments on the fine and coarse sides can produce a gap across the refinement boundary. The reconstruction may also generate ``dangling'' markers, such as $p_4$ and $p_5$, which are not connected consistently to interface segments on the coarse side. These markers prevent the recomputation of unaligned-marker positions during the one-dimensional advection step.
A consistent reconstruction would therefore require additional treatments. For example, the dangling markers on the fine side, $p_4$ and $p_5$, would need to be repositioned appropriately to eliminate the gap, while additional ghost markers may also be required on the coarse side to maintain connectivity. Such treatments would considerably increase the complexity of both the reconstruction and advection algorithms.
To preserve the simplicity of the present 3D extension, we adopt the strategy used in the 2D method \cite{Pan_2024_508}: all cells within the $3\times3\times3$ stencil surrounding each interfacial cell are refined to the maximum allowable level.
Combined with an additional restriction on the $\textrm{CFL}$ number,
set to unity, this approach prevents the interface from being advected between
two grid cells at different refinement levels, and avoids the occurrence of
inconsistent configurations.
Further implementation details are provided in the
online repository (\url{https://basilisk.fr/sandbox/jieyun/src/ebit-3d.h}).

\subsection{Topology ambiguities in 3D}

\begin{figure}[!htb]
\begin{center}
\begin{tabular}{cc}
\includegraphics[width=0.4\textwidth]{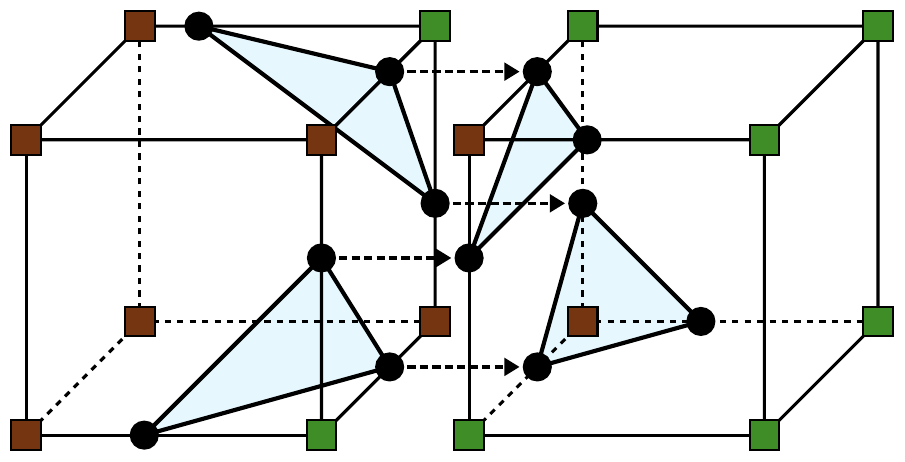}&
\includegraphics[width=0.4\textwidth]{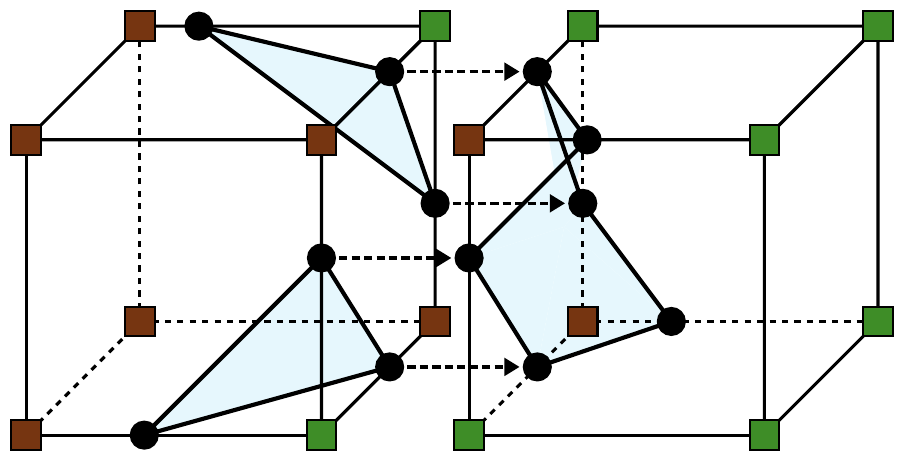}\\
(a) & (b)\\
\end{tabular}
\end{center}
\caption{Face ambiguity between adjacent cubic cells:
(a) inconsistent connectivity producing holes between adjacent cells;
(b) consistent connectivity across the shared face.}
\label{Fig_EBIT_face_ambiguity}
\end{figure}

\begin{figure}[!htb]
\begin{center}
\begin{tabular}{cc}
 \includegraphics[width=0.4\textwidth]{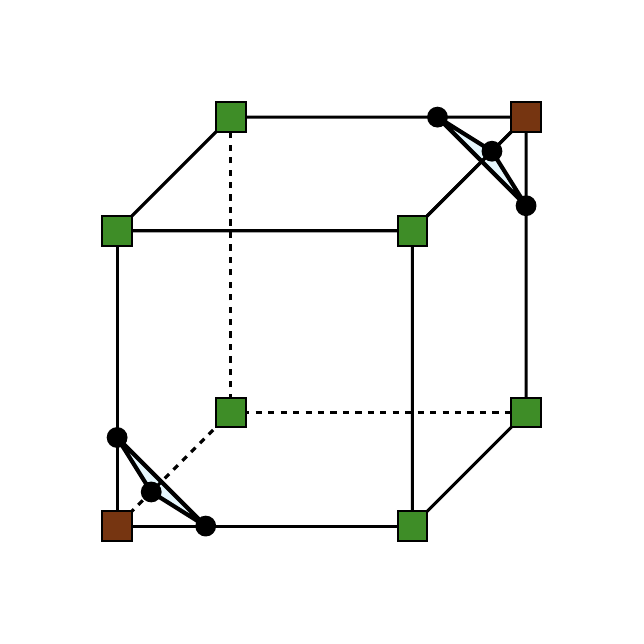}&
\includegraphics[width=0.4\textwidth]{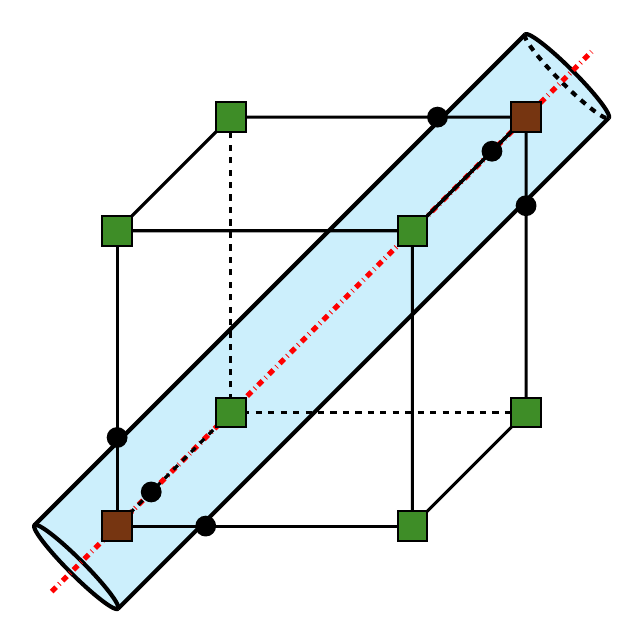}\\
(a) & (b) \\
\end{tabular}
\end{center}
\caption{Internal ambiguity in a cubic cell:
(a) configuration with two separate interface segments;
(b) configuration with a thin ligament structure spanning the diagonal direction.}
\label{Fig_EBIT_internal_ambiguity}
\end{figure}

For interface reconstructions on a simplex, such as triangular cells in 2D and tetrahedral cells in 3D, the connectivity can be uniquely determined once the marker positions are known, as in the LCRM \cite{Yoon_2010_08} and LFRM \cite{Shin_2011_230} coupled with a tetra-marching procedure.
In contrast, reconstructions on rectangular or cuboid cells introduce topology
ambiguities, as illustrated in Fig.~\ref{Fig_color_vertex}. In 2D, these
ambiguities can be resolved by introducing a color vertex field.
A direct extension of the color vertex approach from 2D to 3D simplifies the
representation of topology configurations and the implementation of topology changes, but inevitably results in additional topology ambiguities. 

Topology ambiguities arising in surface reconstructions have long been recognized in the field of computer graphics.
Originally, Lorensen and Cline \cite{Lorensen_1987_21} proposed the marching 
cubes algorithm to visualize CT and MRI data, where a polygonal mesh is generated from a 3D scalar field defined at the cell vertices of a Cartesian grid.
In the marching cubes algorithm, markers are first positioned on cell edges using a
linear or high-order interpolation.
Subsequently, the connectivity of these markers will be determined based on the 
sign of the scalar variable at the cell corners.
However, marker positions together with the corner-based indicator or
level-set functions are not always sufficient to uniquely define connectivity,
giving rise to ambiguous configurations that were not recognized in the
original work of Lorensen and Cline.

Two types of ambiguities are typically identified in this problem: 
the face ambiguity and the internal ambiguity, as shown in Figs.~\ref{Fig_EBIT_face_ambiguity} and \ref{Fig_EBIT_internal_ambiguity}, respectively.
The face ambiguity, which originates from the 2D topology ambiguity, occurs 
when a cell face contains one marker on each edge. 
In this case, two possible configurations exist, and inconsistent choices
between adjacent cells produce spurious holes on the reconstructed interface, as shown in Fig.~\ref{Fig_EBIT_face_ambiguity}a.
This issue has been effectively addressed by introducing an indicator field defined both at the cell corners and face centers, analogous to the color vertex field used in the 2D EBIT method, which yields a unique configuration on each cell face.

Internal ambiguity may arise even in the absence of face ambiguities.
For instance, as shown in Fig.~\ref{Fig_EBIT_internal_ambiguity}, at most two 
markers are present on each cell face, which does not result in face ambiguity. 
However, the same distribution of color vertex values corresponds to two distinct configurations:
either two separate interface segments, as shown in Fig.~\ref{Fig_EBIT_internal_ambiguity}a, or a thin ligament structure spanning the diagonal direction, as shown in Fig.~\ref{Fig_EBIT_internal_ambiguity}b.
The same color vertex pattern may correspond to more than two admissible configurations
as the number of markers within one cell increases.
A complete enumeration of these configurations
was provided in Ref.~\cite{Chemyaev_1995_RP}, and a brief review of
the ambiguity was recently presented by Gennari et al. \cite{Gennari_2025_540}.

Similar to face ambiguities, internal ambiguities can be resolved by defining  
the color vertex also at the cell center.
This additional value effectively subdivides the cubic cell into 24 tetrahedra.
Within each tetrahedron, the marker connectivity is uniquely determined, so that the ambiguity is removed in the same spirit as in the marching-tetrahedra procedure.
However, incorporating this additional level of complexity is beyond the scope of
the present work, whose primary objective is to propose a simple extension of
the EBIT method while laying the groundwork for future improvements. 
In the 3D EBIT method of this work, internal ambiguities are resolved
by consistently selecting the configuration corresponding to separated interface segments.
This choice provides a new type of automatic topology change mechanism in addition to the face-based mechanism.
Moreover, it is consistent with the current marker layout,
which does not include interior markers within the cell to explicitly track the 
ligament-like structures in 3D.
The introduction of such markers is therefore deferred to future development.

\subsection{Governing equations}

In the present work, we solve the Navier--Stokes equations for incompressible immiscible two-phase flow in the one-fluid formulation:
\begin{gather}
\frac{\partial \rho}{\partial t}
+ \U \cdot \nabla \rho= 0, \\
\frac{\partial \rho \U}{\partial t}
+ \nabla \cdot (\rho \U\U) = 
-\nabla p + \nabla \cdot \left[ \mu \left( \nabla \U + \nabla \U^T\right) \right] + \rho \G + \F,
\label{Eq_NS}
\\
\nabla \cdot \U = 0,
\end{gather}
where $\rho$ and $\mu$ denote fluid density and viscosity, respectively.
The gravitational force is accounted for by the term $\rho \G$. Surface tension is
modeled by the term $\F = \sigma \kappa \N \delta_S$, where $\sigma$ is the
surface tension coefficient, $\kappa$ the interface curvature, $\N$ the unit normal,
and $\delta_S$ the surface Dirac delta function. 
The physical properties are calculated as
\begin{gather}
\rho = H \rho_1 + (1 - H) \rho_2, \qquad \mu = H \mu_1 + (1 - H) \mu_2,
\label{Eq_physical}
\end{gather}
where $H$ is the Heaviside function, which is equal to 1 inside the reference phase and 0 elsewhere.

\begin{figure}[!htb]
\begin{center}
\begin{tabular}{cc}
\includegraphics[width=0.4\textwidth]{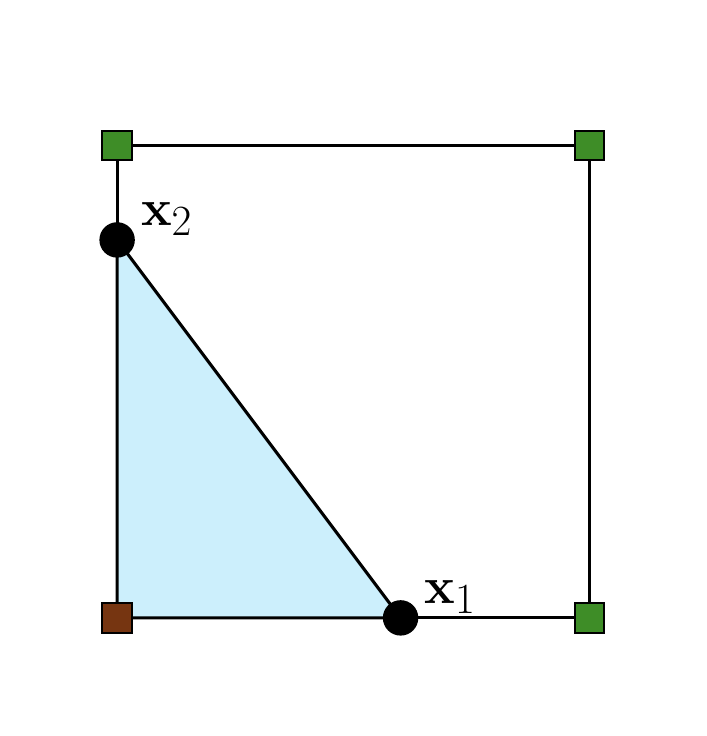} &
\includegraphics[width=0.4\textwidth]{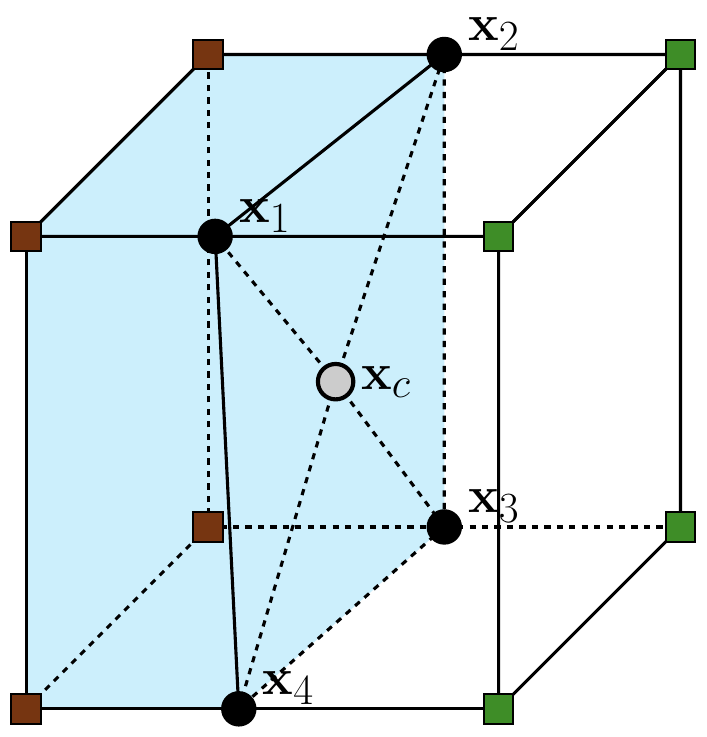}\\
(a) & (b) 
\end{tabular}
\end{center}
\caption{Computation of volume fractions using a geometric method: (a) 2D scenario, (b) 3D scenario.}
\label{Fig_front2vof}
\end{figure}

The numerical implementation of the 3D EBIT method has been developed
using the Basilisk C framework \cite{Popinet_2015_302, Basilisk_website}.
Therefore, the same coupling algorithm previously adopted for the 2D EBIT
method with the Navier--Stokes equations \cite{Pan_2024_508} is retained.
A volume-fraction field is first computed using a geometric approach and then used to approximate the Heaviside function in Eq.~\eqref{Eq_physical}.
Furthermore, the surface tension term is discretized using a well-balanced
Continuum Surface Force (CSF) formulation based on the generalized height
function method \cite{Popinet_2009_228, Popinet_2018_50}.
The discretization of the convection and diffusion terms follows the numerical schemes implemented in Basilisk, which are described in detail in Refs.~\cite{Popinet_2003_190, Popinet_2009_228}.

In our previous 2D work \cite{Pan_2024_508}, we could easily compute the equation of the straight line connecting the markers and then the volume fraction $C$ using the analytical formulae proposed by Scardovelli and Zaleski \cite{Scardovelli00}.
In three dimensions, however, we cannot directly apply those formulae,
which relate the volume fraction to the equation of a given plane,
as the markers within a cell are not necessarily coplanar.
Instead, we use an algorithm, the so-called F2V method \cite{Pan_2026_0}, to compute the volume fraction
\textcolor{red}{(\url{https://basilisk.fr/sandbox/jieyun/src/ebit-3d.h\#volume-fraction-computation})}.
In the F2V method, the volume enclosed by a cubic cell and triangulated
interfaces is computed exactly through a geometric approach.
First, the triangular elements are segmented by the cube faces into several polygons, and only those lying within the cube are retained.
Subsequently, the volume of the reference phase is computed using Gauss's theorem.
Importantly, the F2V method naturally handles cases where a cell
is intersected by multiple interfaces, which is consistent with
the EBIT framework that allows multiple interface segments within a single cell.

For its application within the EBIT method, the original segmentation step 
is replaced by a triangulation procedure applied to each interface segment within
the cell.
First, based on the color vertex distribution, we identify the markers that form a closed loop. Second, the centroid of these markers is computed as
\begin{equation}
\X_c = \frac{\sum_{i=1}^{N_m} \X_i}{{N_m}},
\end{equation}
where $\X_i$ denotes the coordinates of marker $i$ and $N_m$ is the number of markers.
Each interface segment is then decomposed into $N_m$ triangles 
by connecting the centroid to consecutive marker points.
Finally, the F2V algorithm is employed to compute the volume fraction.
A schematic of the simplest case, with only one interface segment within a cell, is presented in Fig.~\ref{Fig_front2vof}b.

Since the surface tension model used in the present EBIT method is identical to
that employed in VOF methods, we also report numerical results obtained with
the PLIC-VOF method implemented in Basilisk for comparison. 
A split scheme proposed by Weymouth and Yue \cite{Weymouth_2010_229} is used to advect the volume-fraction field to ensure exact mass conservation in multi-dimensional problems.

\section{Numerical results and discussion} \label{Numerical results and discussion}

\subsection{Translation with uniform velocity}

\begin{figure}[!htb]
\begin{center}
\includegraphics[width=0.7\textwidth]{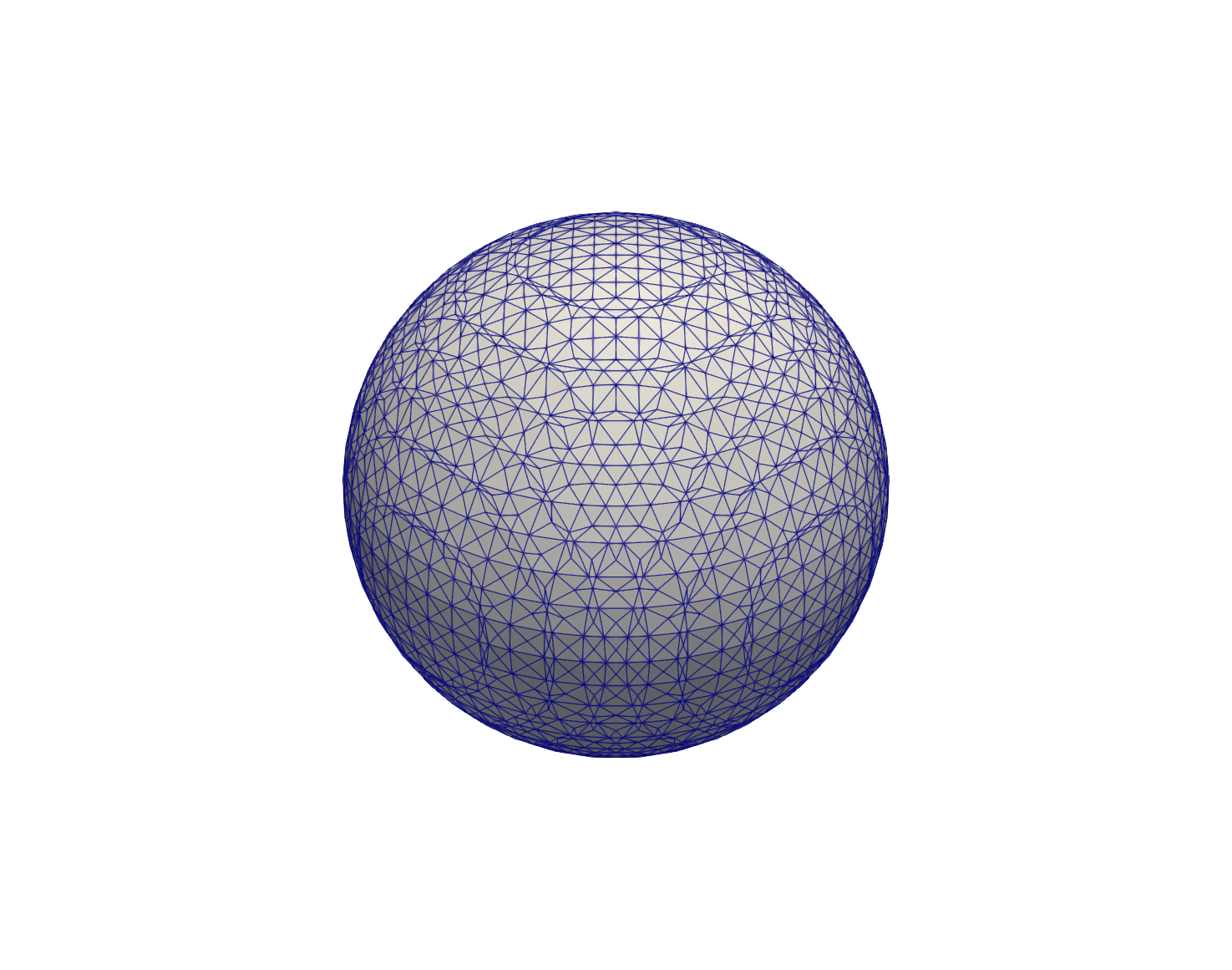}\\
\end{center}
\caption{Interface segments reconstructed on a spherical interface in the translation test at mesh resolution ($N_x=64$). For visualization purposes, the polygonal segments within each cell have been subdivided into triangular elements.}
\label{Fig_translation_intf}
\end{figure}

In this test, a spherical interface of radius $R=0.15$ and center located at
$ \X_c = (x_c, y_c, z_c) = (0.25, 0.25, 0.25)$ is placed inside the unit cubic domain. 
The domain is meshed with $N_x \times N_x \times N_x$ cubic cells
with uniform grid spacing $h =1/N_x$, where $N_x = 32, 64, 128, 256$.
A uniform and constant velocity field $\U = (u, v, w) = (1, 1, 1)$ is applied, so that the interface is advected along a diagonal direction. At half period $t = 0.5\,T$ the center reaches the position $(0.75, 0.75, 0.75)$. 
The velocity field is then reversed, and the spherical interface is expected to
return to its initial position at $t = T = 1$ without distortion.

To initialize the markers on the grid lines, we first compute the signed distance, $\phi = R^2 - (x - x_c)^2 - (y - y_c)^2 - (z - z_c)^2$, on the cell vertices, then we use linear interpolation, when the sign of the distance is opposite on the two endpoints of a cell edge, to calculate the position of a marker.
A representative interface, together with the reconstructed interface segments at mesh resolution $N_x = 64$, is shown in Fig.~\ref{Fig_translation_intf}.
For visualization purposes, polygon segments with more than three edges have been subdivided into triangular elements.
We employ a relatively small $\textrm{CFL}$ number $\textrm{CFL} = (u\,\Delta t)/h = 0.125$. 
The corresponding numerical setup is available at \url{https://basilisk.fr/sandbox/jieyun/test/translation_3d_ebit.c}.

\subsubsection{Scalability}

\begin{table}[hbt!]
\caption{Performance in weak scaling on IRENE.}
\centering
\begin{tabular}{ccccc}
\hline 
 $N_{\text{cell}}$ & $N_{\text{CPU}}$  & $T_s$ (s/step) & $S$/Ideal  & $E$ \\ 
\hline 
$128^3$ & $8$ & $1.48$ & - & - \\ 
$256^3$ & $64$ & $2.17$ & $1 / 1$ & $1$ \\ 
$512^3$ & $512$ & $2.25$ & $7.7 / 8$ & $0.97$ \\ 
$1024^3$ & $4096$ & $2.30$ & $60.4 / 64$ & $0.94$ \\ 
\hline 
\end{tabular}
\label{Tab_weak_scaling}

\end{table}

Similar to our previous work \cite{Pan_2024_508}, we first consider this simple case
to evaluate the scalability of the EBIT method. The scalability tests were performed on the ROME partition of the IRENE supercomputer, which contains 2286 nodes, each equipped with 128 CPU cores. A pure MPI parallelization strategy is adopted, with one MPI process mapped to each CPU core.
For Cartesian meshes, a common domain decomposition strategy is employed to subdivide the computational domain and distribute one subdomain to each MPI process. Markers are associated with grid cells during the reconstruction step and are automatically redistributed among different processes as the interface moves across different sub-domain boundaries.

Here, we focus on the ``weak scaling'' performance of the EBIT method,
where the number of CPU cores is varied while keeping the number of cells per
core fixed. Each sub-domain contains a $64^3$ grid, corresponding to $262144$ cells per core. Only a minor difference in performance is observed when we further increase the mesh resolution of the subdomain. The scalability is quantified using the speedup $S(N_\text{CPU})$ and efficiency $E(N_\text{CPU})$, defined as
\begin{equation}
S(N_\text{CPU}) = \frac{N_\text{CPU} T_s (N_\text{CPU, ref})}{N_\text{CPU, ref} T_s (N_\text{CPU})}, \quad 
E(N_\text{CPU}) = \frac{T_s (N_\text{CPU, ref})}{T_s (N_\text{CPU})},
\label{Eq_speedup}
\end{equation}
where $N_\text{CPU, ref} = 64$ is the reference number of CPU cores and $T_s$ the wall time per advection step.
The weak scaling results for four different runs are summarized in Table~\ref{Tab_weak_scaling}.

In an ideal weak-scaling scenario without the communication and
synchronization overhead, the wall time should remain constant as the
number of CPU cores increases. 
In practice, however, communication and synchronization costs are unavoidable, 
leading to an increase in wall time with increasing core counts, as shown in Table~\ref{Tab_weak_scaling}. Consequently, the overheads incur a degradation of the speedup and efficiency and will eventually affect the overall scalability.

In our previous scalability tests of the 2D EBIT code, the efficiency decreased smoothly as $N_\text{CPU}$ increased beyond 4. 
In contrast, for the present 3D implementation, we observe a noticeable drop in
efficiency when $N_\text{CPU}$ increases from 8 to 64.
Given the fact that each compute node consists of two 64-core CPUs located on
separate sockets, this sudden decrease of efficiency is most likely attributable to the saturation of the memory-access bandwidth.
Therefore, we adopt 64 instead of 8 as the reference number of CPU cores to better exclude memory-competing effects.
Once the solver becomes memory-bound, the efficiency only slightly
decreases as we further increase the number of cores up to $N_\text{CPU} = 4096$, demonstrating excellent weak scalability and
the automatic parallelization capability of the EBIT method.

\subsubsection{Advection errors}

Additionally, this case is used to test the simple extension algorithm 
from 2D to 3D, in which a 3D advection problem within a cell is decomposed into 2D
advection problems on cell faces, while the subroutines developed for the 2D
EBIT method are entirely reused. 
The accuracy and mass conservation of the method are measured by the mass and shape errors. 
The mass error, $E_\text{mass}$, is defined as the time-averaged absolute relative variation of the volume occupied by the reference phase 
\begin{equation}
E_\text{mass} = \frac{\int_0 ^{T}|V(t) - V(0)| dt}{V(0)} = \frac{\sum_{j=1}^{N_t} \left| \sum_{i=1}^{N_\text{cell}} \left[ C_i(t_j) - C_i(0) \right] \right|}{N_{t}\sum_{i=1} ^{N_\text{cell}} C_i(0)}.
\label{Eq_error_surface}
\end{equation}
where $N_{t}$ and $N_\text{cell}$ denote the numbers of time steps and computational cells, respectively.

The relative shape error, in an $L_1$ norm, is defined as
\begin{equation}
E_\text{shape} = \frac{\sum_{i=1}^{N_\text{cell}} \left| C_i(T) - C_i(0)\right|}{\sum_{i=1}^{N_\text{cell}} C_i(0)}.
\label{Eq_error_shape}
\end{equation}
The shape error is meaningful only when the interface at $t = T$ has
returned to its initial position, thus it is evaluated at the end of the advection.

\begin{table}[hbt!]
\footnotesize
\caption{Mesh convergence study for the translation test.}
\centering
\begin{tabular}{cc|cccc}
\hline 
 &$N_x$ &32 & 64 & 128 & 256\\ 
\hline 
EBIT&$E_\text{mass}$ & $5.03\times 10^{-4}$ & $6.13\times 10^{-5}$ & $1.35\times 10^{-5}$ & $2.81\times 10^{-6}$ \\ 
&$E_\text{shape}$ & $1.56\times 10^{-3}$ & $3.68\times 10^{-4}$ & $1.25 \times 10^{-4}$ & $2.95 \times 10^{-5}$\\
\hline 
VOF&$E_\text{shape}$ & $3.73\times 10^{-2}$ & $1.09\times 10^{-2}$ & $2.47 \times 10^{-3}$ & $1.94 \times 10^{-3}$\\
\hline 
\end{tabular}
\label{Tab_translation_error}
\normalsize
\end{table}

\begin{figure}[!htb]
\begin{center}
\begin{tabular}{cc}
\includegraphics[width=0.45\textwidth]{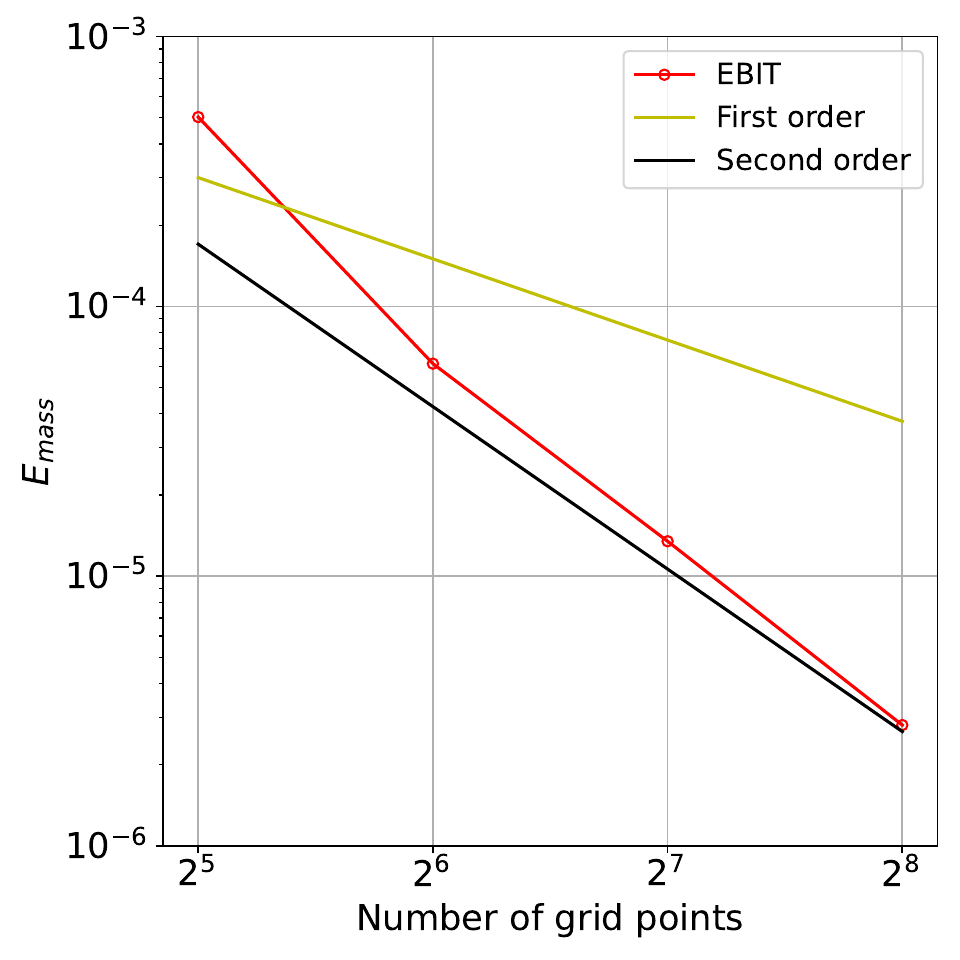} &
\includegraphics[width=0.45\textwidth]{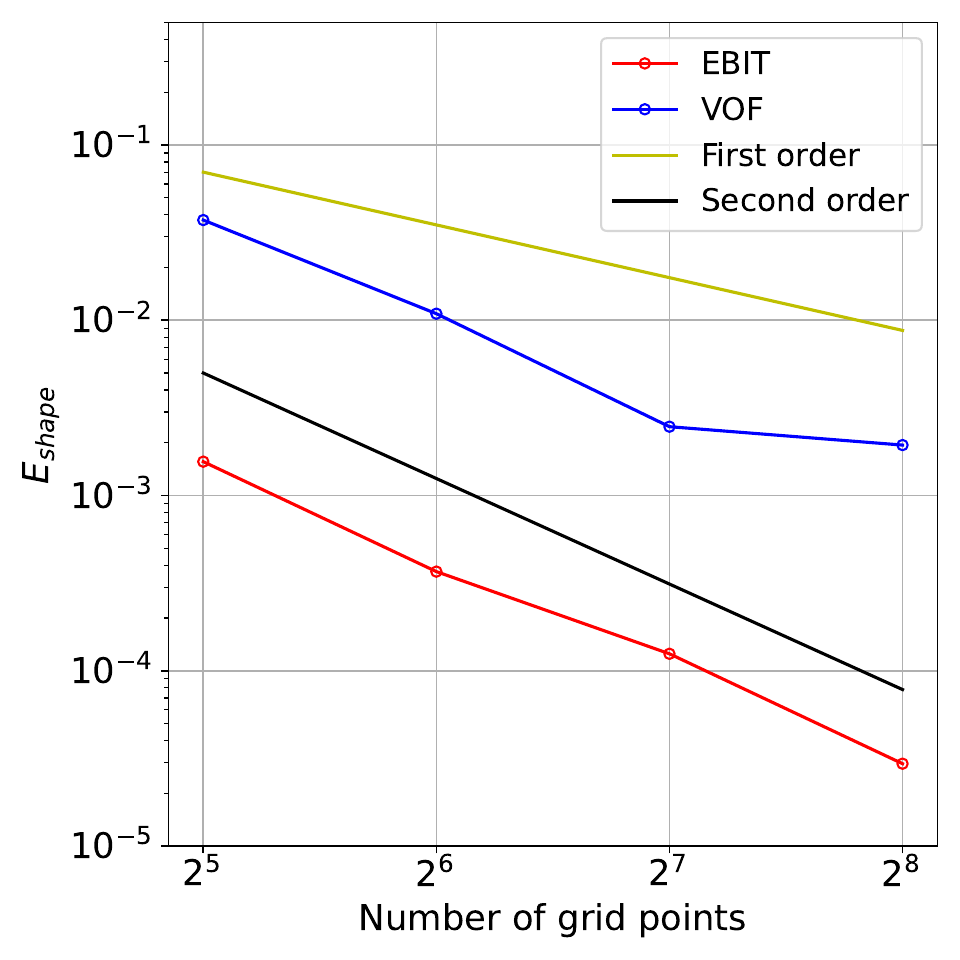}\\
(a) & (b)
\end{tabular}
\end{center}
\caption{Errors in the 3D translation test for different methods as a function of grid resolution: (a) mass error $E_\text{mass}$; (b) shape error $E_\text{shape}$.}
\label{Fig_translation_error}
\end{figure}

The mass error $E_\text{mass}$ and shape error $E_\text{shape}$ are listed in Table~\ref{Tab_translation_error} and shown in Fig.~\ref{Fig_translation_error} for the different methods considered here. 
For the EBIT method, a second-order convergence rate is observed for both the mass error and shape error. 
Since the mass error of the PLIC-VOF \cite{Weymouth_2010_229} always reaches machine precision, only the
shape error is presented in Fig.~\ref{Fig_translation_error}b. The PLIC-VOF
method demonstrates a second-order convergence rate at coarse mesh resolution
but reaches a saturation regime at finer mesh resolution, a behavior that has also
been observed in our previous 2D study \cite{Pan_2024_508}.

Moreover, the shape error with the EBIT method is around one order of magnitude
smaller than that of the PLIC-VOF method across all mesh resolutions considered.
For a spherical interface, the intersection lines with the $x$-, $y$-, and $z$-coordinate planes are circles. The circle fit used in the 2D EBIT method can thus preserve the interface shape exactly. This leads to smaller shape errors compared with the PLIC-VOF scheme, which may introduce perturbations to the interface during the advection.

\subsection{3D deformation test}

This test, analogous to the 2D single vortex test \cite{Rider_1998_141}, was first proposed by Enright et al. \cite{Enright_2002_183} to assess the capability of an interface tracking method to accurately capture the temporal evolution of a highly stretched and deformed interface.

An initial spherical interface of radius $R=0.15$ and centered at $(0.35, 0.35, 0.35)$ is placed inside the unit cubic domain. 
A divergence-free velocity field proposed by LeVeque \cite{LeVeque_1996_33}
\begin{equation}
\U (\X, t) = \left(u, v, w \right) = \cos(t / T)\left[ \begin{matrix}
2 \sin^2(\pi x) \sin(2\pi y) \sin(2\pi z) \\ 
-\sin(2\pi x) \sin^2(\pi y) \sin(2\pi z)\\ 
-\sin(2\pi x) \sin(2\pi y) \sin^2(\pi z)
\end{matrix} \right]^T,
\end{equation}
is imposed throughout the domain, which combines deformation in the $x$-$y$
plane with deformation in the $x$-$z$ plane. The flow is modulated by a
cosine function with period $T=3$,  which gradually slows down and 
subsequently reverses the motion. As a result, the interface reaches its maximum deformation at $t = 0.5\,T$ and returns to its initial position without distortion at $t = T$.

The timestep is kept constant throughout the simulation and is computed from the maximum value of the $y$-velocity, $v_\text{max}$, at time $t=0$, such that $\textrm{CFL} = v_\text{max} \Delta t / h = 0.125$. 
The corresponding numerical setup is available at \url{https://basilisk.fr/sandbox/jieyun/test/vortex_3d_ebit.c}.

\begin{figure}[!htb]
\begin{center}
\begin{tabular}{cc}
\includegraphics[width=0.45\textwidth]{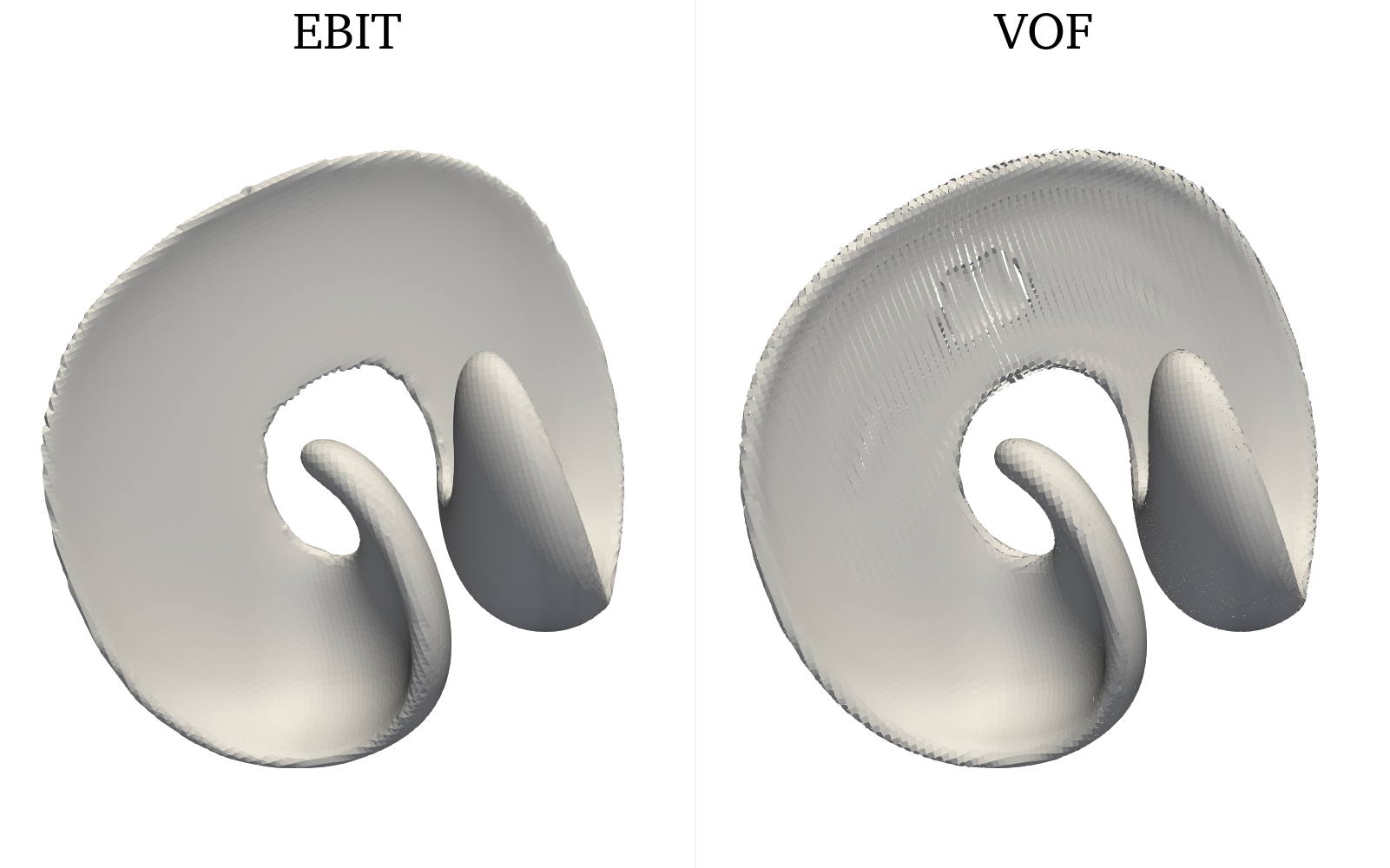} &
\includegraphics[width=0.45\textwidth]{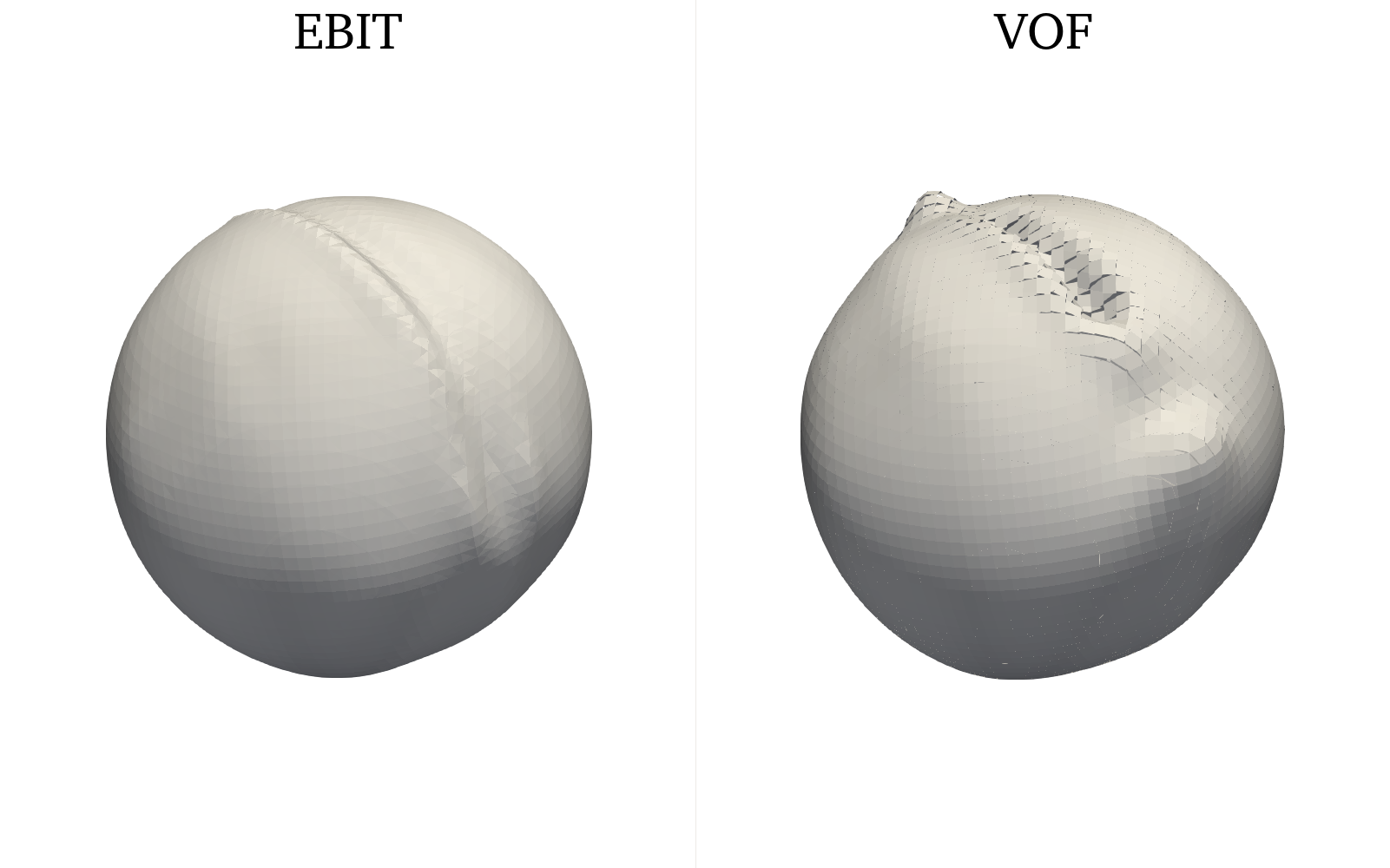}\\
(a) & (b) \\
\includegraphics[width=0.45\textwidth]{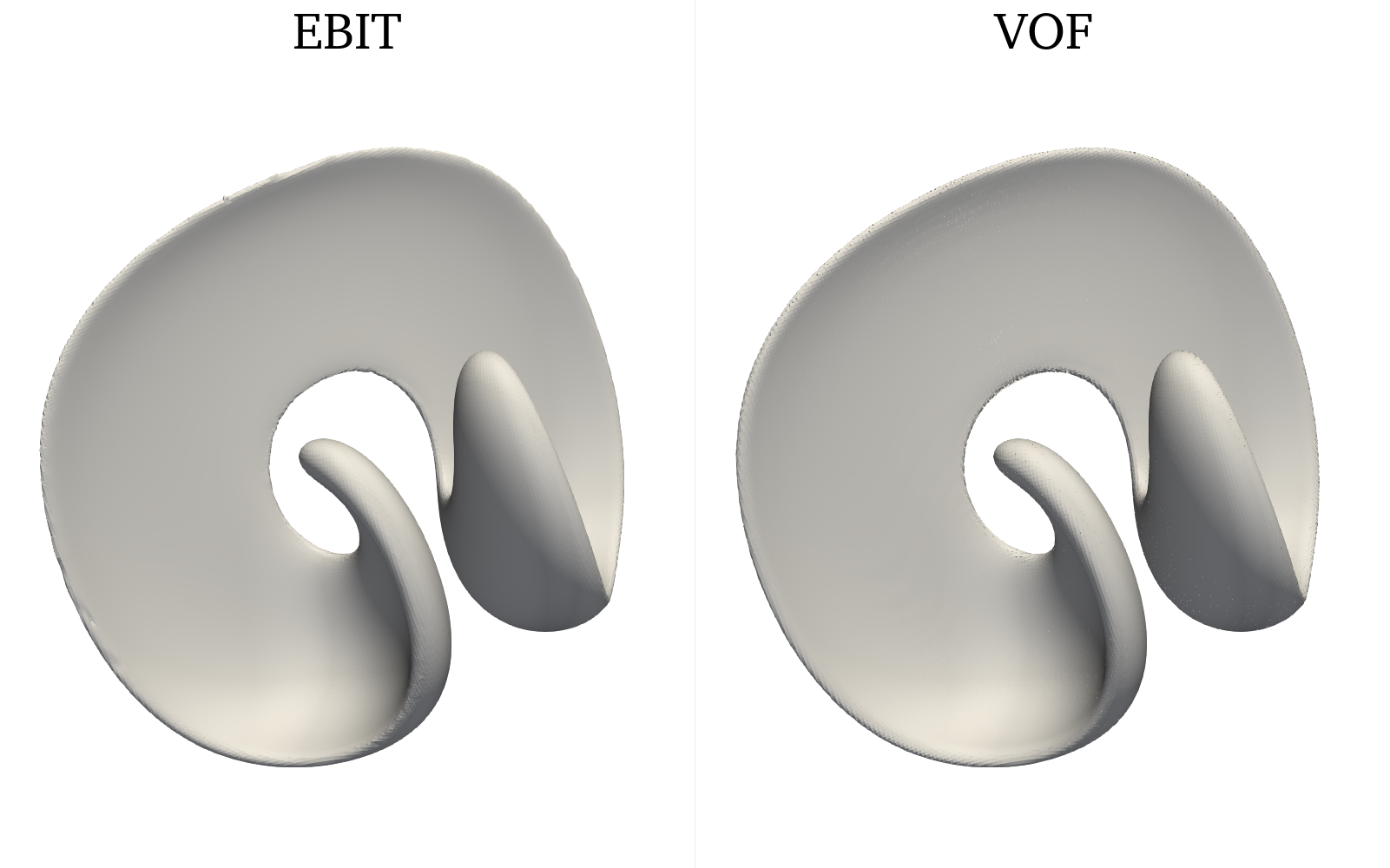} &
\includegraphics[width=0.45\textwidth]{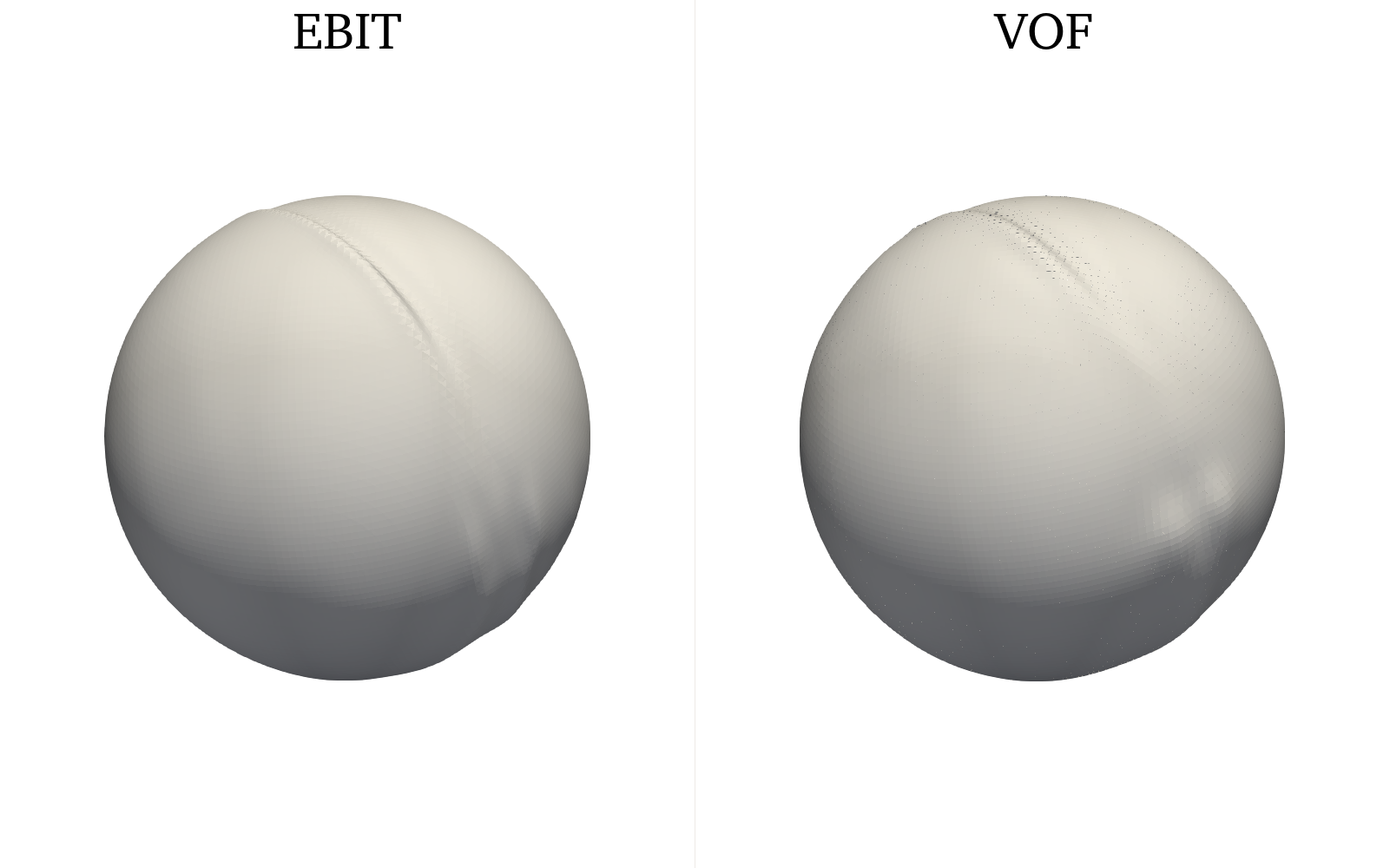}\\
(c) & (d)
\end{tabular}
\end{center}
\caption{3D deformation test with period $T = 3$: (a) interfaces at $t = 0.5\,T$, $N_x = 128$; (b) interface at $t = T$, $N_x = 128$; (c) interfaces at $t = 0.5\,T$, $N_x = 256$; (d) interface at $t = T$, $N_x = 256$.}
\label{Fig_vortex_intf}
\end{figure}
\begin{table}[!hbt]
\footnotesize
\caption{Mesh convergence study for the 3D deformation test.}
\centering
\begin{tabular}{cc|cccc}
\hline 
 &$N_x$ &32 & 64 & 128 & 256\\ 
\hline 
EBIT&$E_\text{mass}$ & $3.06\times 10^{-1}$ & $9.78\times 10^{-2}$ & $3.22\times 10^{-3}$ & $1.49\times 10^{-3}$ \\ 
&$E_\text{shape}$ & $5.66\times 10^{-1}$ & $2.23\times 10^{-1}$ & $3.82 \times 10^{-2}$ & $1.58 \times 10^{-2}$\\
\hline 
VOF&$E_\text{shape}$ & $5.70\times 10^{-1}$ & $1.91\times 10^{-1}$ & $4.80 \times 10^{-2}$ & $1.11 \times 10^{-2}$\\
\hline 
\end{tabular}
\label{Tab_vortex_error}
\normalsize
\end{table}
\begin{figure}[!htb]
\begin{center}
\begin{tabular}{cc}
\includegraphics[width=0.45\textwidth]{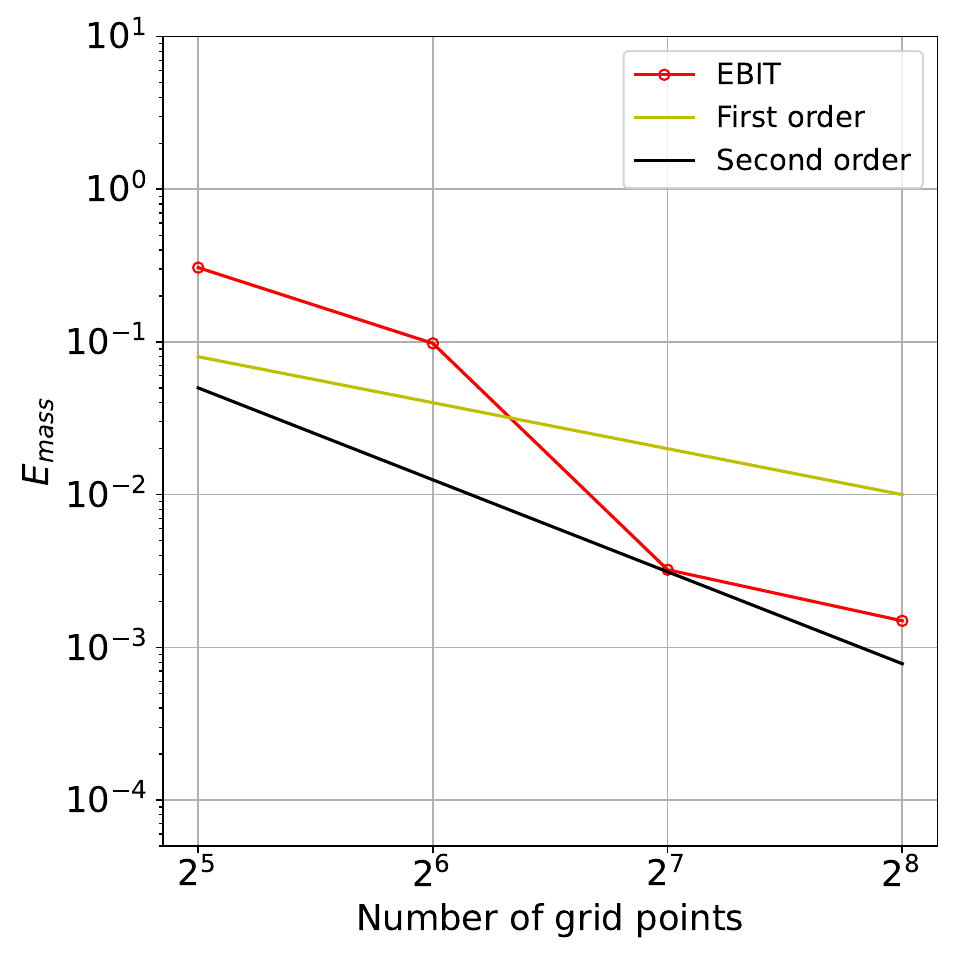} &
\includegraphics[width=0.45\textwidth]{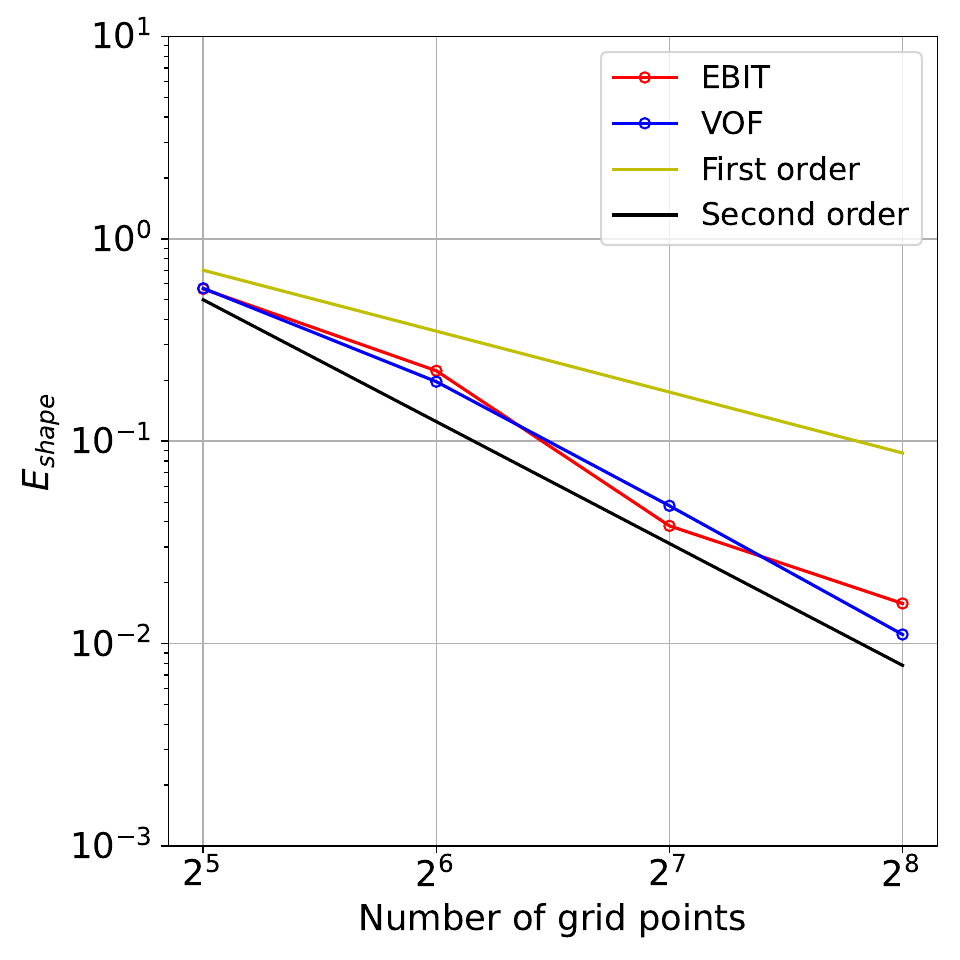}\\
(a) & (b)
\end{tabular}
\end{center}
\caption{Errors in the 3D deformation test for different methods as a function of grid resolution: (a) mass error $E_\text{mass}$; (b) shape error $E_\text{shape}$.}
\label{Fig_vortex_error}
\end{figure}

The interface at maximum deformation and after returning to its initial position
is shown in Fig.~\ref{Fig_vortex_intf} for two different mesh resolutions, $N_x = 128, 256$. 
At the intermediate mesh resolution $N_x = 128$, the PLIC-VOF method exhibits holes on the interface at maximum deformation, as shown in Fig.~\ref{Fig_vortex_intf}(a), whereas the corresponding result with the EBIT method is noticeably smoother.
At the end of the advection, the two methods are in good qualitative agreement,
both showing a ridge-like structure that has been reported in previous studies
using a variety of interface tracking methods \cite{Gorges_2023_491, Shin_2011_230}. However, the PLIC-VOF method leaves small numerical artifacts far from the initial interface position, whereas such artifacts are absent in the EBIT results. 
In EBIT, subgrid-scale interface structures are eliminated during the topology-change process, which effectively smooths the interface. Nevertheless, this process, together with the interface reconstruction at each advection time step, also introduces mass conservation errors.
As the mesh resolution is increased, the results obtained with both methods
converge and show improved agreement, as illustrated in Figs.~\ref{Fig_vortex_intf}(c) and (d). 
It is worth noting that the EBIT method tends to generate small oscillations along the outer edge of thin sheets, suggesting that the circle fit procedure becomes less accurate in regions characterized by large curvature gradients.

The mass error $E_\text{mass}$ and the shape error $E_\text{shape}$ are listed in Table~\ref{Tab_vortex_error} and are shown in Fig.~\ref{Fig_vortex_error} for the different methods considered. 
For the shape error, a second-order convergence rate is observed for both methods, and a minor discrepancy is observed for the values at different mesh resolutions.
For the mass error, the EBIT method indicates a second-order convergence rate. However, a noticeable drop in the error is observed at the mesh resolution $N_x = 128$,
which corresponds to the critical resolution at which the sign of the mass difference changes, i.e., a mass loss is observed for $N_x = 32$ and $64$,
whereas a slight mass gain occurs for $N_x = 128$ and $256$ at the end of the simulation.

\subsection{Oscillating drop}

To validate the efficacy of the EBIT method in accurately predicting capillary effects,
we perform simulations for a 3D spherical drop undergoing a small-amplitude oscillation.
Lamb's analytic solution \cite{Lamb_1932_book} is commonly adopted as a
reference, in which the normal modes corresponding to small-amplitude oscillation in zero gravity are described by the $n$-th order Legendre polynomial $P_n$. Therefore, the radial position of the interface evolves as
\begin{gather}
R(\theta, t) = R_0 + a_n(t) P_n(\cos\theta),\\
a_n(t)  = a_0 e^{-t/\tau} \sin(\omega_n t), \quad \tau = \frac{R_0^2}{(n - 1) (2n + 1)\nu_1},
\end{gather}
where $\theta \in \left[0, \pi\right]$ is the polar angle measured with respect to 
the $z$-axis, implying an axisymmetric configuration about this axis. $R_0 = D_0 / 2$ is the radius of the spherical interface at the equilibrium state, and $\omega_n$ is the angular frequency of the $n$-th normal mode, given by
\begin{equation}
\omega_n^2 = \frac{n (n + 1) (n - 1) (n + 2)\sigma}{\Gamma_n R_0^3},
\qquad \Gamma_n = (n + 1)\rho_2 + n \rho_1,
\end{equation}
where $\rho_1$ and $\rho_2$ are the densities of the liquid and gas phases, respectively.
The viscous damping is taken into account through the exponential
decay term in the amplitude function, with $\nu_1$ representing the kinematic 
viscosity of the liquid phase.

It should be noted, however, that Lamb's solution does not rigorously describe
the oscillating motion of a deformed drop initially released in a quiescent velocity field, which is the
initial condition commonly adopted in numerical simulations reported in 
the literature.
Under such an initial condition, the oscillating drop should instead be
more accurately considered as an initial-value problem, as first pointed out
and solved by Prosperetti \cite{Prosperetti_1980_100} using Laplace transforms, and later validated numerically by one of us \cite{Popinet_thesis_2000, Popinet_2002_464} and Cheng et al. \cite{Cheng_2020_408} through direct
simulations of the Navier--Stokes equations.
Lamb's solution corresponds in fact to the asymptotic limit of Prosperetti's solution as $t \rightarrow \infty$.
Moreover, for strong damping cases, this asymptotic regime may never be
observed, as the oscillation decays to negligible levels before the long-time behavior becomes established.

For this initial-value problem, the Laplace-transformed functions of the
amplitude, $\tilde{a}_n (p)$, can be expressed as
\begin{equation}
\tilde{a}_n (p) = \frac{1}{p}\left[a_n(0) + \frac{\dot{a}_n(0)p - \omega_n^2 a_n(0)}{p^2 + \Gamma_n^{-1}p \tilde{D}_n(p) + \omega_n^2}\right],
\end{equation}
where
\begin{gather}
\tilde{D}_n (p) = \frac{D_{n,\, 1} D_{n,\, 2}}{\mu_2 \mathcal{K}_{n - 1/2} (q_2) + \mu_1 \mathcal{I}_{n + 3/2} (q_1) + 2 (\mu_2 - \mu_1)},\\
D_{n,\, 1} = (2n + 1) \mu_1 \mathcal{I}_{n + 3/2} (q_1) + 2 n (n + 2) (\mu_2 - \mu_1),\\
D_{n,\, 2} = (2n + 1) \mu_2 \mathcal{K}_{n - 1/2} (q_2) - 2 (n - 1) (n + 1) (\mu_2 - \mu_1),\\
q_{1,\, 2} = R_0 (p/\nu_{1,\, 2}), \\
\mathcal{I}_{n + 3/2} (q) = q I_{n+1/2}(q) / I_{n+3/2}(q), \quad
\mathcal{K}_{n - 1/2} (q) = q K_{n+1/2}(q) / K_{n-1/2}(q).
\end{gather}
where $I$ and $K$ are the modified Bessel functions of the first and second kinds, respectively.
The time-dependent amplitude $a_n(t)$ is subsequently obtained by performing 
inverse Laplace transforms numerically. In this study, the inversion is carried out
using the \textit{mpmath} Python library.

\begin{table}[hbt!]
\caption{Physical properties for the 3D oscillating drop test.}
\centering
\begin{tabular}{cccccccc}
\hline 
Test case&$\rho_1$& $\rho_2$& $\mu_1$ & $\mu_2$ & $\sigma$ &  $\textrm{La}$ & $a_0$\\ 
\hline 
$1$ & $10 $ & $0.1$ & $5 \times 10^{-2}$ & $5 \times 10^{-4}$  & $0.1$ & $8 \times 10^2$ & $0.025/0.005$\\ 
$2$ & $10 $ & $0.1$ & $5 \times 10^{-2}$ & $5 \times 10^{-4}$  & $10$ & $8 \times 10^4$ & $0.025/0.005$\\
$3$ & $10 $ & $10$ & $5 \times 10^{-2}$ & $5 \times 10^{-2}$  & $0.1$ & $8 \times 10^2$ & $0.005$\\ 
$4$ & $10 $ & $10$ & $5 \times 10^{-2}$ & $5 \times 10^{-2}$  & $10$ & $8 \times 10^4$ & $0.005$\\ 
\hline 
\end{tabular}
\label{Tab_para_oscillating}
\end{table}
To demonstrate the discrepancy between the solutions of Lamb and Prosperetti,
we investigated this configuration under four sets of physical parameters
characterized by different surface tension coefficients.
The values of physical properties employed in the simulations, together with the corresponding Laplace numbers, 
$\textrm{La} = (\rho_1 \sigma D_0) \big/ \mu_1^2$, are listed in Table~\ref{Tab_para_oscillating}.
The second case with a larger surface tension and hence a higher Laplace number is adopted from Ref. \cite{Shin_2011_230}.

Following the setup in \cite{Shin_2011_230}, we place a drop of radius
$R_0 = D_0 / 2 = 1$ at the center of a cubic computational domain of size $2 D_0$.
Symmetric boundary conditions are imposed on all six faces of the domain. 
Here, we first consider the oscillation corresponding to the second normal mode with an initial amplitude of $a_n(0) = a_0 = 0.025$.
In all subsequent dynamic simulations, the timestep is restricted by both the CFL number, with $\textrm{CFL}=0.1$, and the capillary timestep constraint:
\begin{gather}
\Delta t = \min \left( \Delta t_u, \Delta t_\sigma \right),\\
\Delta t_u = \mathrm{CFL}\frac{\, h}{\max\left(u_{\max},\, v_{\max},\, w_{\max}\right)}, \quad 
\Delta t_\sigma = \sqrt{\frac{(\rho_1 + \rho_2) h^3}{2 \pi\sigma}},
\label{Eq_timestep}
\end{gather}
where $u_{\max}$, $v_{\max}$ and $w_{\max}$ denote the maximum values of the three 
velocity components over the entire computational domain.
The detailed numerical setup is available at \url{https://basilisk.fr/sandbox/jieyun/test/oscillation_3d_ebit.c}.

\begin{figure}[!htb]
\begin{center}
\begin{tabular}{cc}
\includegraphics[width=0.45\textwidth]{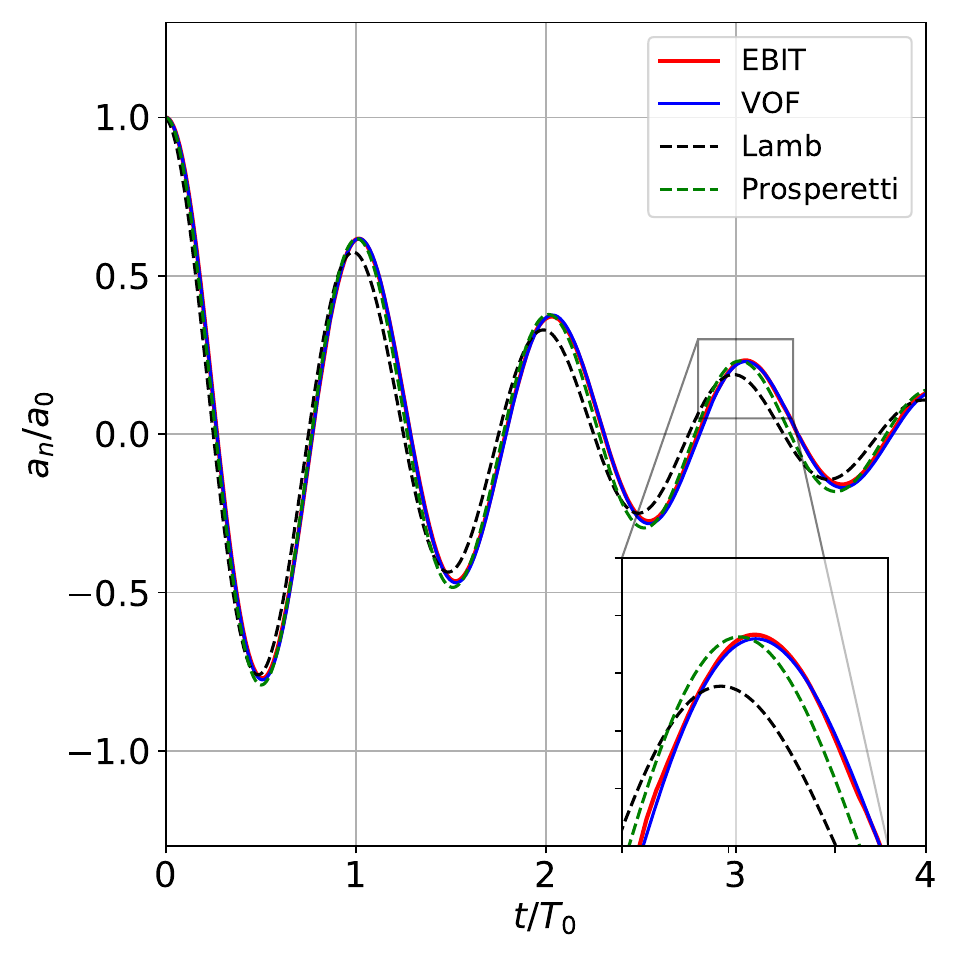} &
\includegraphics[width=0.45\textwidth]{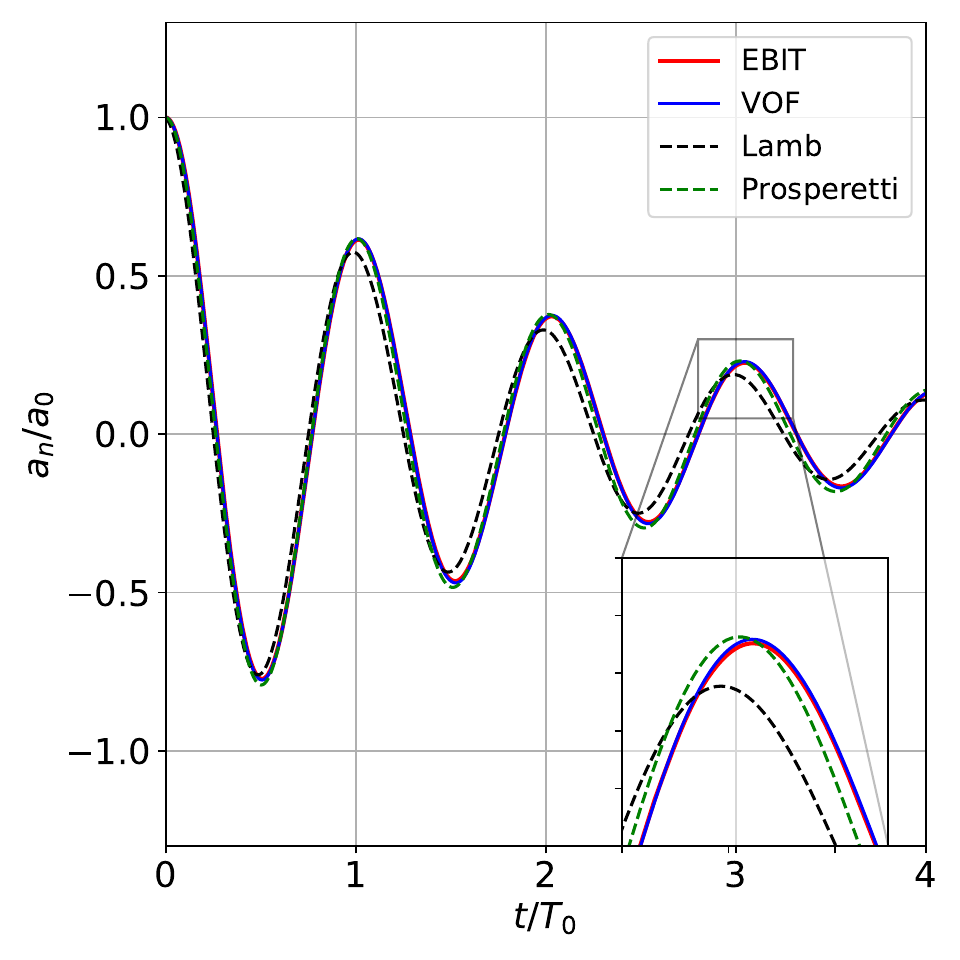}\\
(a) & (b) 
\end{tabular}
\end{center}
\caption{Oscillating drop with $\text{La} = 800, \rho_1/\rho_2 = 100, \mu_1 / \mu_2 = 100, a_0 = 0.025$: 
time evolution of the amplitude of the interface oscillation: (a) $N_x = 64$, (b) $N_x = 128$.}
\label{Fig_oscillating_amp_800}
\end{figure}
\begin{figure}[!htb]
\begin{center}
\begin{tabular}{cc}
\includegraphics[width=0.45\textwidth]{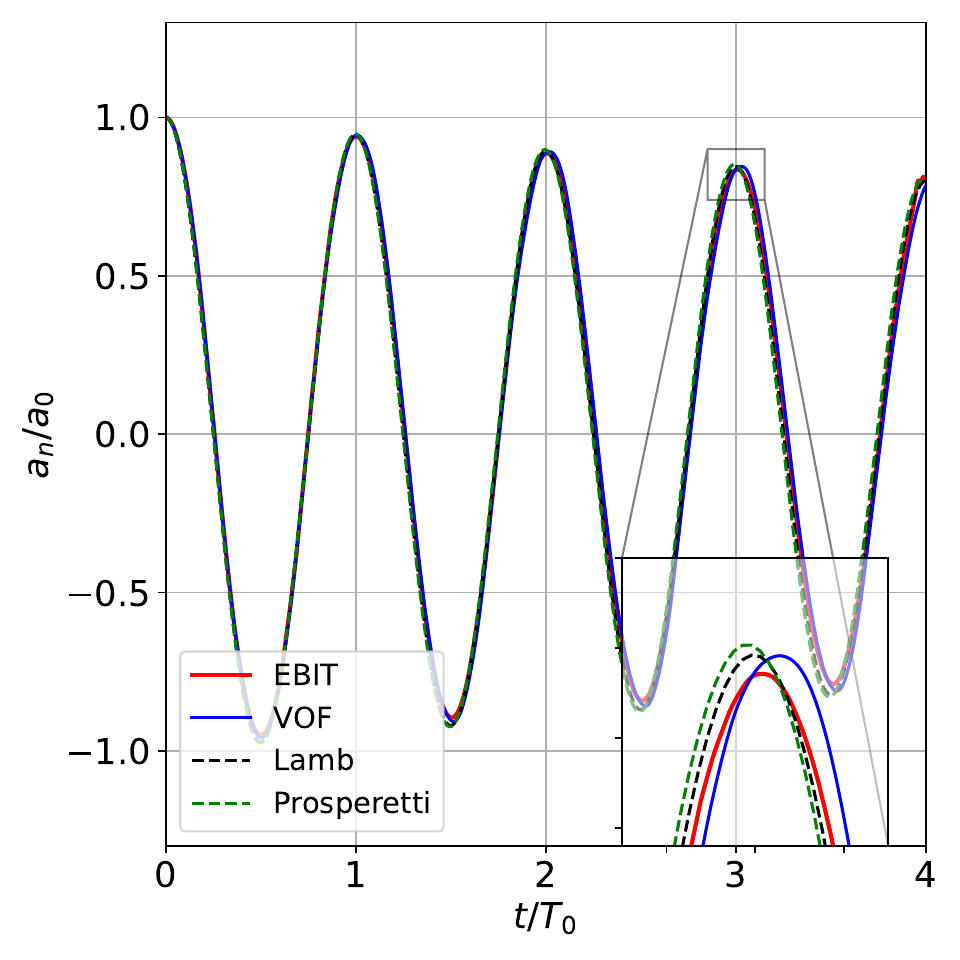} &
\includegraphics[width=0.45\textwidth]{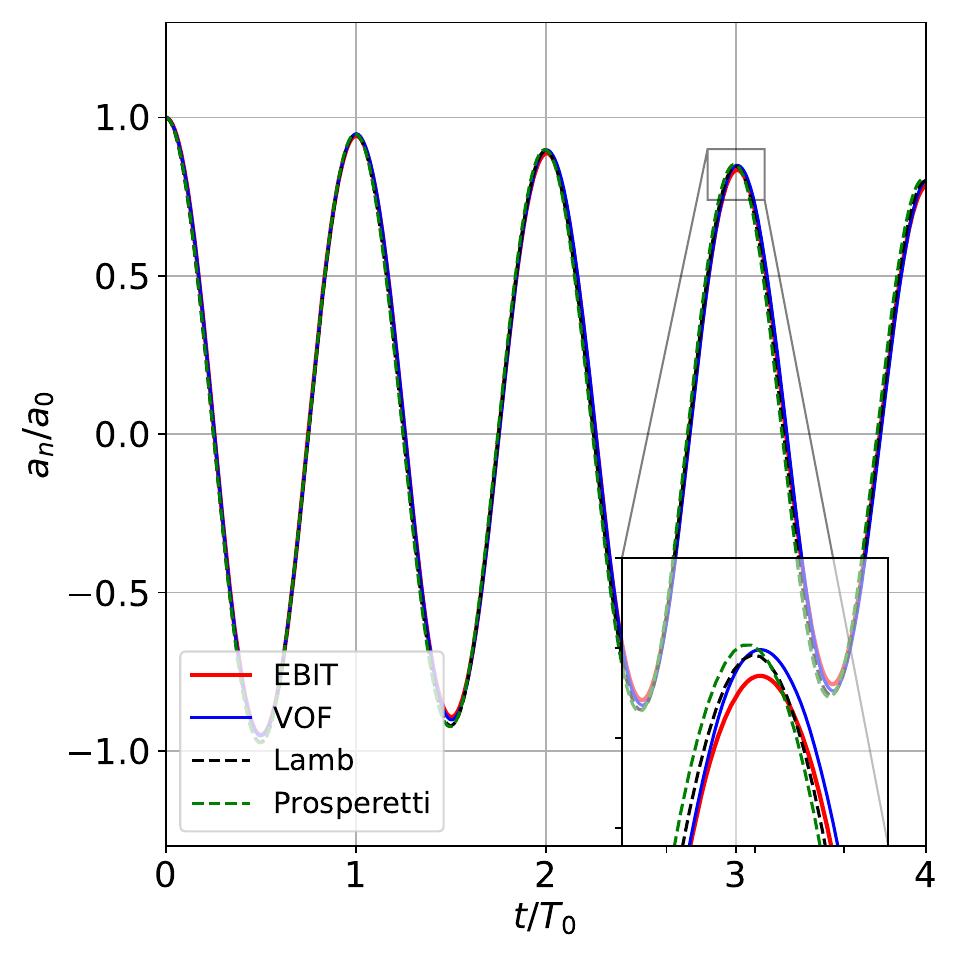}\\
(a) & (b) 
\end{tabular}
\end{center}
\caption{Oscillating drop with $\text{La} = 80000, \rho_1/\rho_2 = 100, \mu_1 / \mu_2 = 100, a_0 = 0.025$: 
time evolution of the amplitude of the interface oscillation: (a) $N_x = 64$, (b) $N_x = 128$.}
\label{Fig_oscillating_amp_80000}
\end{figure}

The time evolution of the amplitude is shown in Figs.~\ref{Fig_oscillating_amp_800} and \ref{Fig_oscillating_amp_80000}
for different mesh resolutions, together with the theoretical solutions and
the results obtained with the PLIC-VOF method. Time is normalized by
the oscillating period based on the normal-mode analysis $T_0 = 2\pi/\omega_2$.

For the first case with a lower Laplace number ($\textrm{La} = 800$), pronounced
discrepancies are observed between the Lamb and Prosperetti solutions.
The solution of the initial value problem no longer yields
a single-frequency oscillation, as evidenced by the increasing distance between
successive crests. Furthermore, the amplitude predicted by Prosperetti's solution decays more slowly.
The numerical results obtained with both the EBIT and PLIC-VOF methods are in
much better agreement with Prosperetti's solution than with Lamb's solution.
Moreover, although the two numerical methods agree well with each other,
they do not fully converge to Prosperetti's solution as the mesh resolution 
increases, as indicated by a persistent phase shift relative to the theoretical prediction.

For the second test case with a higher Laplace number ($\textrm{La} = 80000$), 
the discrepancy between the Lamb and Prosperetti solutions is considerably
reduced, indicating that the oscillation predicted by the initial-value problem can
be well approximated by a harmonic damped motion.
However, compared with the low Laplace number case, the differences between
the numerical results obtained with the EBIT and PLIC-VOF methods become
more prominent at the same mesh resolution. The EBIT method presents a slightly
stronger numerical damping.
As in the low Laplace number case, the EBIT results cease to converge toward
the theoretical solution once the mesh resolution reaches $N_x = 64$ (32 cells per radius).
By comparison, the PLIC-VOF method continues to approach Prosperetti's solution as the mesh is further refined.
 
\begin{figure}[!htb]
\centering
\begin{tabular}{cc}
\includegraphics[width=0.45\textwidth]{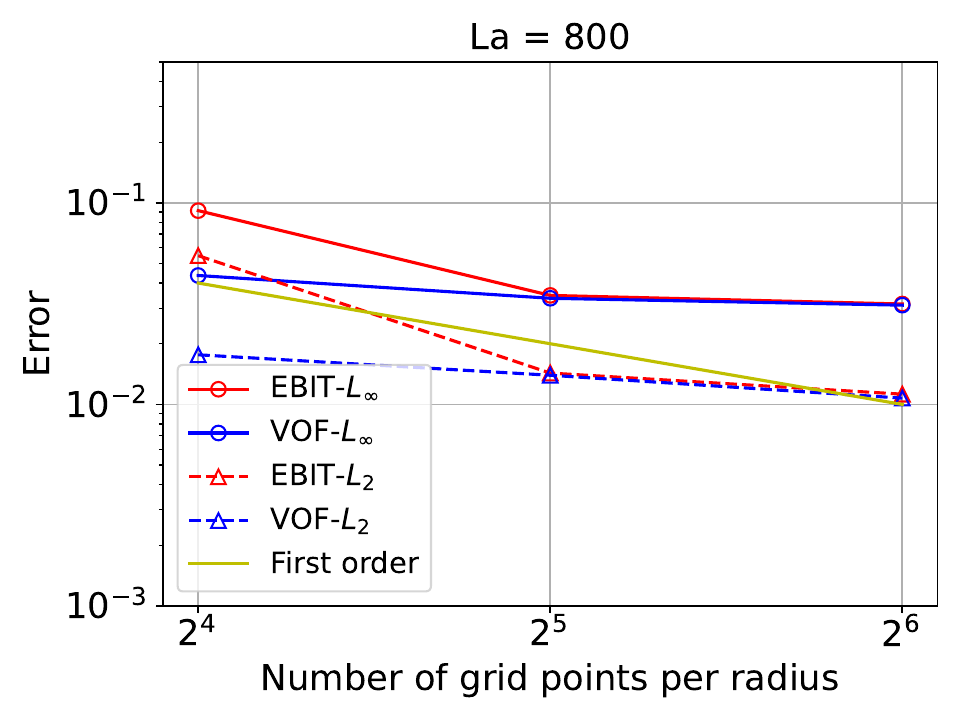} &
\includegraphics[width=0.45\textwidth]{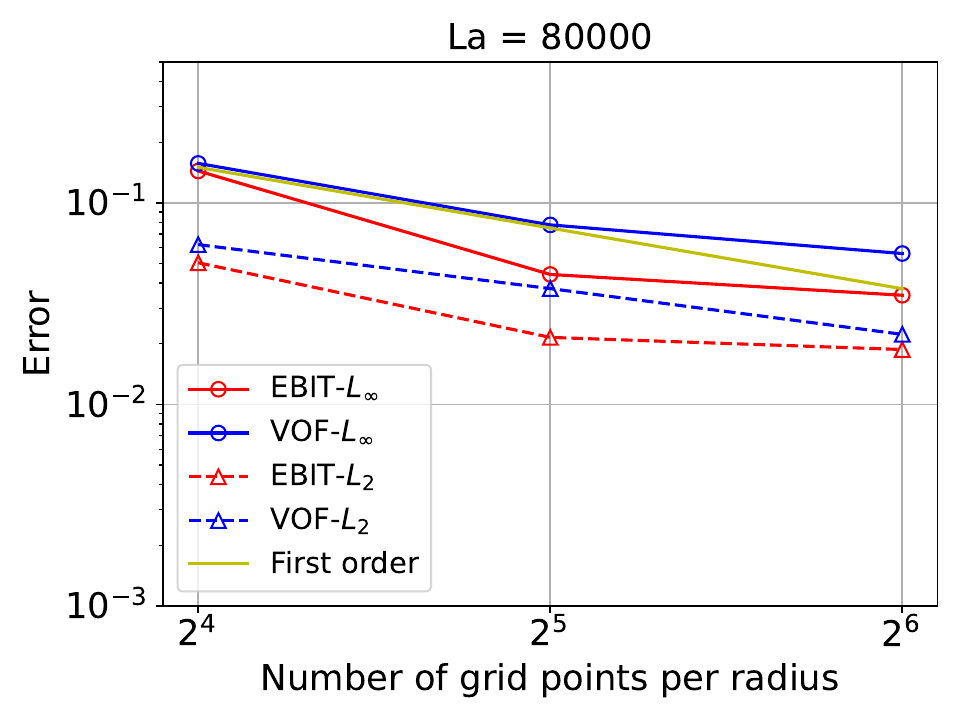}\\
(a) & (b) 
\end{tabular}
\caption{Errors $E_\infty$ and $E_2$ in the oscillating drop test as functions of grid resolution for different methods: (a) case 1, $\text{La} = 800, \rho_1/\rho_2 = 100, \mu_1 / \mu_2 = 100, a_0 = 0.025$, (b) case 2, $\text{La} = 80000, \rho_1/\rho_2 = 100, \mu_1 / \mu_2 = 100, a_0 = 0.025$.}
\label{Fig_oscillating_acc_r100_25}
\end{figure}

\begin{figure}[!htb]
\centering
\begin{tabular}{cc}
\includegraphics[width=0.45\textwidth]{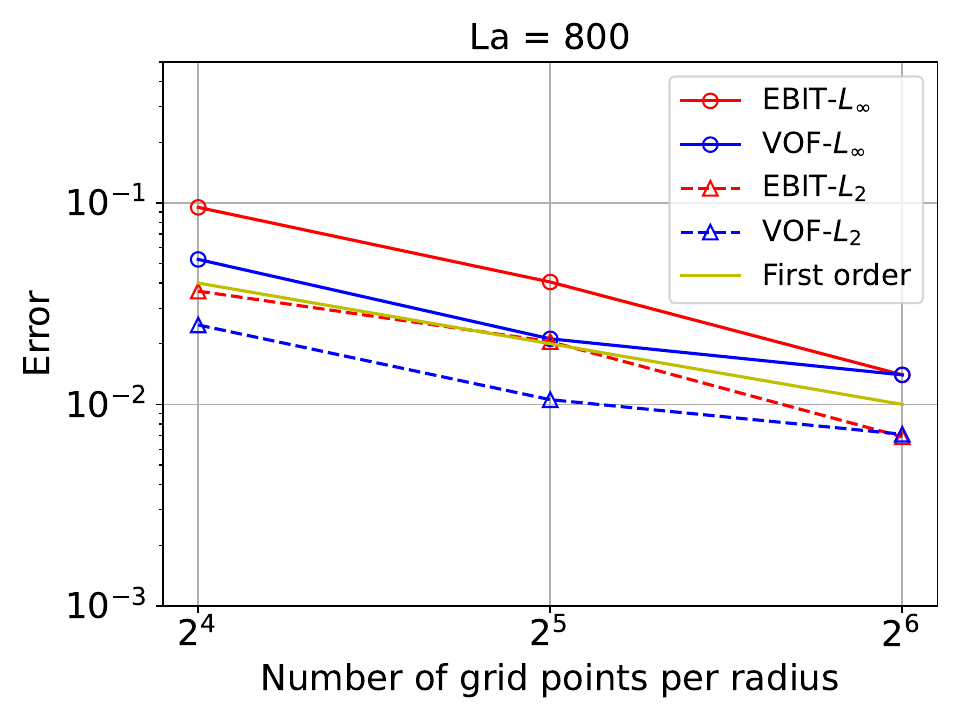} &
\includegraphics[width=0.45\textwidth]{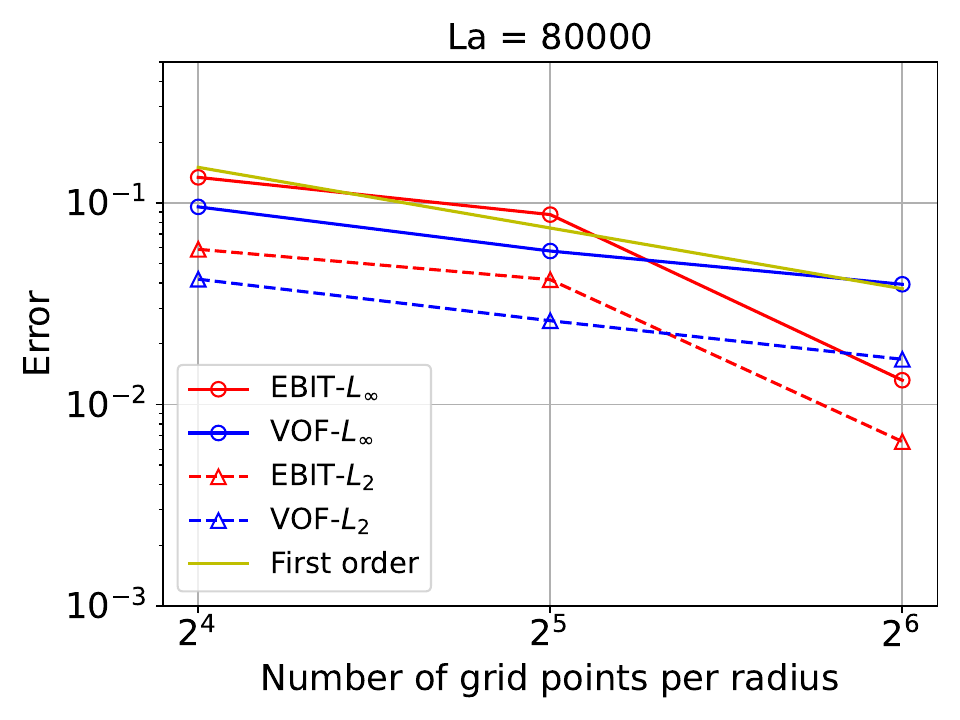}\\
(a) & (b) 
\end{tabular}
\caption{Errors $E_\infty$ and $E_2$ in the oscillating drop test as functions of grid resolution for different methods: (a) case 1, $\text{La} = 800, \rho_1/\rho_2 = 100, \mu_1 / \mu_2 = 100, a_0 = 0.005$, (b) case 2, $\text{La} = 80000, \rho_1/\rho_2 = 100, \mu_1 / \mu_2 = 100, a_0 = 0.005$.}
\label{Fig_oscillating_acc_r100_5}
\end{figure}

\begin{figure}[!htb]
\centering
\begin{tabular}{cc}
\includegraphics[width=0.45\textwidth]{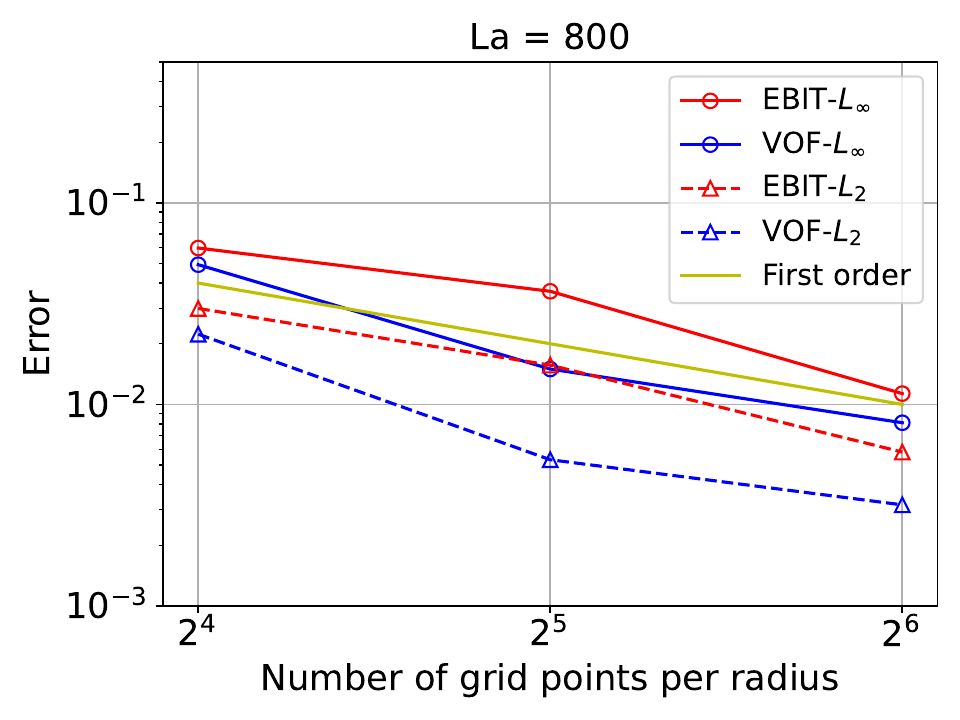} &
\includegraphics[width=0.45\textwidth]{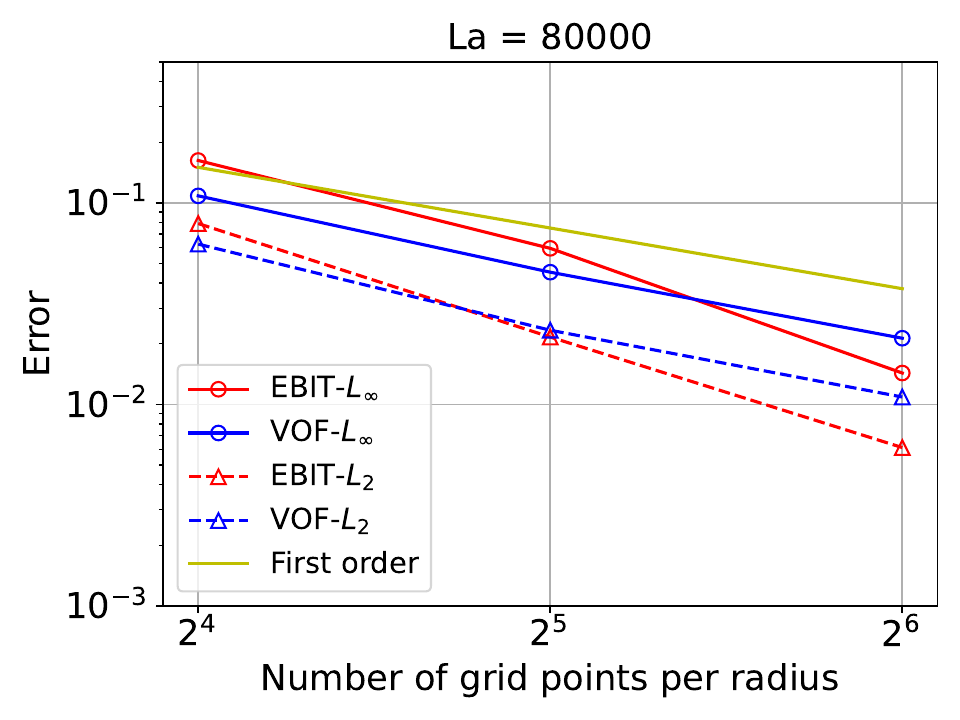}\\
(a) & (b) 
\end{tabular}
\caption{Errors $E_\infty$ and $E_2$ in the oscillating drop test as functions of grid resolution for different methods: (a) case 3, $\text{La} = 800, \rho_1/\rho_2 = 1, \mu_1 / \mu_2 = 1, a_0 = 0.005$, (b) case 4, $\text{La} = 80000, \rho_1/\rho_2 = 1, \mu_1 / \mu_2 = 1, a_0 = 0.005$.}
\label{Fig_oscillating_acc_r1_5}
\end{figure}

The discrepancies between Prosperetti's solution and the two numerical solutions are further analyzed with the $L_2$ and $L_\infty$ norms, defined as
\begin{gather}
E_\infty = \frac{\textrm{max}_i \left(\left| a_{\text{numerical}, i} - a_{\text{theory}, i}\right|\right)}{a_0},\\
E_2 = \frac{1}{a_0}\sqrt{\frac{\sum_{i=1}^{N_i} \left(a_{\text{numerical}, i} - a_{\text{theory}, i}\right)^2}{N_i}},
\label{Eq_E2}
\end{gather}
where $a_{\text{numerical}, i}$ and $a_{\text{theory}, i}$ denote numerical and 
theoretical amplitudes at time instant $t_i$, respectively, and $N_i$ is the total
number of sampling points. A total of 3072 uniformly distributed time samples up to time $t/T_0 = 3$ are used to compute the errors.

For the case with $\textrm{La} = 800$, as shown in Fig.~\ref{Fig_oscillating_acc_r100_25}a, a convergence rate considerably smaller than 1 is
observed for both methods. Moreover, the errors nearly approach a plateau
as the mesh resolution increases, indicating that the numerical solutions converge
to a solution slightly different from Prosperetti's solution. This behavior is consistent
with the persistent phase shift observed in Fig.~\ref{Fig_oscillating_amp_800}.

For the case with $\textrm{La} = 80000$, as shown in Fig.~\ref{Fig_oscillating_acc_r100_25}b, the error plateaus are
more noticeable for the EBIT method while a first-order convergence rate is identified for the PLIC-VOF method. However, the convergence rate of the
PLIC-VOF method already begins to deteriorate at the highest mesh resolution, as indicated by the $L_\infty$ errors. At sufficiently fine mesh resolutions, the
EBIT method yields slightly smaller errors than the PLIC-VOF method. 

It is important to note that differences exist between the models used for
the numerical simulations and those employed in the theoretical analysis.
First, the theoretical solution relies on a linearization assumption, which is 
strictly valid only for infinitesimal oscillation amplitudes, while the
direct numerical simulations account for nonlinear effects associated with
finite amplitudes. Second, while the theoretical analysis is based on a two-fluid
formulation, the present study employs a one-fluid formulation of the Navier-Stokes equations.
A clear identification of discretization errors, reflected by a well-defined convergence
rate, is possible only when these two modeling errors are negligible.

To assess the contribution of the finite-amplitude effect, we first performed two
additional tests using the same physical properties as in cases 1 and 2, but with
a smaller initial amplitude of $a_0=0.005$. We then further reduced both the
density and viscosity ratios to unity, corresponding to cases 3 and 4 in
Table~\ref{Tab_para_oscillating}, to examine the influence of the one-fluid
formulation. The corresponding mesh-convergence results are presented in
Figs.~\ref{Fig_oscillating_acc_r100_5} and \ref{Fig_oscillating_acc_r1_5}, respectively.

As shown in Fig.~\ref{Fig_oscillating_acc_r100_5}, reducing the initial amplitude removes the error plateaus observed at $a_0=0.025$ for both Laplace numbers, indicating that finite-amplitude effects contribute significantly to the discrepancy from the linear theoretical solution. The errors decrease consistently with mesh refinement for both interface-tracking methods, although the observed convergence rates differ somewhat between EBIT and VOF.

When the density and viscosity ratios are additionally reduced to unity, the convergence behavior becomes even clearer for both methods, as presented in Fig.~\ref{Fig_oscillating_acc_r1_5}. In particular, the VOF results recover an approximately first-order convergence rate, while the EBIT results exhibit a comparable or slightly higher rate over the resolutions considered. These observations indicate that both the finite-amplitude effect and the discrepancy associated with the one-fluid formulation contribute to the departure of the numerical results from Prosperetti's linear two-fluid solution. As these two sources of modeling error are reduced, the discretization error becomes more clearly identifiable through systematic mesh convergence.

Despite the discrepancy between the converged numerical results and
the theoretical predictions, the excellent agreement between the EBIT
and PLIC-VOF results verifies the coupling algorithm to the Navier--Stokes equations proposed in this work.
In addition, we emphasize that, although the oscillating drop problem is
widely used as a benchmark for verifying interface tracking methods,
careful consideration is required when comparing numerical results with
theoretical solutions. First, simulations initialized from a quiescent velocity
field should be compared with the solution of the initial-value problem, rather than with the normal-mode analysis, especially in the low Laplace number regime.
Second, a small enough initial amplitude should be employed to minimize nonlinear effects resulting from finite-amplitude oscillations.

\subsection{Rising bubble}

In this test, we examine a single bubble rising under buoyancy inside a heavier fluid. 
The 3D configuration of this test case has been widely used to verify interface tracking methods in previous studies \cite{Unverdi_1992_100, Shin_2002_180, Shin_2011_230}. 
Following the setup in \cite{Unverdi_1992_100}, we consider a cuboid computational domain $[0, 2\,D] \times [0, 4\,D] \times [0, 2\,D]$, with $D=0.5$. At the beginning of the simulation, a spherical bubble of radius $R=D/2$ is positioned in the bottom part of the domain with center at $(D, D, D)$.
Free-slip wall boundary conditions are enforced at all the boundaries of the computational domain. 

\begin{table}[hbt!]
\caption{Physical properties for the rising bubble test.}
\centering
\begin{tabular}{ccccccccc}
\hline 
 Test case &$\rho_1$& $\rho_2$& $\mu_1$ & $\mu_2$ & $g$ & $\sigma$ & $\textrm{M}$ & $\textrm{Bo}$ \\ 
\hline 
$1$ &$1000 $ & $25$ & $ 35$ & $0.233$ & $0.98$ &$24.5$ & $0.1$ & $10$ \\ 
$2$ &$1000 $ & $25$ & $ 11.07$ & $0.234$ & $0.98$ &$2.45$ & $1$ & $100$ \\ 
\hline 
\end{tabular}
\label{Tab_para_rising}
\end{table}
Two representative setups, corresponding to different flow regimes, are
selected from \cite{Unverdi_1992_100}. The relevant physical properties
are summarized in Table~\ref{Tab_para_rising},
where the Morton number is defined as $\textrm{M} = \big( g \mu_1^4  \big) \big/ \big( \rho_1 \sigma^3 \big) $ and the Bond number as
$\textrm{Bo} = \big( \rho_1 g D^2 \big) \big/ \sigma$.
The corresponding numerical setup is provided at \url{https://basilisk.fr/sandbox/jieyun/test/rising_3d_ebit.c}.

\begin{figure}[!htb]
\begin{center}
\begin{tabular}{cc}
\includegraphics[width=0.45\textwidth]{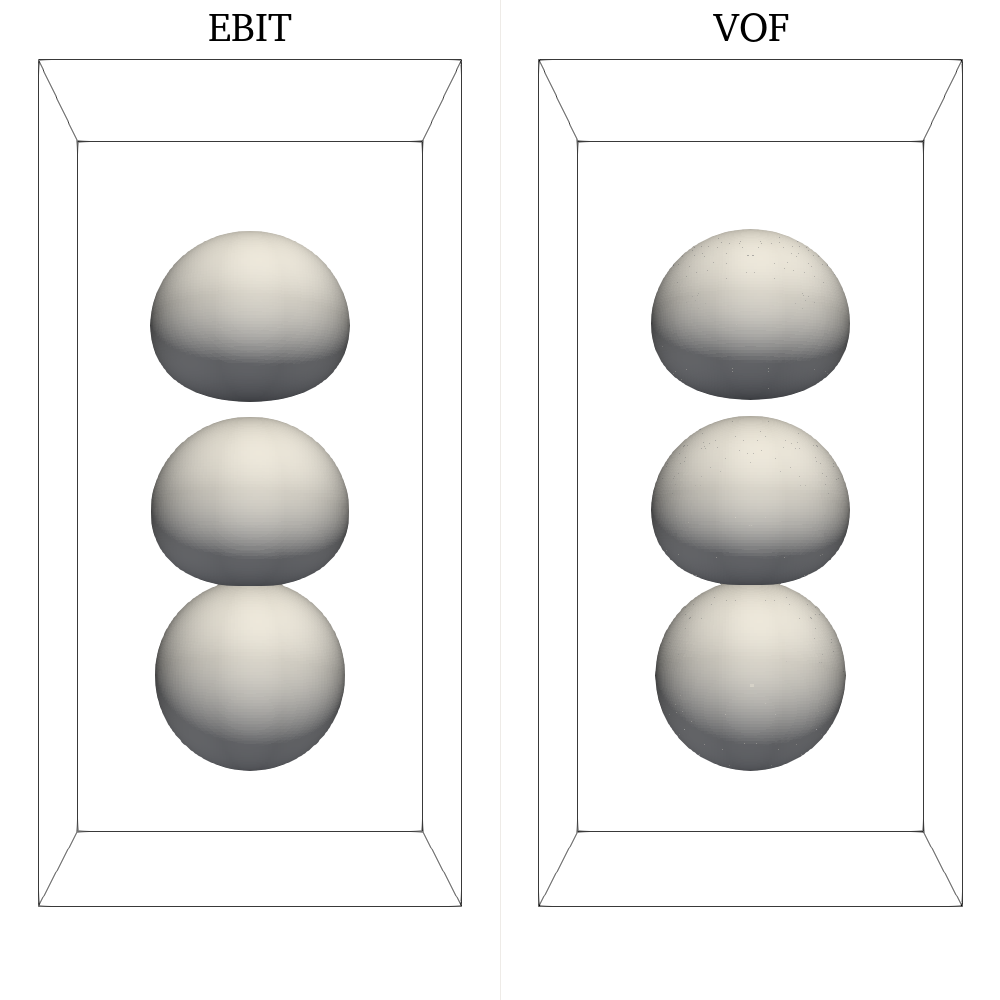} &
\includegraphics[width=0.45\textwidth]{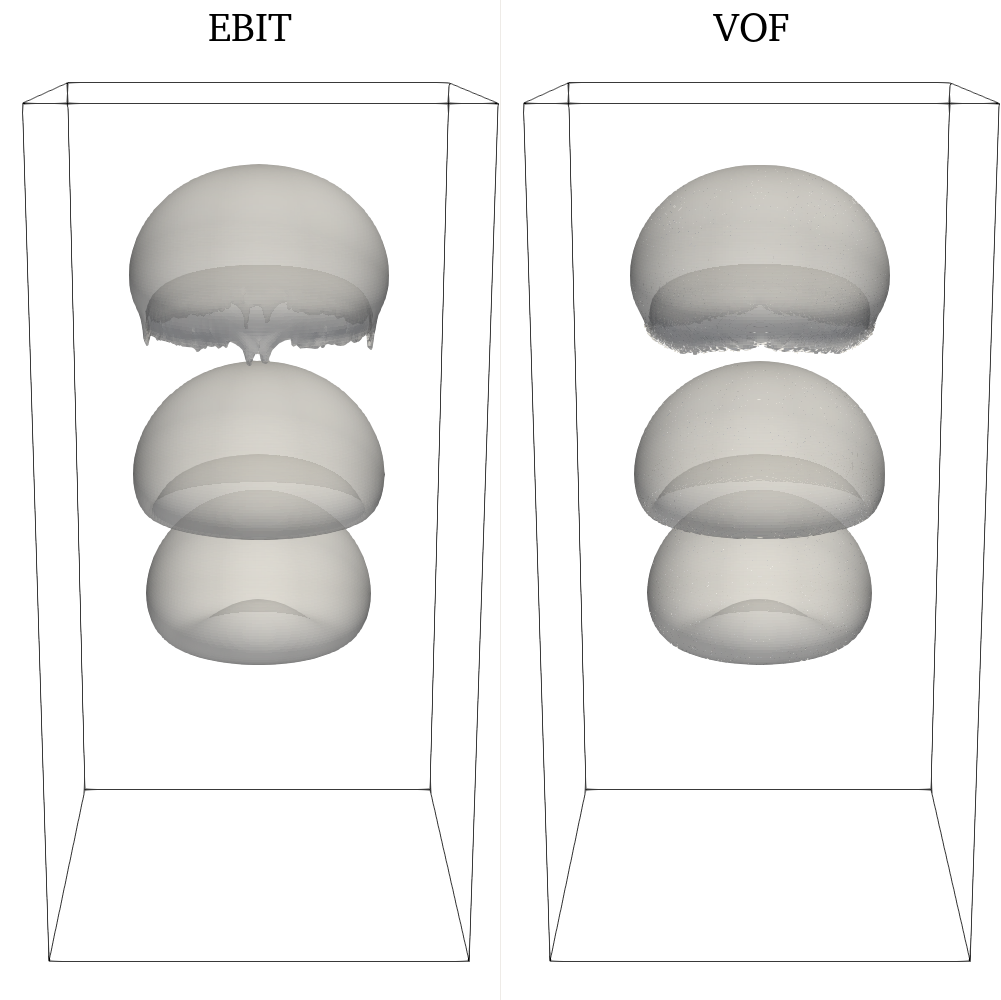}\\
(a) & (b) 
\end{tabular}
\end{center}
\caption{Rising bubble test. Interfaces at the different time instants ($128 \times 256 \times 128$): (a) test case 1, bubbles from bottom to top correspond to time instants $t = 0, 2, 4$, respectively; (b) test case 2, bubbles from bottom to top correspond to time instants $t = 1, 2, 3.5$, respectively.}
\label{Fig_rising_intfs}
\end{figure}
\begin{figure}[!htb]
\begin{center}
\begin{tabular}{cc}
\includegraphics[width=0.4\textwidth]{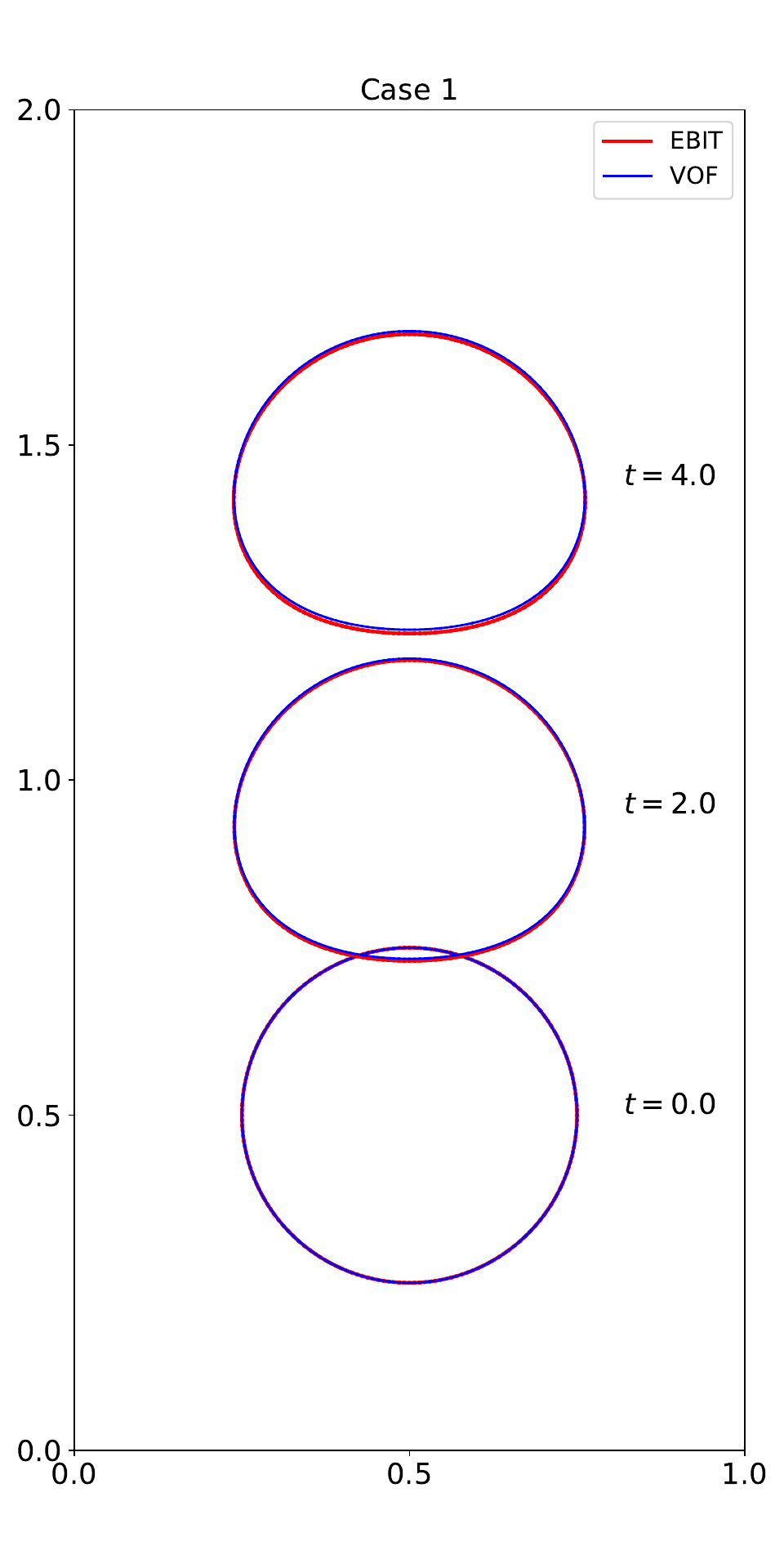} &
\includegraphics[width=0.4\textwidth]{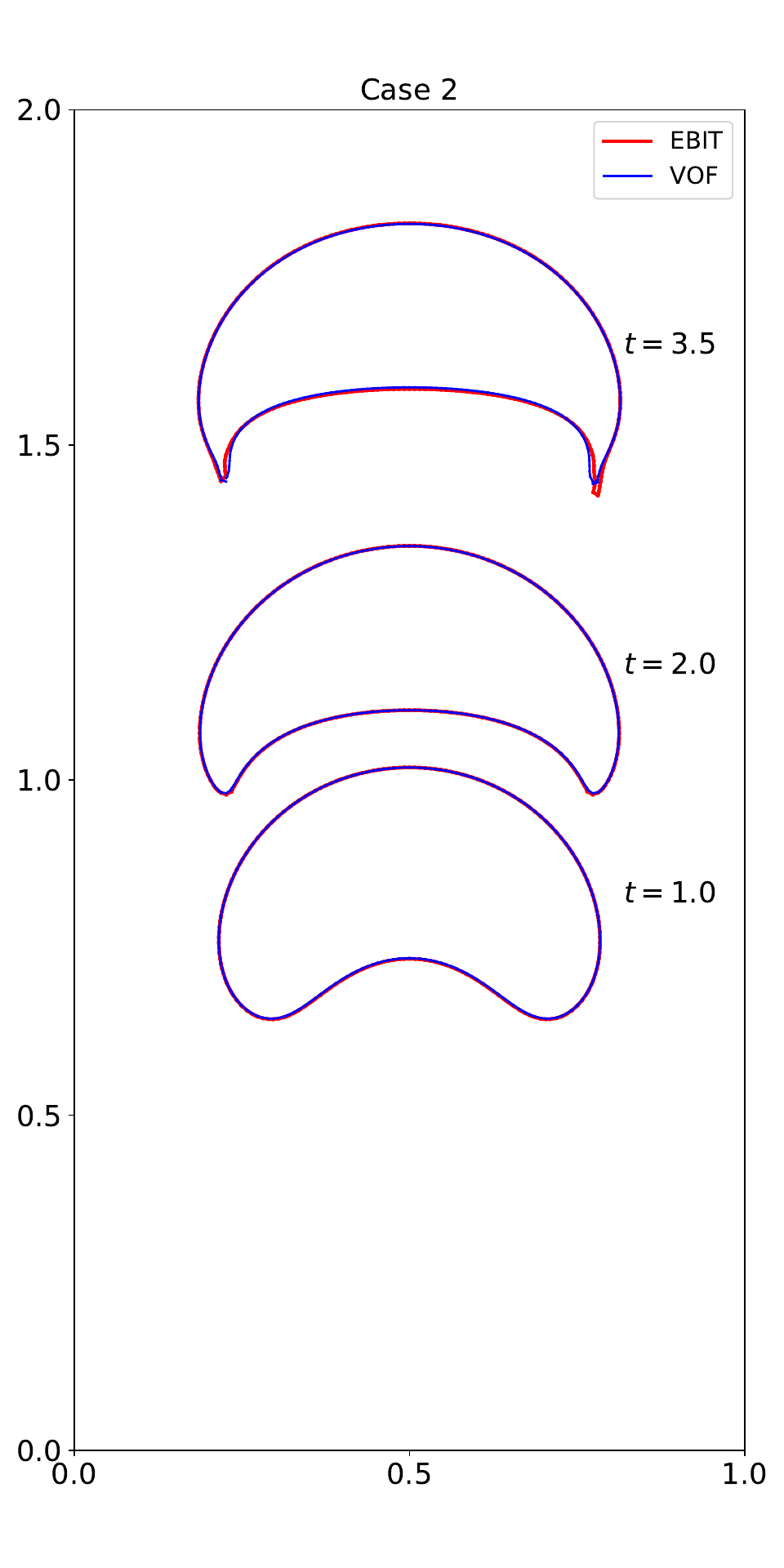}\\
(a) & (b)
\end{tabular}
\end{center}
\caption{Rising bubble test. Intersection profiles of interface at different time instants with different methods ($128 \times 256 \times 128$): (a) test case 1, bubbles from bottom to top correspond to time instants $t = 0, 2, 4$, respectively; (b) test case 2, bubbles from bottom to top correspond to time instants $t = 1, 2, 3.5$. Red solid line: EBIT, blue solid line: PLIC-VOF.}
\label{Fig_rising_profiles}
\end{figure}

\begin{figure}[!htb]
\begin{center}
\begin{tabular}{cc}
\includegraphics[width=0.45\textwidth]{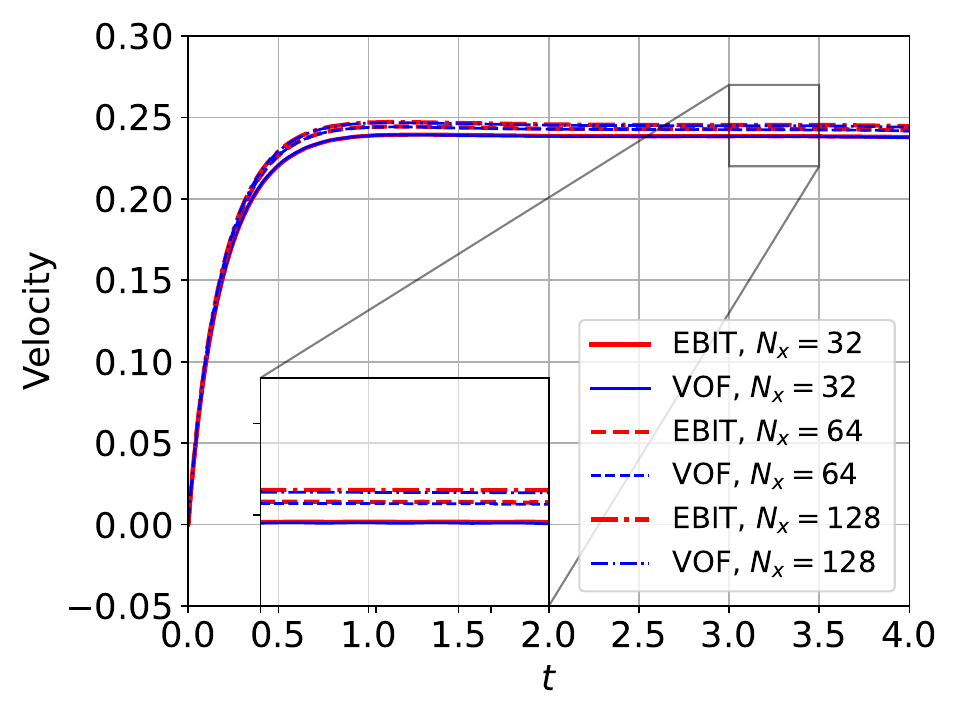} &
\includegraphics[width=0.45\textwidth]{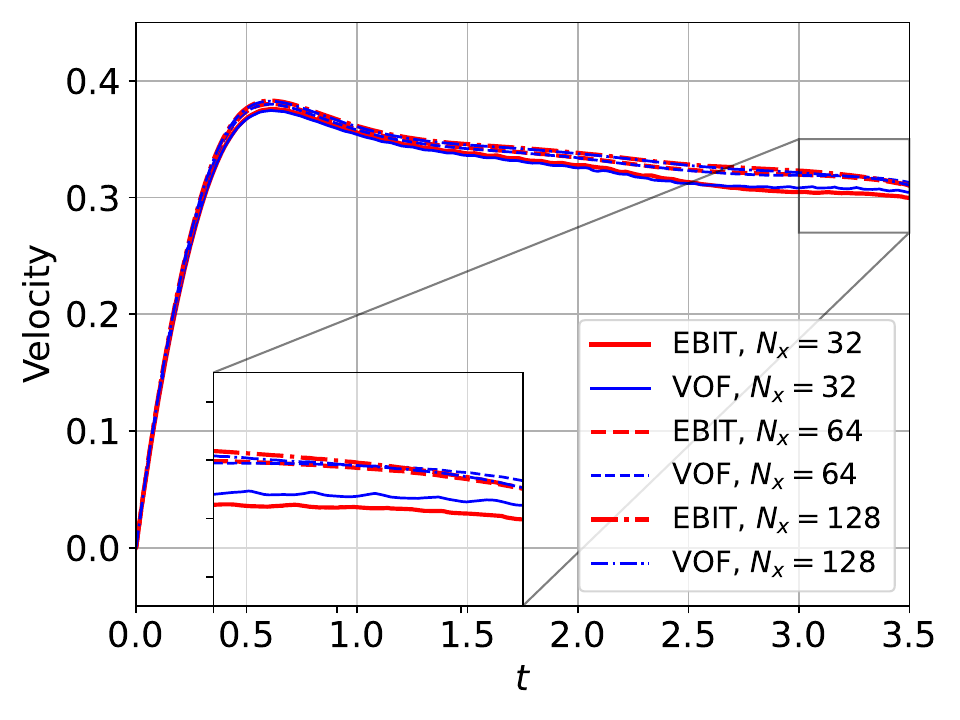}\\
(a) & (b) 
\end{tabular}
\end{center}
\caption{Rising bubble test. Bubble velocities as a function of time with different methods: (a) test case 1; (b) test case 2.}
\label{Fig_rising_vel}
\end{figure}

In the first test, the bubble ends up in the ellipsoidal regime \cite{Clift_1978}, as the surface tension forces are strong enough to maintain its integrity, and no breakup is present. 
The interfaces at different time instants $t = 0, 2, 4$ are shown in Fig.~\ref{Fig_rising_intfs}a for a mesh resolution $N_x \times N_y \times N_z = 128 \times 256 \times 128$. Owing to the strong surface tension, the bubbles only slightly deform from a spherical shape.
The intersection profiles of bubbles obtained by slicing the computation domain with the plane $x = 0.5$ are shown in Fig.~\ref{Fig_rising_profiles}a. 
Qualitatively, both the interface shapes and intersection profiles indicate a good agreement between the EBIT and PLIC-VOF methods.
It is noted that the bubble predicted by the PLIC-VOF method rises slightly
faster than that predicted by the EBIT method, exhibiting a similar tendency to that observed in our previous 2D rising bubble simulations \cite{Pan_2024_508}.
The slower bubble rise velocity obtained with the EBIT method may be attributed
to the linear element reconstruction used for color function computation, which leads to a smaller bubble volume and consequently reduced buoyancy.

For quantitative comparison, the bubble rise velocity is computed as
\begin{equation}
u_r = \frac{\sum_{i=1}^{N_\text{cell}} (1 - C_i) v_{i}}{\sum_{i=1}^{N_\text{cell}}(1 - C_i)}.
\label{Eq_rising_vel}
\end{equation}
The rising velocities as a function of time for test 1 are shown in Fig.~\ref{Fig_rising_vel}a at different mesh resolutions, demonstrating mesh convergence for both methods.
Since there is no topology change in this case, the results obtained with the EBIT and PLIC-VOF methods agree well even on a coarse mesh $N_x = 32$.
The bubble reaches a steady state shortly after release at around $t = 1$ and maintains a nearly constant rising velocity thereafter.

In the second test, the bubble falls between the skirted and the dimpled cap regimes, where a thin skirted structure forms at the trailing edge.
The interfaces at different instants $t = 1, 2, 3.5$ are presented in Fig.~\ref{Fig_rising_intfs}b for the mesh resolution $N_x \times N_y \times N_z = 128 \times 256 \times 128$. 
During the early stage of the rise, $t = 1, 2$, the bottom of the bubble deforms inward, forming an axisymmetric dimpled shape. At this stage, good qualitative agreement is observed between the EBIT and PLIC-VOF results.
As the bubble continues to rise and approaches the upper boundary of the computational domain, a skirt arises at the trailing edge.
Notably, the skirt loses axisymmetry, and anisotropy becomes apparent in
the cross-sectional planes $x = 0.5$ and $y = 0.5$.
The EBIT method yields spike-like structures near these planes, whereas the PLIC-VOF method produces shallow indentations at the corresponding locations.

For the PLIC-VOF results, the asymmetry is probably caused by the HF method
used for curvature computation.
Even at the finest mesh resolution in our simulations, only two to three cells span
the thin skirt, preventing the construction of a consistent HF stencil in this region. Consequently, a less accurate paraboloid fit is employed to compute the curvature, resulting in more prominent orientation-dependent surface tension forces at the skirt tip.
For the EBIT method, the non-axisymmetric behavior is further amplified by the orientation-dependent topology changes that occur at the skirt edge.

The corresponding intersection profiles are presented in Fig.~\ref{Fig_rising_profiles}b. 
Before the formation of the skirted tip ($t < 2$), the results with the EBIT and PLIC-VOF methods agree closely, with discrepancies even smaller than those observed in the first test case.
Despite differences at the skirt tip at later times, the bulk shape of the bubble
remains accurately captured by the EBIT method compared with the PLIC-VOF method.

The rising velocities as a function of time for the second setup 
are shown in Fig.~\ref{Fig_rising_vel}b at different mesh resolutions, again demonstrating mesh convergence for both methods.
On the coarse mesh $N_x = 32$, the discrepancy between the EBIT and PLIC-VOF results is more noticeable than in the first case. Furthermore, strong oscillations appear in the results with the PLIC-VOF method at the end of the simulation.
This discrepancy is due to the different topology changes predicted by the two methods, as illustrated in Fig.~\ref{Fig_rising_intfs}b for simulations performed at higher mesh resolutions.
 
\subsection{Bubble merging}

\begin{table}[!hbt]
\caption{Physical properties for the bubble merging test.}
\centering
\begin{tabular}{cccccccc}
\hline 
$\rho_1$& $\rho_2$& $\mu_1$ & $\mu_2$ & $g$ & $\sigma$ & $\textrm{M}$ & $\textrm{Bo}$ \\ 
\hline 
$1000 $ & $50$ & $ 18.61$ & $0.74$ & $0.98$ &$4.9$ & $1$ & $50$ \\ 
\hline 
\end{tabular}
\label{Tab_para_bubble_merging}
\end{table}

In the last test case, we investigate the merging of two rising bubbles to demonstrate the automatic topology change capability of the 3D EBIT method. 
This configuration was first proposed by Unverdi and Tryggvason \cite{Unverdi_1992_100} to illustrate the topology change algorithm in their classical front-tracking code, and has since been widely adopted to verify multiphase flow solvers \cite{Sussman_2000_162, Shin_2002_180, de_Sousa_2004_198,  Shin_2011_230}. 

We consider a cuboid computational domain defined as $[0, 2\,D] \times [0, 4\,D] \times [0, 2\,D]$, with $D=0.5$.
Free-slip boundary conditions are imposed at all domain boundaries. Both the domain size and boundary conditions are identical to those used in the 3D rising bubble test case.
Initially, two spherical bubbles with equal radius, $R=D/2$, and offset centers are positioned at the bottom of the domain.
The exact initial positions are not reported in most previous studies \cite{Unverdi_1992_100, Shin_2002_180, de_Sousa_2004_198,  Shin_2011_230}. 
Although Sussman and Puckett \cite{Sussman_2000_162} specified the offset
distances, the precise coordinates of the bubble centers were not explicitly given.
Following their setup, we offset the bubble centers in $x$-direction by one bubble radius and in $y$-direction by $2.3$ radii. Accordingly, the initial centers are located at $(0.75\,D, D, D)$ and $(1.25\,D, 2.15\,D, D)$, respectively.

The physical parameters and the corresponding dimensionless numbers used in the simulations are listed in Table~\ref{Tab_para_bubble_merging}.
Under these conditions, the bubble shape is located between the ellipsoidal and dimpled-cap regimes \cite{Clift_1978}, indicating that, in the absence of interaction, each bubble will exhibit only a slightly dimpled bottom.
The corresponding numerical setup is available at \url{https://basilisk.fr/sandbox/jieyun/test/merging_3d_ebit.c}.

\begin{figure}[!htb]
\begin{center}
\begin{tabular}{cc}
\includegraphics[width=0.45\textwidth]{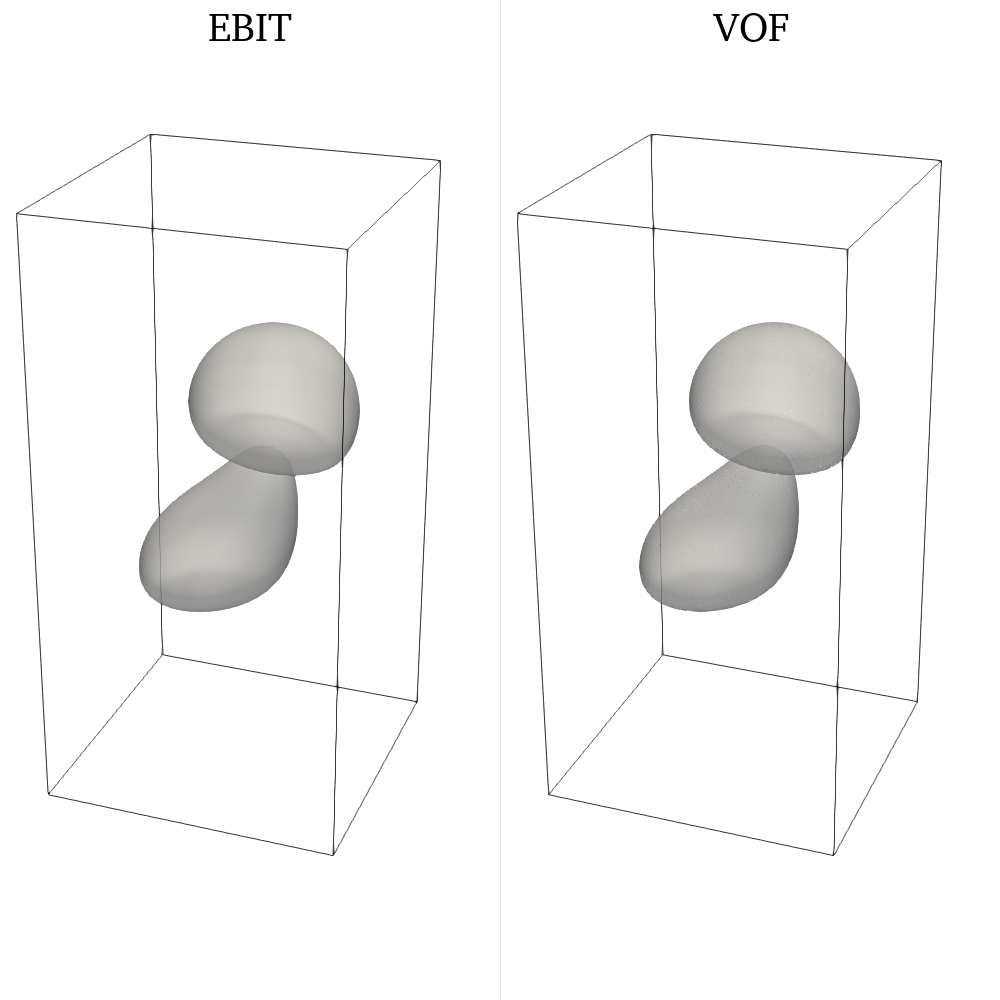} &
\includegraphics[width=0.45\textwidth]{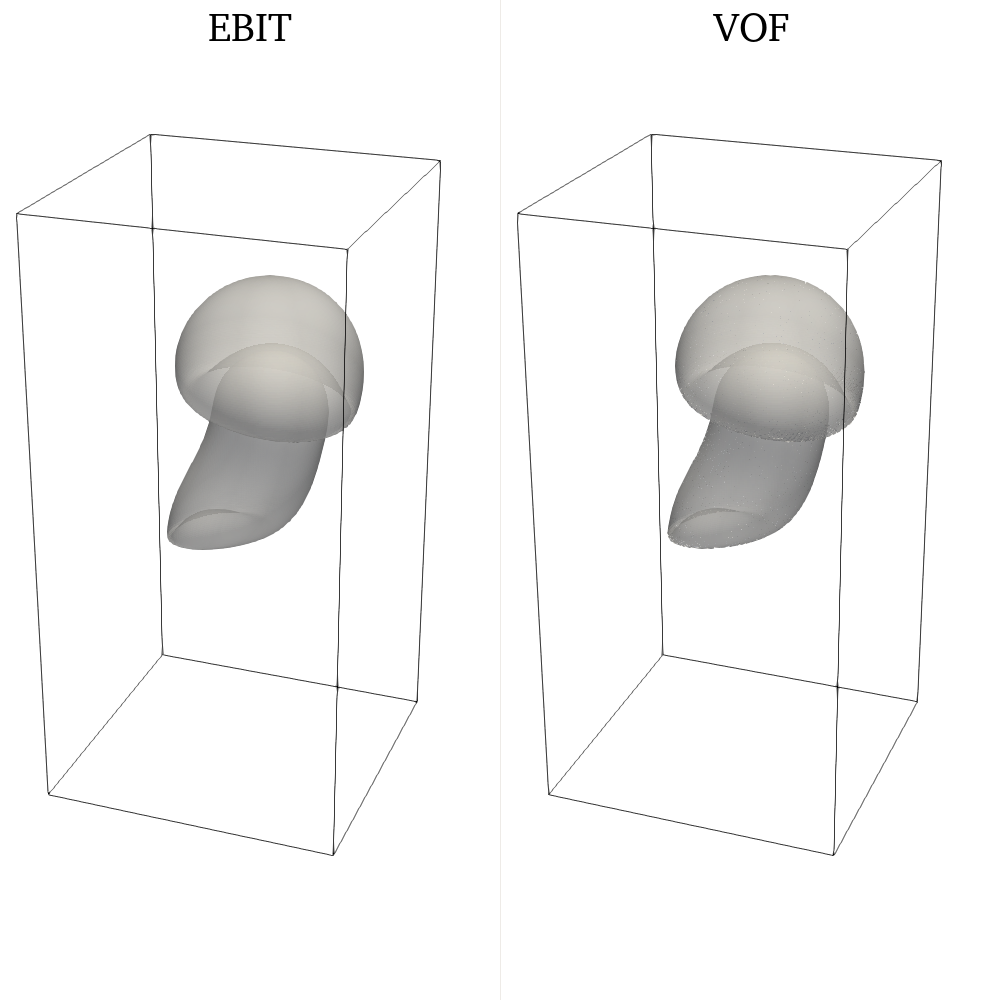}\\
(a) & (b) \\
\includegraphics[width=0.45\textwidth]{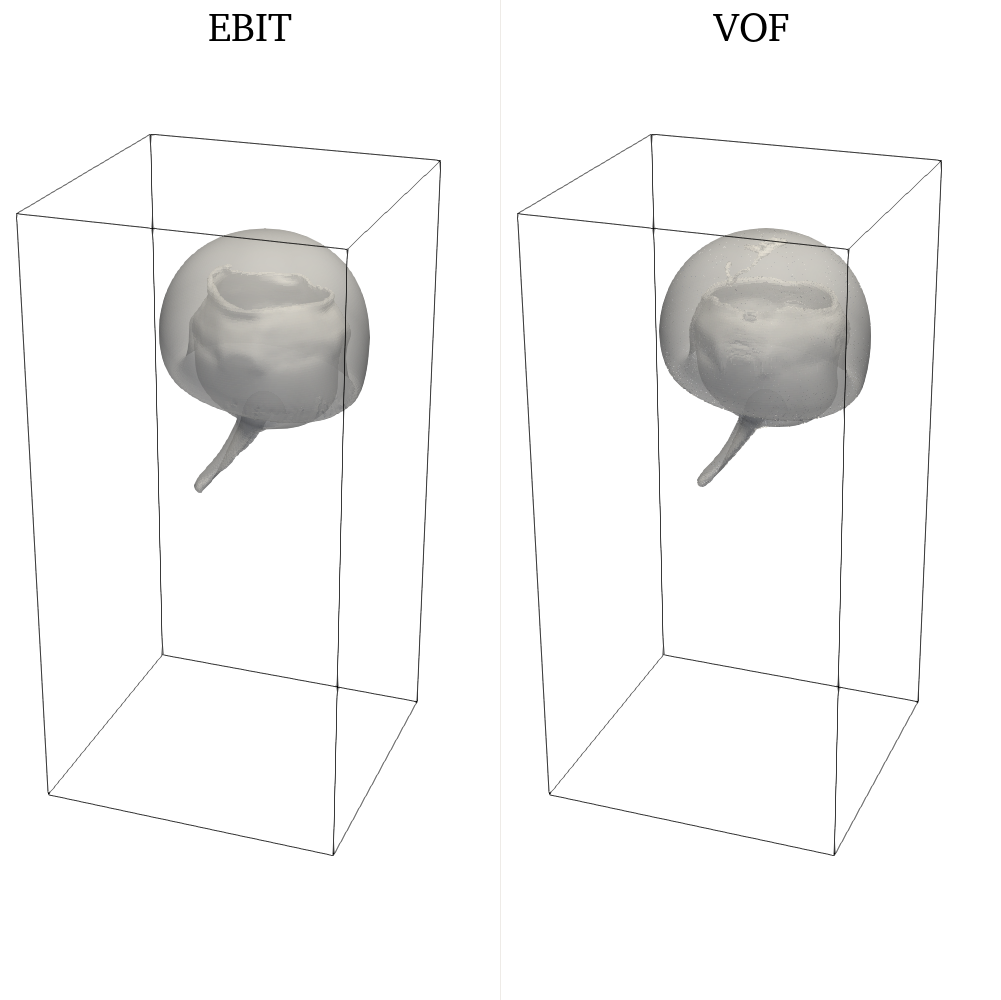} &
\includegraphics[width=0.45\textwidth]{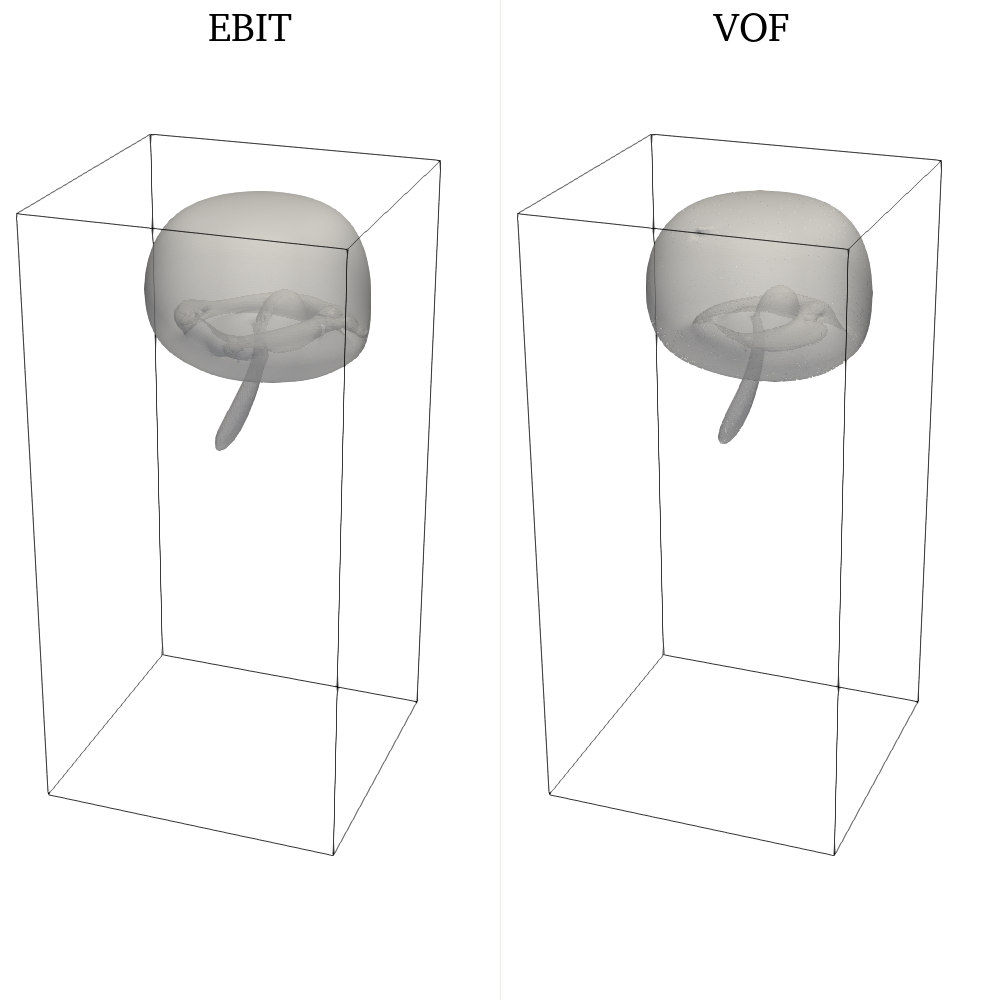}\\
(c) & (d) 
\end{tabular}
\end{center}
\caption{Bubble merging. Interfaces at different time instants with different methods ($128 \times 256 \times 128$): (a) $t = 1.0$; (b) $t = 1.5$; (c) $t = 2.0$; (d) $t = 2.5$.}
\label{Fig_bubble_merging_intfs}
\end{figure}
\begin{figure}[!htb]
\begin{center}
\begin{tabular}{cccc}
\includegraphics[width=0.22\textwidth]{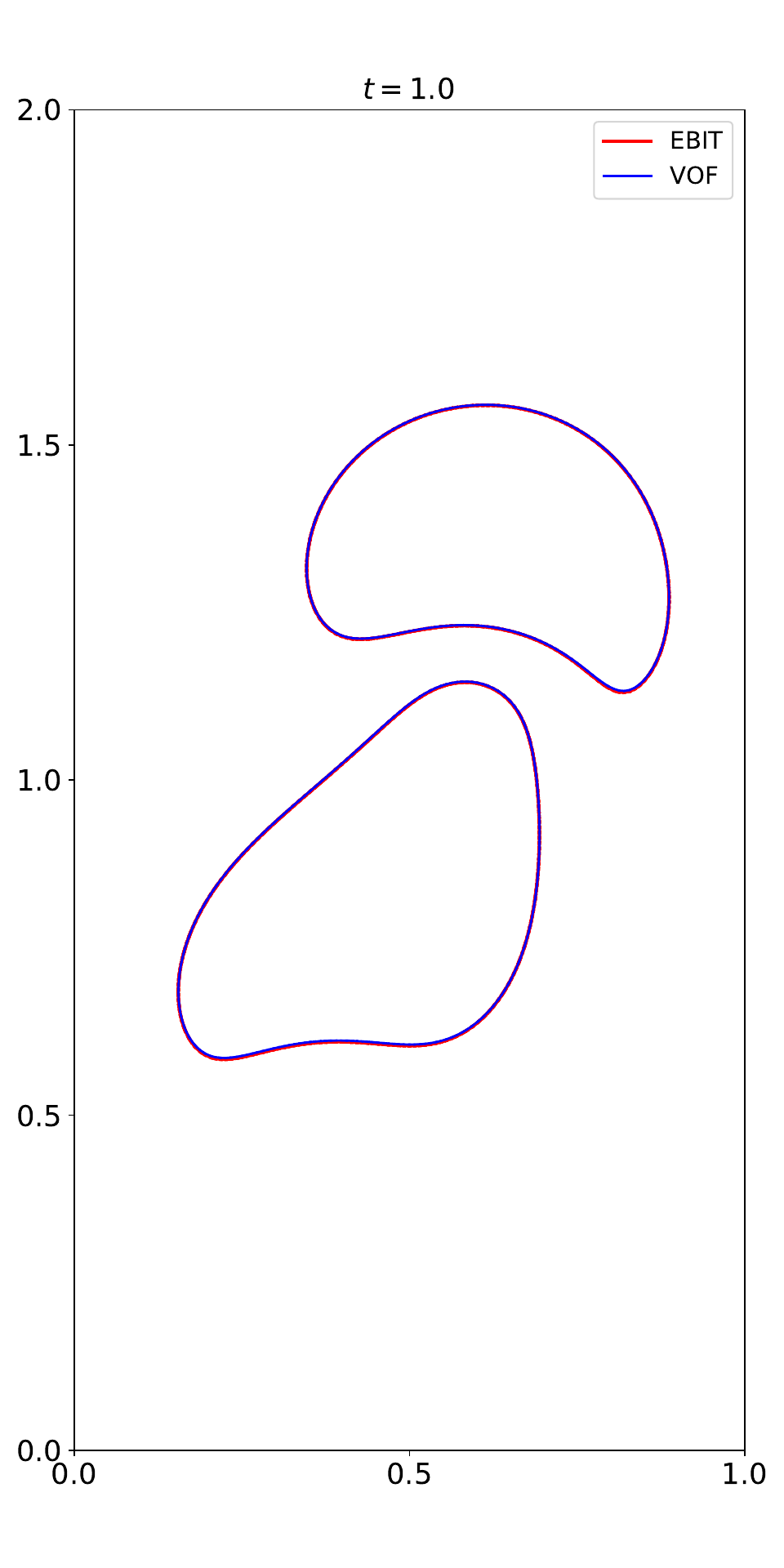} &
\includegraphics[width=0.22\textwidth]{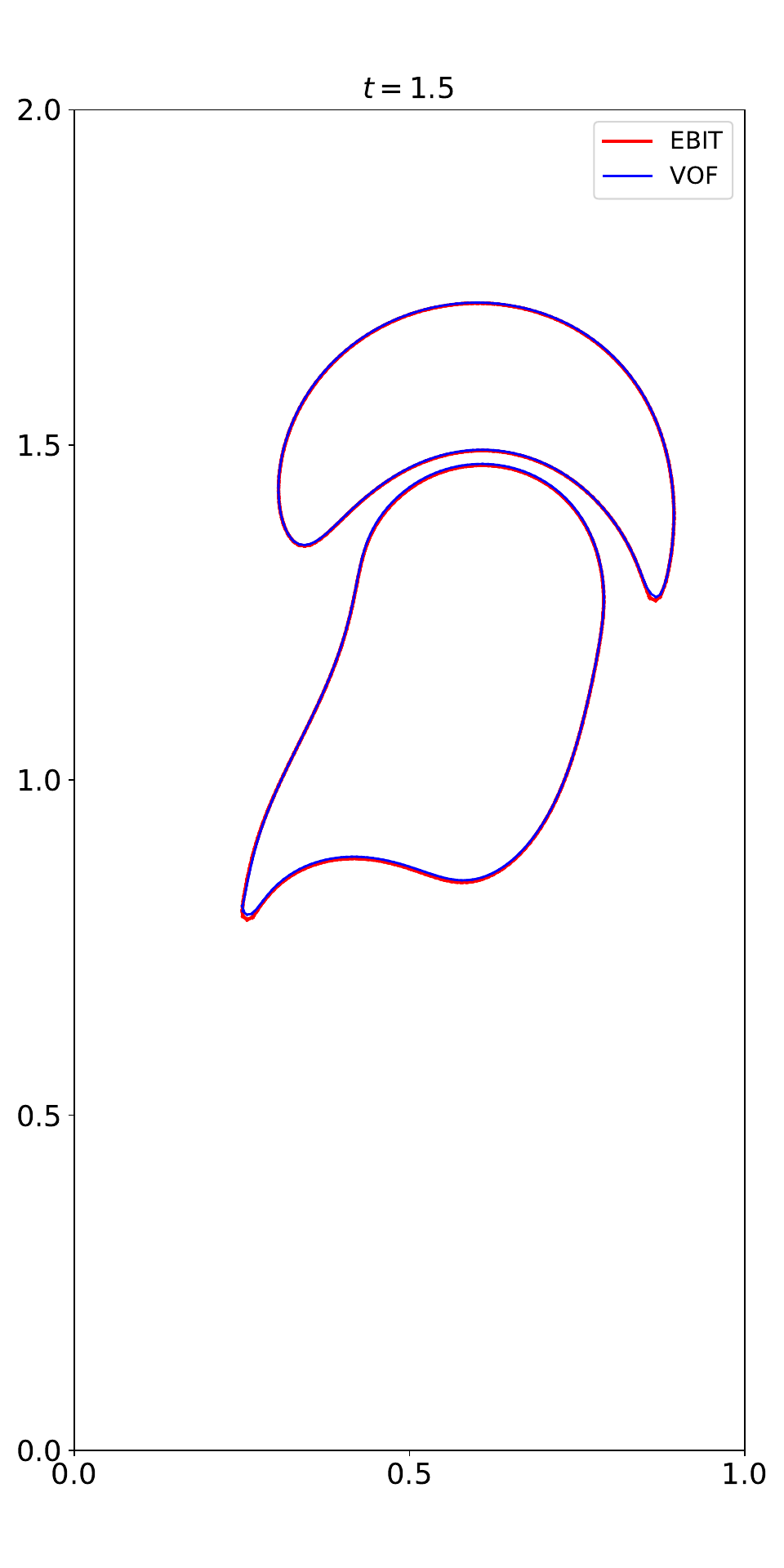}&
\includegraphics[width=0.22\textwidth]{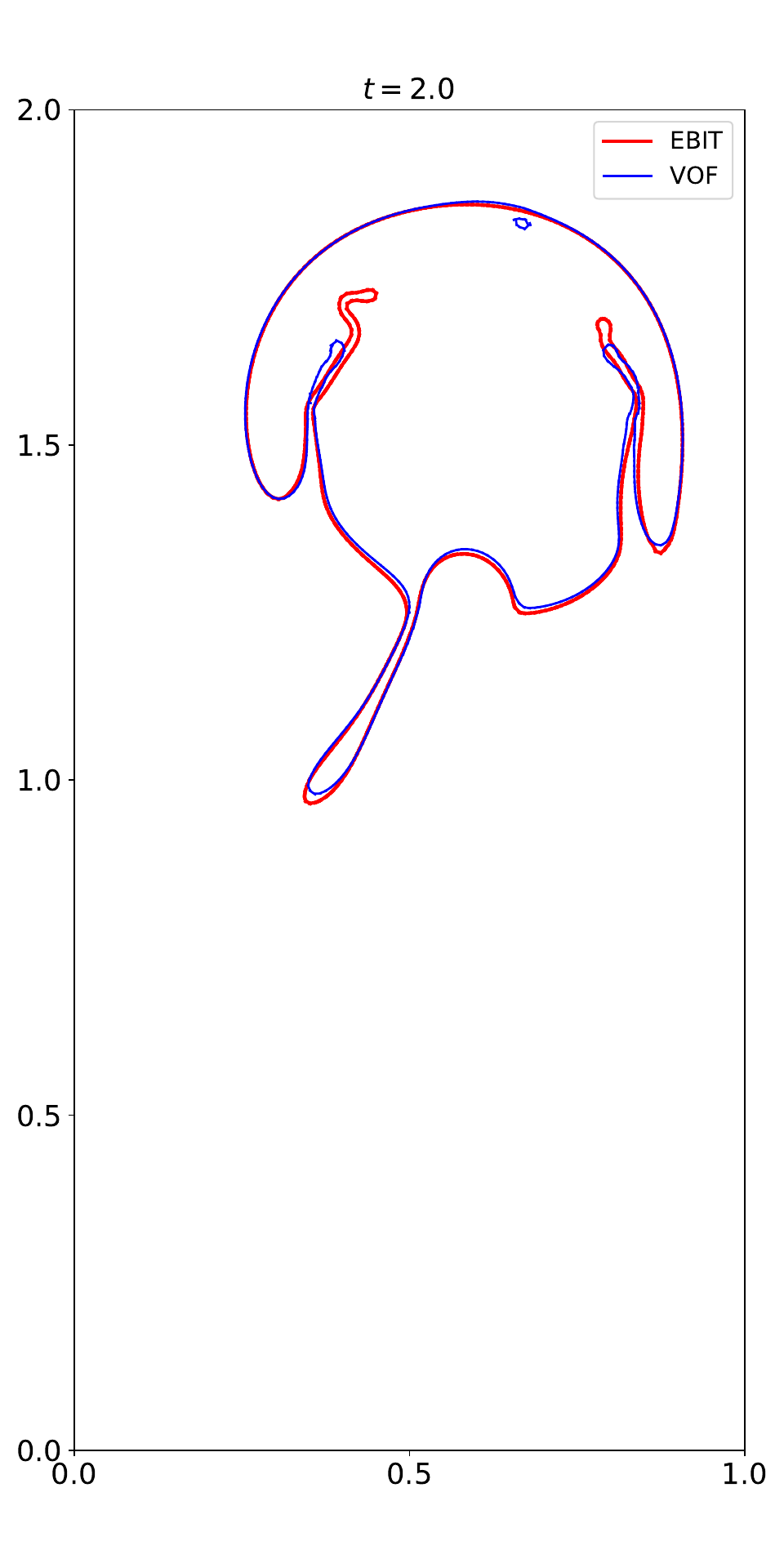} &
\includegraphics[width=0.22\textwidth]{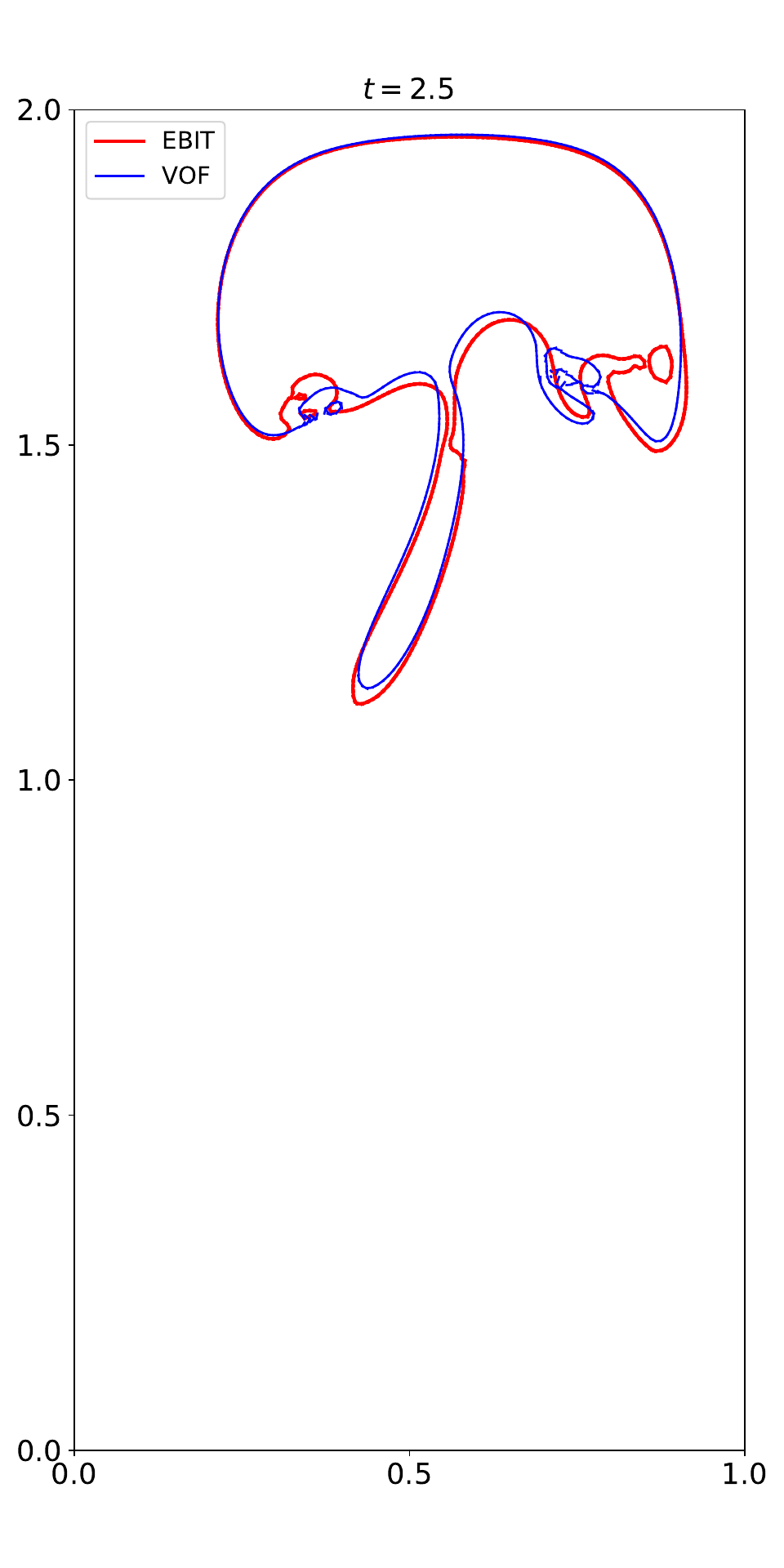}\\
(a) & (b) & (c) & (d) 
\end{tabular}
\end{center}
\caption{Bubble merging. Intersection profiles of interface at different time instants with different methods ($128 \times 256 \times 128$): (a) $t = 1.0$; (b) $t = 1.5$; (c) $t = 2.0$; (d) $t = 2.5$.}
\label{Fig_bubble_merging_profile}
\end{figure}

The interfaces at different time instants are shown in Fig.~\ref{Fig_bubble_merging_intfs} for mesh resolution $N_x \times N_y \times N_z = 128 \times 256 \times 128$.
As the bubbles rise, the bottom of the upper bubble deforms inward, while the top of the lower bubble is drawn toward the upper bubble, forming a pear-shaped
structure, as exhibited in Fig.~\ref{Fig_bubble_merging_intfs}a. 
As the bubbles approach each other, a thin liquid film develops between them before coalescence, as shown in Fig.~\ref{Fig_bubble_merging_intfs}b.
Before merging occurs, the bubble shapes predicted by the EBIT and PLIC-VOF methods agree well with each other.

The coalescence first takes place between the bottom of the upper bubble and the top of the lower bubble, after which the tip of the trapped liquid film starts to retract. At this stage, the tail of the lower bubble is significantly deformed into a cylindrical shell-like structure, as shown in Fig.~\ref{Fig_bubble_merging_intfs}c. Noticeable differences between the two methods can be observed near the tip of the liquid film. 
In the PLIC-VOF simulation, ligament-like artifacts are injected into the upper
region of the merged bubble. Overall, however, the shapes of the thin liquid film between the bubbles and the tail of the lower bubble show good qualitative agreement between the two methods.

At the end of the simulations ($t = 2.5$), the merged bubble reaches the top boundary of the computational domain. Its upper region is flattened due to the wall effect, while the retracting tip impinges on the bottom of the upper bubble. Moreover, the tail of the lower bubble has not been entirely absorbed into the merged bubble and remains as a slender shape.
Since the precise positions of the initial bubbles are not
reported in the published literature, a detailed time-resolved comparison of the interfaces is not feasible. Nonetheless, the overall bubble dynamics are qualitatively consistent with those reported by Unverdi and Tryggvason \cite{Unverdi_1992_100}.
To further support the comparison between the EBIT and PLIC-VOF methods, we 
plot the intersection profiles of the bubbles with the plane $z = 0.5$ in Fig.~\ref{Fig_bubble_merging_profile}, where discrepancies at the thin liquid film and tip region can be more clearly identified.

\begin{figure}[!htb]
\begin{center}
\begin{tabular}{c}
\includegraphics[width=0.8\textwidth]{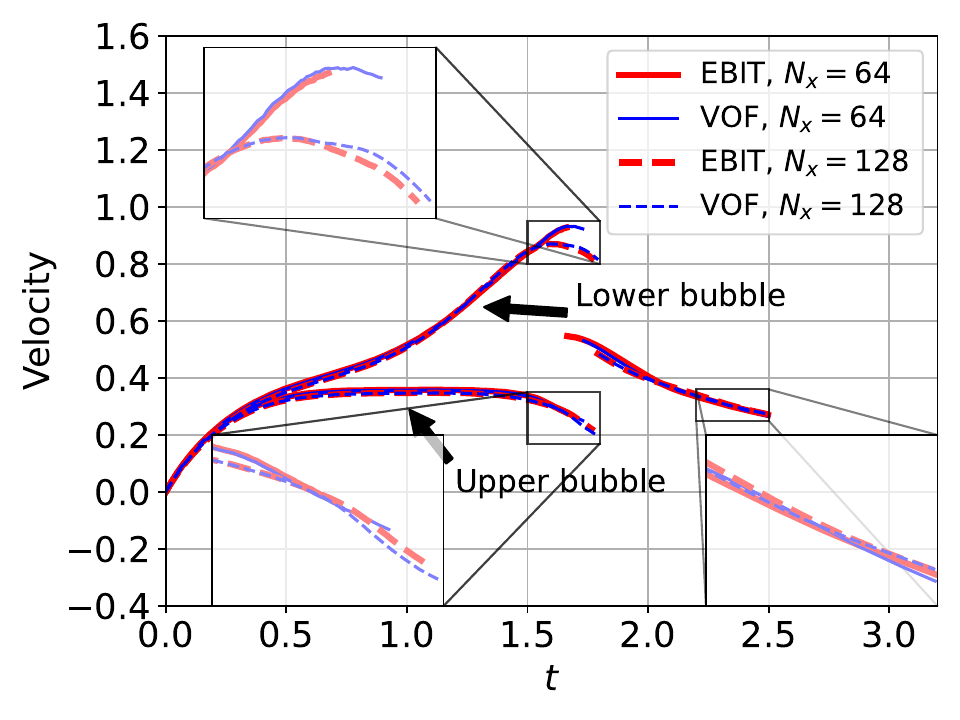}
\end{tabular}
\end{center}
\caption{Bubble merging. Bubble velocities as a function of time with different methods.}
\label{Fig_bubble_merging_vel}
\end{figure}

The rising velocities as a function of time are shown in Fig.~\ref{Fig_bubble_merging_vel} for different mesh resolutions.
We have not shown the results for the coarse mesh $N_x = 32$, as mesh convergence is not achieved even in the simpler single rising bubble case at
this resolution.
Note that the rising velocity of each bubble is computed separately when two separate gas regions can be identified. After coalescence, the rising velocity
is evaluated for the whole merged bubble, potentially including numerical artifacts, using Eq.~\ref{Eq_rising_vel}.
Therefore, the onset of bubble merging can be identified by a discontinuity in the velocity curves. 

The acceleration of the lower bubble and the deceleration of the upper bubble due to their interaction can be observed. Before merging, the velocities predicted by
the EBIT and PLIC-VOF methods exhibit excellent agreement. 
Once topology changes occur (around $t=1.7$), small discrepancies arise for
the coarse mesh resolution $N_x = 64$, due to the under-resolved thin film between the bubbles.
As the mesh resolution is increased to $N_x = 128$, the rising velocities predicted by the two methods agree closely, despite remarkable differences in tip retraction dynamics of the trapped liquid film (see Figs.~\ref{Fig_bubble_merging_intfs}c and d).

At a given mesh resolution, only minor differences are observed in the predicted 
onsets of topology change between the EBIT and PLIC-VOF methods, and
this discrepancy diminishes as the mesh is refined.
For each method, the onset of merging is delayed with increasing mesh resolution. This behavior is expected, since in both the current EBIT and PLIC-VOF
frameworks, topology changes arise only when two interface segments reside
within the same cell, implying a mesh-dependent prediction of the merging event.

In summary, this test case clearly demonstrates the efficacy of the EBIT method in
handling bubble interactions and automatic topology changes in 3D.
At the same time, the mesh-dependent prediction of the merging onset
highlights a limitation of the current EBIT method and suggests a direction 
for further improvement.

\section{Conclusions}

We have presented a straightforward extension of the EBIT method to 3D. 
The proposed algorithm relies on a split scheme for the interface advection,
where a 3D advection problem along one direction is decomposed into two 2D advection problems on the two corresponding cell faces.
This strategy enables the direct reuse of the 2D circle fit algorithm for interface reconstruction.
The 3D interface topology is represented by the color vertex distribution on 
the six faces of each cell, thereby preserving the evolution rules of the 2D color vertex field.
This simplified extension algorithm may introduce internal topology ambiguities,
even though face ambiguities are addressed using the color vertex defined at the face center.
In such cases, the configuration consistent with the marker arrangement is uniquely selected.
Nevertheless, this extension approach, together with the close correspondence between the 2D and 3D data structures, significantly reduces both the algorithmic and implementation complexity.

For coupling with the Navier--Stokes equations, volume fractions are first
computed using the F2V algorithm, from which the physical properties are calculated, and the surface tension forces are evaluated using the Height-Function method.
The 3D EBIT method has been implemented in the free Basilisk framework, with the documented source code available in the online repository.
Its performance is verified through a series of kinematic and dynamic benchmark problems.
Comparisons with the PLIC-VOF method available in Basilisk demonstrate
good agreement across all test cases, confirming the efficacy and robustness of
the proposed extension.

The use of quad-/octree AMR is particularly beneficial when
studying flows with a wide range of scales, especially when combined with high-performance computing.
The compatibility of EBIT with AMR, together with its scalability, 
makes it well-suited for multiscale multiphase flow simulations and establishes a solid foundation for future improvements.

As anticipated in the 2D EBIT study, potential improvements can be categorized
into kinematic and dynamic enhancements.
On the kinematic side, relaxing the restriction on the number of markers per
edge would allow the representation of more complex sub-grid structures and
enable fully controllable topology changes, bringing EBIT on par with traditional Front-Tracking methods in this regard. 
Thanks to the 3D extension algorithm presented here, such improvements in 2D can be readily generalized to 3D.
However, resolving internal topology ambiguities becomes more difficult, as the
number of admissible configurations increases when multiple markers per edge
are allowed.

On the dynamic side, a promising extension would be the direct computation of
surface tension forces from the EBIT representation, bypassing the intermediate
volume-fraction field. This could improve the accuracy of SGS dynamics,
including triple points, small droplets and bubbles, and contact line motion.
However, such a dynamic extension is significantly more challenging than
the kinematic improvements, as it cannot be simplified through directional splitting.
Regarding surface tension models, further accuracy may be achieved by 
implementing the momentum-conserving, well-balanced methods proposed by Popinet and Al-Saud et al. \cite{abu2018conservative}. 
However, these formulations are currently fully developed only for 2D and axisymmetric configurations, and their extension to 3D is being actively pursued \cite{Gennari_2026}.
Finally, recent advances in contact line modeling on complex 3D solid surfaces within the FT framework \cite{Zeng_2025_539}
could be incorporated into the EBIT framework, further broadening its applicability to challenging multiphase flow problems.

\section{CRediT authorship contribution statement}

\textbf{J. Pan}: Writing - review \& editing, Writing - original draft, Validation, 
Software, Methodology, Formal analysis, Data curation, Conceptualization.

\textbf{T. Long}: Writing - review \& editing, Software, Methodology, Formal analysis, Conceptualization.

\textbf{R. Scardovelli}: Writing - review \& editing, Writing - original draft, 
Software, Methodology, Formal analysis, Conceptualization.

\textbf{S. Popinet}: Writing - review \& editing, Writing - original draft,
Software, Methodology, Formal analysis

\textbf{S. Zaleski}: Writing - review \& editing, Writing - original draft, 
Supervision, Resources, Project administration, Funding acquisition, 
Methodology, Formal analysis, Conceptualization.

\section{Declaration of competing interest}
The authors declare that they have no known competing financial interests or personal relationships that could have appeared to influence the work reported in this paper.

\section{Acknowledgements}
St\'{e}phane Zaleski and St\'{e}phane Popinet recall meeting Sergei Semushin in March 1995 and learning about his method. They thank him for the explanation of the method.
This project has received funding from the European Research Council (ERC) under the European Union's Horizon 2020 research and innovation programme (grant agreement number 883849). We thank the European PRACE group, the Swiss supercomputing agency CSCS, the French national GENCI supercomputing agency and the relevant supercomputer centers for their grants of CPU time on massively parallel machines, and their teams for assistance and the use of Irene-Rome at TGCC.






 \bibliographystyle{model1-num-names}
\bibliography{multiphase-jieyun,multiphase-stephane}
\end{document}